\documentstyle[12pt,epsf]{article} 
\catcode`@=11
\newcount\@tempcntc
\def\@citex[#1]#2{\if@filesw\immediate\write\@auxout{\string\citation{#2}}\fi
  \@tempcnta\z@\@tempcntb\m@ne\def\@citea{}\@cite{\@for\@citeb:=#2\do
    {\@ifundefined
       {b@\@citeb}{\@citeo\@tempcntb\m@ne\@citea\def\@citea{,}{\bf ?}\@warning
       {Citation `\@citeb' on page \thepage \space undefined}}%
    {\setbox\z@\hbox{\global\@tempcntc0\csname b@\@citeb\endcsname\relax}%
     \ifnum\@tempcntc=\z@ \@citeo\@tempcntb\m@ne
       \@citea\def\@citea{,}\hbox{\csname b@\@citeb\endcsname}%
     \else
      \advance\@tempcntb\@ne
      \ifnum\@tempcntb=\@tempcntc
      \else\advance\@tempcntb\m@ne\@citeo
      \@tempcnta\@tempcntc\@tempcntb\@tempcntc\fi\fi}}\@citeo}{#1}}
\def\@citeo{\ifnum\@tempcnta>\@tempcntb\else\@citea\def\@citea{,}%
  \ifnum\@tempcnta=\@tempcntb\the\@tempcnta\else
   {\advance\@tempcnta\@ne\ifnum\@tempcnta=\@tempcntb \else \def\@citea{--}\fi
    \advance\@tempcnta\m@ne\the\@tempcnta\@citea\the\@tempcntb}\fi\fi}
\catcode`@=12
\def\theequation{\arabic{section}.\arabic{equation}}

\begin{document}
\vskip -1.5cm
\begin{flushright}
ANL-HEP-PR-00-030 \\[-0.1cm]
CERN-TH/2000-082\\[-0.1cm]
EFI-2000-10 \\[-0.1cm]
FERMILAB-Pub-00/063-T\\[-0.1cm]
hep-ph/0003180\\[-0.1cm]
March 2000
\end{flushright}

\begin{center}
{\Large {\bf Renormalization-Group-Improved Effective Potential\\[0.3cm]
\hspace{-0.15cm}for the 
MSSM  Higgs Sector with Explicit CP Violation}}\\[1.0cm]
{\large M. Carena$^{\,a,b}$, J. Ellis$^{\, a}$, A. Pilaftsis$^{\, a,b}$ 
and C.E.M. Wagner$^{\,a,c,d}$}\\[0.4cm]
$^a${\em Theory Division, CERN, CH-1211 Geneva 23, Switzerland}\\[0.2cm]
$^b${\em Fermilab, P.O. Box 500, Batavia IL 60510, U.S.A.}\\[0.2cm]
$^c${\em High Energy Physics Division, Argonne National Lab., Argonne
  IL 60439, U.S.A.}\\[0.2cm]
$^d${\em Enrico Fermi Institute, University of Chicago, 5640 Ellis Ave.,
Chicago
  IL 60637, U.S.A.}
\end{center}
\vskip0.7cm \centerline{\bf ABSTRACT} 
We     perform    a     systematic    study     of     the    one-loop
renormalization-group-improved  effective  potential  of  the  minimal
supersymmetric extension  of the  Standard Model (MSSM),  including CP
violation induced  radiatively by soft  trilinear interactions related
to  squarks of  the third  generation.  We calculate  the charged  and
neutral  Higgs-boson  masses  and  couplings, including  the  two-loop
logarithmic corrections that  arise from QCD effects, as well as those
associated with  the top- and  bottom-quark Yukawa couplings.  We also
include  the potentially  large  two-loop non-logarithmic  corrections
induced  by one-loop threshold  effects on  the top-  and bottom-quark
Yukawa  couplings,  due  to  the  decoupling  of  the third-generation
squarks.  Within this minimal  CP-violating framework, the charged and
neutral  Higgs sectors become  intimately related  to one  another and
therefore  require a  unified  treatment.   In the  limit  of a  large
charged Higgs-boson mass, $M_{H^{\pm}}  \gg M_Z$, the lightest neutral
Higgs  boson  resembles  that  in  the Standard  Model  (SM),  and  CP
violation occurs  only in the  heavy Higgs sector. Our  analysis shows
that sizeable radiative effects of CP violation in the Higgs sector of
the MSSM may lead to significant modifications of previous studies for
Higgs-boson searches at LEP~2, the Tevatron and the LHC. In particular,
CP violation could enable a relatively light Higgs boson to escape
detection at LEP~2.

\newpage

\setcounter{equation}{0}
\section{Introduction}

The  violation of CP  was first  observed in  the neutral  kaon system
\cite{CPkaon},  where  a  deviation  from  the  superweak  theory  has
recently  been confirmed  \cite{eps},  and CP  violation in  $B$-meson
decays~\cite{CPB}     is      strongly     suggested     by     recent
experiments~\cite{Bexpts}.  In  addition to its  interest for particle
physics,   CP   non-conservation  provides   a   key  ingredient   for
cosmological  baryogenesis,  namely   for  explaining  the  underlying
mechanism  which caused  matter to  dominate over  anti-matter  in our
observable  universe~\cite{ADS}. Although a  fundamental understanding
of the origin of CP violation  is still lacking, most of the scenarios
proposed in  the existing literature indicate  that Higgs interactions
play a key role in mediating CP violation.  For instance, CP violation
is  broken explicitly  in the  Standard Model  (SM) by  complex Yukawa
couplings of the  Higgs boson to quarks.  Another  appealing scheme of
CP violation occurs in models  with an extended Higgs sector, in which
the CP  symmetry of the theory  is broken spontaneously  by the ground
state  of  the Higgs  potential  \cite{TDLee}.  Supersymmetric  (SUSY)
theories,  including  the  minimal  supersymmetric  extension  of  the
Standard Model (MSSM), predict an extended Higgs sector, and therefore
may  realize either  or both  the above  two schemes  of  explicit and
spontaneous CP violation.  However, the Higgs potential of the MSSM is
invariant under CP at the  tree level, and any explicit or spontaneous
breakdown of CP symmetry can arise only via radiative corrections. The
case of purely spontaneous CP violation in the MSSM \cite{NM} leads to
an unacceptably  light CP-odd scalar  \cite{APNH}, as a result  of the
Georgi-Pais theorem \cite{GP}, and hence  such a scenario is ruled out
experimentally.

It  has  recently  been   shown~\cite{APLB}  that  the  tree-level  CP
invariance of  the MSSM  Higgs potential may  be violated  sizeably by
loop effects involving soft CP-violating trilinear interactions of the
Higgs bosons  to top and  bottom squarks.  A  detailed study~\cite{PW}
has shown  that significant CP-violating effects of  level crossing in
the Higgs  sector can take  place in such  a minimal SUSY  scenario of
explicit CP violation, which may  lead to drastic modifications of the
tree-level   couplings   of    the   Higgs   particles   to   fermions
\cite{PW,Demir}  and to  the $W^\pm$  and $Z$  bosons  \cite{PW}.  The
latter  can  have   important  phenomenological  consequences  on  the
production rates of the lightest Higgs particle, even though the upper
bound on its mass was found~\cite{PW} to be very similar to that found
previously              in              the              CP-conserving
case~\cite{Mh,HH,KYS,CEQR,CEQW,CQW,HHW,Zhang}.   The MSSM  predicts an
upper  bound on  the lightest  Higgs boson  mass of  approximately 110
(130) GeV  for small  (large)  values  of the  ratio  of Higgs  vacuum
expectation values $\tan\beta \approx  2$ $(>15)$.  On the other hand,
experiments at LEP~2, running  at center-of-mass energies $\sqrt{s} =$
196--202 GeV,  have placed a  severe lower bound of  approximately 108
GeV on the mass of the SM Higgs boson~\cite{expMH}.  LEP~2 is expected
to run  at center-of-mass energies up  to $\sqrt{s} =  206$ GeV during
the year  2000.  Consequently, for  stop masses smaller than  1~TeV, a
significant portion of the  parameter space spanned by $\tan\beta$ and
the CP-odd scalar mass $M_A$  can be tested for the CP-conserving case
in this  next round of  experiments at LEP~2~\cite{PW}.   However, the
explorable region of parameters is  smaller for larger amounts of stop
mixing and/or larger  CP-violating phases.  A decisive test  of such a
scenario can  only be provided  by the upgraded Tevatron  collider and
the LHC.

The earlier study of the renormalization-group (RG) improved effective
potential  of the  MSSM with  explicit CP  violation  in~\cite{PW} was
based on an expansion of the Higgs quartic couplings in inverse powers
of  the arithmetic  average  of  stop and  sbottom  masses, under  the
assumption that the mass splittings of the left- and right-handed stop
and  sbottom masses are small.   The  mass expansion  of the  one-loop
effective potential  was truncated  up to renormalizable  operators of
dimension  four.   Although  the  above approach  captures  the  basic
qualitative  features of the  underlying dynamics  under study,  it is
known~\cite{CQW,PW} that  such a  mass expansion has  limitations when
the  third-generation  squark mixing  is  large.   Since the  dominant
CP-violating  loop contributions to  the effective  Higgs-boson masses
and  mixing angles  occur  for large  values  of the  third-generation
squark-mixing  parameters, it  is  necessary to  provide  a more  {
  complete} one-loop computation of the effective MSSM Higgs potential
with   explicit  radiative  breaking   of  CP   invariance,  including
non-renormalizable  operators below the  heavy scalar-quark  scale and
without       resorting       to       any       other       kinematic
approximations.\footnote[1]{We    recall    that   the    diagrammatic
  computation   of  scalar-pseudoscalar   transitions  is   the  focus
  of~\cite{APLB}, whilst  sbottom contributions and  relevant $D$-term
  effects   on   the    effective   potential   are   not   considered
  in~\cite{Demir}.}   

To this end, we consider  here the two-loop leading logarithms induced
by top- and bottom-quark Yukawa  couplings as well as those associated
with QCD corrections,  by means of RG methods.   In the calculation of
the  RG-improved  effective potential,  we  also  include the  leading
logarithms  generated   by  one-loop  gaugino   and  higgsino  quantum
effects~\cite{HH}.   Finally,  we   implement  the  potentially  large
two-loop corrections that are induced by the one-loop stop and sbottom
thresholds in  the top- and  bottom-quark Yukawa couplings,  which may
become particularly relevant in  the large-$\tan\beta$ regime.  On the
basis of  the RG-improved effective potential,  we present predictions
for the Higgs-boson mass spectrum and the effective Higgs couplings to
fermions and gauge  bosons.  The results of the  analysis are compared
with those obtained in  the CP-conserving case~\cite{CQW,CMW} and also
with  the  results obtained  by  truncating  the one-loop  RG-improved
effective potential up to renormalizable operators~\cite{PW}.

In this analysis,  it is important to consider  the constraints on the
low-energy  CP-violating parameters  of the  MSSM that  originate from
experimental upper limits on  the electron~\cite{commins} and
neutron~\cite{PGH} electric dipole moments (EDMs)
\cite{EFN,DGH,PN,GD,IN,EFlores,Bartl,AF,DDLPD,CKP,APino}. Most  of the
EDM  constraints affect the  CP-violating couplings  of the  first two
generations \cite{EFN,DGH}.  Thus, making the first two generations of
squarks  rather heavy,  much above  the TeV  scale \cite{PN,GD},  is a
possibility that can drastically  reduce one-loop contributions to the
neutron  EDM,  without  suppressing  the CP-violating  phases  of  the
theory. Another  interesting possibility for avoiding  any possible CP
crisis in the MSSM is to arrange for cancellations among the different
EDM terms, either at the level of short-distance diagrams \cite{IN} or
(for the  neutron EDM) at  the level of the  strong-interaction matrix
elements    for   operators    with   $s$,    $u$   and    $d$   quark
flavours~\cite{EFlores,Bartl}.   Alternatively,  one  might require  a
specific  form  of  non-universality  in  the  soft  trilinear  Yukawa
couplings \cite{AF,CKP}.

However, third-generation squarks can  also give rise by themselves to
observable  effects  on the  electron  and  neutron  EDMs through  the
three-gluon operator  \cite{DDLPD}, through the  effective coupling of
the `CP-odd' Higgs  boson to the gauge bosons  \cite{CKP}, and through
two-loop  gaugino/higgsino-mediated  EDM  graphs~\cite{APino}.   These
different  EDM contributions  of the  third generation  can  also have
different  signs and  add destructively  to the  electron  and neutron
EDMs.  In  our phenomenological discussion,  we take into  account the
relevant EDM constraints related to the CP-violating parameters of the
stop and sbottom sectors.

This  paper is organized  as follows:  in Section  2 we  calculate the
complete  one-loop CP-violating  effective potential,  and  derive the
analytic expressions for the effective charged and neutral Higgs-boson
mass matrices.   Technical details  are given in  Appendices A  and B.
Section  3  describes  our  approach to  determining  the  RG-improved
Higgs-boson  mass  matrices,  after  including  the  leading  two-loop
logarithms associated  with Yukawa and  QCD corrections. Section  4 is
devoted  to the  calculation of  the effective  top-  and bottom-quark
Yukawa  couplings,   in  which  one-loop  threshold   effects  of  the
third-generation squarks  are implemented.   In Section 5,  we discuss
the  phenomenological   implications  of  representative  CP-violating
scenarios compatible with EDM constraints for direct Higgs searches at
LEP~2 and the upgraded Tevatron  collider. We also compare the results
of  our   analysis  with  those  obtained   using  the  mass-expansion
method~\cite{PW}.  Finally, in Section 6 we summarize our conclusions.

\setcounter{equation}{0}
\section{CP-Violating One-Loop Effective Potential}

In this Section,  we first describe the basic  low-energy structure of
the  MSSM that contains  explicit CP-violating  sources, such  as soft
CP-violating  trilinear interactions.  Then  we calculate  the general
one-loop    CP-violating   effective   potential.     Finally,   after
implementing the minimization tadpole  conditions related to the Higgs
ground state, we derive  the effective charged and neutral Higgs-boson
mass matrices.

CP  violation is    introduced  into  the  MSSM  through  the    Higgs
superpotential and the soft supersymmetry-breaking Lagrangian:
\begin{eqnarray}
  \label{Wpot}
W &=& h_l\, \widehat{H}^T_1 i\tau_2 \widehat{L} \widehat{E}\: +\: 
h_d\, \widehat{H}^T_1 i\tau_2 \widehat{Q} \widehat{D}\: +\: 
h_u\, \widehat{Q}^T i\tau_2 \widehat{H}_2 \widehat{U}\: -\:
\mu\,\widehat{H}^T_1 i\tau_2 \widehat{H}_2\, ,\\
  \label{Lsoft}
-{\cal L}_{\rm soft} & = & 
-\, \frac{1}{2}\, \Big( m_{\tilde{g}}\,
\lambda^a_{\tilde{g}}\lambda^a_{\tilde{g}}\:+\: 
m_{\widetilde{W}}\,\lambda^i_{\widetilde{W}}\lambda^i_{\widetilde{W}}\: 
+\: m_{\widetilde{B}}\,\lambda_{\widetilde{B}}\lambda_{\widetilde{B}}
\: +\: {\rm h.c.}\Big)\:
+\: \widetilde{M}^2_L\, \widetilde{L}^\dagger 
\widetilde{L}\: +\: \widetilde{M}^2_Q\, \widetilde{Q}^\dagger 
\widetilde{Q}\nonumber\\
&& +\, \widetilde{M}^2_U\,\widetilde{U}^*\widetilde{U}\: 
+\: \widetilde{M}^2_D\,\widetilde{D}^* \widetilde{D}\: 
+\: \widetilde{M}^2_E\, \widetilde{E}^* \widetilde{E}\: +\: 
m^2_1\, \widetilde{\Phi}^\dagger_1\widetilde{\Phi}_1\: +\:
m^2_2\, \Phi^\dagger_2\Phi_2\: -\: \Big(B\mu\,\widetilde{\Phi}^T_1 i\tau_2
\Phi_2\nonumber\\
&& +\: {\rm h.c.}\Big)\: 
+\: \Big(\, h_l A_l\,\Phi_1^\dagger \widetilde{L} \widetilde{E}\: +\:
h_dA_d\,\Phi_1^\dagger \widetilde{Q} \widetilde{D}\: -\:  
h_uA_u\,\Phi_2^T i\tau_2 \widetilde{Q} \widetilde{U}\ +\ {\rm h.c.}\, \Big)\,,
\end{eqnarray}
where $\widetilde{\Phi}_1 = i\tau_2  \Phi^*_1$ is the scalar component
of  the Higgs chiral  superfield $\widehat{H}_1$  and $\tau_2$  is the
usual Pauli  matrix.  The  conventions followed throughout  this paper,
including the  quantum-number assignments of  the fields under  the SM
gauge  group,   are  displayed  in  Table~\ref{tab1}.  

\begin{table}[b]

\begin{center}

\begin{tabular}{|c||c|c|c|}
\hline
Superfields & Bosons & Fermions &SU(3)$_c\otimes$SU(2)$_L\otimes$U(1)$_Y$ \\
\hline\hline
\underline{Gauge multiplets} & & & \\
$\widehat{G}^a$ & $ G^a_\mu\, \frac{1}{2} \lambda^a$ & $\tilde{g}^a$ & 
                                                                $(8,1,0)$ \\
$\widehat{W}$ & $ W^i_\mu\, 
  \frac{1}{2} \tau_i$ & $\widetilde{W}^i$ & $(1,3,0)$ \\
$\widehat{B}$ & $B_\mu$ & $\widetilde{B}$ & $(1,1,0)$ \\
 &&& \\
\hline
\underline{Matter multiplets} & & & \\
$\widehat{L}$ & $\widetilde{L}^T = (\tilde{\nu}_l,\tilde{l})_L$ &
$L^T = (\nu_l, l)_L$ & $(1,2,-1)$ \\
$\widehat{E}$ & $\widetilde{E} = \tilde{l}^*_R$ &
$(e_R)^C = (e^C)_L$ & $(1,1,2)$ \\
$\widehat{Q}$ & $\widetilde{Q}^T = (\tilde{u},\tilde{d})_L$ &
$Q^T = (u, d)_L$ & $(3,2,\frac{1}{3})$ \\
$\widehat{U}$ & $\widetilde{U} = \tilde{u}^*_R$ &
$(u_R)^C = (u^C)_L$ & $(3,1,-\frac{4}{3})$ \\
$\widehat{D}$ & $\widetilde{D} = \tilde{d}^*_R$ &
$(d_R)^C = (d^C)_L$ & $(3,1,\frac{2}{3})$\\
$\widehat{H}_1$ & $\widetilde{\Phi}_1^T = (\phi^{0 *}_1,-\phi^-_1)$ &
$(\bar{\psi}^0_{H_1}, \psi^-_{H_1})$  & $(1,2,-1)$ \\
$\widehat{H}_2$ & $\Phi_2^T = (\phi^+_2,\phi^0_2)$ &
$(\psi^+_{H_2}, \psi^0_{H_2} )$  & $(1,2,1)$ \\
 &&& \\
\hline
\end{tabular}
\end{center}

\caption{\it The field content of the MSSM.}\label{tab1}
\end{table}

As can be seen from  (\ref{Wpot}) and (\ref{Lsoft}), the MSSM includes
additional complex parameters with new  CP-odd phases that are  absent
in the SM.    These new CP-odd  phases   may reside in the   following
parameters:  (i) the  mass  parameter  $\mu$  describing the  bilinear
mixing of the two Higgs chiral superfields in the superpotential; (ii)
the   soft supersymetry-breaking     gaugino   masses     $m_{\tilde{g}}$,
$m_{\widetilde{W}}$   and  $m_{\widetilde{B}}$   of the  gauge  groups
SU(3)$_c$, SU(2)$_L$  and  U(1)$_Y$,   respectively; (iii)  the   soft
bilinear Higgs-mixing  mass  $B\mu$; and  (iv) the  soft
supersymmetry-breaking
trilinear couplings $A_f$  of   the Higgs bosons   to  sfermions.   In
addition,  there may exist other large  CP-odd phases, associated with
flavor off-diagonal soft supersymmetry-breaking masses of squarks and
sleptons.
We assume  that these off-diagonal  masses are small  and therefore do
not give sizeable   contributions  to the effective    Higgs potential
\cite{Silv/Mas}.   The number of   independent  CP-odd phases  may  be
reduced if  one prescribes  a  universality condition for  all gaugino
masses at  the unification scale $M_X$; the   gaugino masses will then
have  a  common   phase.  Correspondingly,  the   different  trilinear
couplings $A_f$  may be  considered to   be all equal  at $M_X$,  {\it
i.e.},  $A_f  \equiv  A$.    In this   case, however,  because  of the
different  RG running of the  phases of the trilinear couplings, their
values at low energies will be different.\footnote[2]{For a discussion
of  the RG effects, see~\cite{Wells}.}   Two CP-odd phases may further
be eliminated by  employing the following  two global  symmetries that
govern the dimension-four   operators     in the  MSSM     Lagrangian:
\begin{itemize}
  
\item[ (i)] The U(1)$_Q$ symmetry specified by $Q(\widehat{H}_1) = 1$,
$Q(\widehat{H}_2) = -2$, $Q(\widehat{Q}) = Q(\widehat{L}) = 0$,
$Q(\widehat{U}) = 2$ and $Q(\widehat{D}) = Q(\widehat{E}) = -1$. This
U(1)$_Q$ symmetry is broken by the $\mu$ parameter and the respective
soft supersymmetry-breaking one, $B\mu$.
  
\item[(ii)] The   U(1)$_R$  symmetry acting   on  the Grassmann-valued
coordinates, {\it i.e.},  the  $\theta$  coordinate of    superspace
carries
charge   1. Under the  $R$ transformation,  the matter superfields and
gaugino  fields carry   charge 1,  whilst the  Higgs  superfields are
$R$-neutral.  The $R$ symmetry is  violated by the gaugino masses, the
trilinear couplings $A_f$ and the parameter $\mu$.

\end{itemize}

We  concentrate   on  the  parameters   which  may  have   a  dominant
CP-violating effect on the  MSSM Higgs potential, under the assumption
of a common phase for the gauginos; the latter is made less important by
the
fact that the one-loop gaugino corrections  are subdominant compared to the
ones induced  by the third-generation squarks.  As  has been mentioned
above,  two  CP-odd  phases  of  the complex  parameters  $\{  \mu  ,\ 
m^2_{12},\ m_\lambda ,\  A \}$ may be removed  by employing the global
symmetries (i) and (ii).  Specifically,  one of the Higgs doublets and
the common phase  of the gaugino fields can be rephased  in a way such
that  the  gaugino  masses  and  $B\mu$ become  real  numbers.   As  a
consequence,  arg($\mu$)  and  arg($A_{t,b}$)  are the  only  physical
CP-violating phases  in the  MSSM which affect  the Higgs sector  in a
relevant way.

It  is obvious  from (\ref{Wpot})  and (\ref{Lsoft})  that  the Yukawa
interactions  of the  third-generation quarks,  $Q^T =  (t_L,b_L)$ and
$t_R,\   b_R$,   as  well   as   their   SUSY  bosonic   counterparts,
$\widetilde{Q}^T  =  (\tilde{t}_L,\tilde{b}_L  )$  and  $\tilde{t}_R,\ 
\tilde{b}_R$, play the most  significant role in radiative corrections
to the Higgs sector.  Therefore,  it is useful to give the interaction
Lagrangians related to the $F$ and $D$ terms of the third generation:
\begin{eqnarray}
  \label{LF}
-{\cal L}_F & = & |h_b|^2\, |\Phi^+_1 \widetilde{Q}|^2\, +\, 
|h_t|^2\, |\Phi^T_2 i\tau_2 \widetilde{Q}|^2\, \nonumber\\
&&-\, \Big(\, \mu h_b^*\, \widetilde{Q}^\dagger \Phi_2 \tilde{b}_R\, +\,
\mu h_t^* \widetilde{Q}^\dagger i\tau_2 \Phi_1^* \tilde{t}_R\ +\
{\rm h.c.}\, \Big) \nonumber\\
&& -\, \Big(\, h_b^*\, \tilde{b}_R \Phi_1^T i\tau_2\, +\,
h_t^* \tilde{t}_R \Phi_2^\dagger \Big)\, \Big(\, h_b\, i\tau_2 \Phi_1^*
\tilde{b}^*_R\, - \, h_t\, \Phi_2 \tilde{t}^*_R\, \Big) ,\\
  \label{LD}
-{\cal L}_D & = & \frac{g_w^2}{4}\, \Big[\, 2|\Phi^T_1 i\tau_2
\widetilde{Q}|^2\, +\, 2|\Phi^\dagger_2 \widetilde{Q}|^2\,
-\, \widetilde{Q}^\dagger \widetilde{Q}\, (\Phi^\dagger_1 \Phi_1\,
+\, \Phi_2^\dagger \Phi_2)\, \Big]\nonumber\\
&&+\, \frac{g'^2}{4}\, ( \Phi^\dagger_2 \Phi_2\, -\, \Phi_1^\dagger \Phi_1)\,
\Big[\, {\textstyle \frac{1}{3}}\, (\widetilde{Q}^\dagger \widetilde{Q})\, -\,
{\textstyle \frac{4}{3}}\, (\tilde{t}_R \tilde{t}^*_R)\, +\,
{\textstyle \frac{2}{3}}\, (\tilde{b}_R \tilde{b}^*_R)\, \Big]\, .
\end{eqnarray}
where $g_w$  and   $g'$ are  the  usual  SU(2)$_L$ and  U(1)$_Y$ gauge
couplings. Further, the interaction Lagrangian of the Higgs bosons
to the top and bottom quarks is given by
\begin{equation}
  \label{Lfermions} 
-{\cal L}_{\rm fermions} \ =\  \Big(\, h_b\, \bar{b}_R 
Q^T \Phi^*_1\, +\, {\rm h.c.}\, \Big)\, +\, 
\Big(\, h_t\, \bar{t}_R Q^T i\tau_2 \Phi_2\ +\ {\rm h.c.}\, \Big)\, ,
\end{equation} 
where $h_t$   and  $h_b$ are  the  top and  bottom Yukawa couplings,
respectively.

With the help of the Lagrangians (\ref{LF})--(\ref{Lfermions}), we now
proceed  with the  calculation of  the one-loop  effective  potential. 
More  explicitly,  in the  $\overline{\rm  MS}$  scheme, the  one-loop
CP-violating effective potential is determined by
\begin{equation}
  \label{LVeff}
-{\cal L}_V\ =\ -{\cal L}^0_V\, +\,
\frac{3}{32\pi^2}\, \sum\limits_{q = t,b} \bigg[\,\sum\limits_{i = 1,2} 
\widetilde{m}^4_{q_i}\, \bigg( \ln \frac{\widetilde{m}^2_{q_i}}{Q^2}\, -\,
\frac{3}{2}\,\bigg)\, -\, 2\bar{m}^4_q\, 
\bigg( \ln \frac{\bar{m}^2_q}{Q^2}\, -\, \frac{3}{2}\,\bigg)\, \bigg]\, .
\end{equation}
In (\ref{LVeff}), ${\cal L}^0_V$  is the tree-level Lagrangian of
the MSSM Higgs potential
\begin{eqnarray}
  \label{LVtree}
{\cal L}^0_V &=& \mu^2_1 (\Phi_1^\dagger\Phi_1)\, +\, 
\mu^2_2 (\Phi_2^\dagger\Phi_2)\, +\, m^2_{12} (\Phi_1^\dagger \Phi_2)\, 
+\, m^{*2}_{12} (\Phi_2^\dagger \Phi_1)\, +\, 
\lambda_1 (\Phi_1^\dagger \Phi_1)^2\nonumber\\
&&+\, \lambda_2 (\Phi_2^\dagger \Phi_2)^2\, +\, 
\lambda_3 (\Phi_1^\dagger \Phi_1)(\Phi_2^\dagger \Phi_2)\, +\, 
\lambda_4 (\Phi_1^\dagger \Phi_2)(\Phi_2^\dagger \Phi_1)\, ,
\end{eqnarray}
with
\begin{eqnarray}
  \label{LVpar}
\mu^2_1 &=& -m^2_1 - |\mu|^2\, ,\qquad \mu^2_2\ =\ -m^2_2 - |\mu|^2\,
,\qquad m^2_{12}\ =\ B\mu\, , \nonumber\\
\lambda_1 &=& \lambda_2\ =\ -\, \frac{1}{8}\, (g_w^2 + g'^2)\, ,\qquad
\lambda_3\ =\ -\frac{1}{4}\, (g_w^2 -g'^2)\, ,\qquad 
\lambda_4\ =\ \frac{1}{2}\, g_w^2\, .
\end{eqnarray}
Further, in (\ref{LVeff}),     $\bar{m}^2_i$     (with     $i=t,b$)
and
$\widetilde{m}^2_{q_k}$  (with  ${q_k}  =t_1,b_1,t_2,b_2$) denote  the
eigenvalues of  the quark and  squark mass matrices  ${\cal M}^\dagger
{\cal M}$  and $\widetilde{\cal  M}^2$, respectively, which  depend on
the  Higgs background fields.   Specifically, ${\cal  M}^\dagger {\cal
  M}$ reads
\begin{equation}
  \label{Mf}
{\cal M}^\dagger {\cal M}\ =\ \left( \begin{array}{cc}
|h_t|^2\, |\Phi_2|^2 & h_t h_b^*\, \Phi_1^T i\tau_2 \Phi_2 \\
h_t^* h_b\, \Phi_1^\dagger i\tau_2 \Phi^*_2 & |h_b|^2\, |\Phi_1|^2 \end{array}
\right)\, ,
\end{equation}
with eigenvalues
\begin{equation}
  \label{mudbar}
\bar{m}^2_{t (b)}\: =\: \frac{1}{2}\, \Big[\,
|h_b|^2\, |\Phi_1|^2\, +\, |h_t|^2\, |\Phi_2|^2\ +(-)\
\sqrt{ \big( |h_b|^2\, |\Phi_1|^2\, +\, |h_t|^2\, |\Phi_2|^2\big)^2\, -\,
4|h_t|^2 |h_b|^2 |\Phi^\dagger_1\Phi_2|^2}\ \Big]\, . 
\end{equation}
It  is  easy  to  see  that, for  $\phi^\pm_{1,2}  =0$,  (\ref{mudbar})
simplifies   to   the   known   expressions:  $\bar{m}^2_t   =   |h_t|^2
|\phi^0_2|^2$ and $\bar{m}^2_b = |h_b|^2 |\phi^0_1|^2$.

The  $(4\times  4)$ squark mass  matrix  $\widetilde{\cal
  M}^2$  is more complicated.  In  the weak  basis $\{  \widetilde{Q}^T =
(\tilde{t}_L,    \tilde{b}_L),    \widetilde{U}    =    \tilde{t}^*_R,
\widetilde{D} = \tilde{b}^*_R \}$,  $\widetilde{\cal M}^2$ may be cast
in the form:
\begin{equation}
  \label{Msquark}
\widetilde{\cal M}^2\ =\ \left( \begin{array}{ccc}
(\widetilde{\cal M}^2)_{\widetilde{Q}^\dagger \widetilde{Q}} & 
(\widetilde{\cal M}^2)_{\widetilde{Q}^\dagger \widetilde{U}^*} &
(\widetilde{\cal M}^2)_{\widetilde{Q}^\dagger \widetilde{D}^*} \\
(\widetilde{\cal M}^2)_{\widetilde{U} \widetilde{Q}} & 
(\widetilde{\cal M}^2)_{\widetilde{U} \widetilde{U}^*} &
(\widetilde{\cal M}^2)_{\widetilde{U} \widetilde{D}^*} \\
(\widetilde{\cal M}^2)_{\widetilde{D}\widetilde{Q}} & 
(\widetilde{\cal M}^2)_{\widetilde{D} \widetilde{U}^*} &
(\widetilde{\cal M}^2)_{\widetilde{D} \widetilde{D}^*}\end{array}\right)\, ,
\end{equation}
with
\begin{eqnarray}
  \label{MSelem}
(\widetilde{\cal M}^2)_{\widetilde{Q}^\dagger \widetilde{Q}} &=&
\widetilde{M}^2_Q\, {\bf 1}_2\ +\ |h_b|^2\, \Phi_1\Phi^\dagger_1\ +\ |h_t|^2\, 
\Big( \Phi^\dagger_2 \Phi_2\, {\bf 1}_2\, -\, \Phi_2\Phi^\dagger_2 \Big)\ -\ 
\frac{1}{2}\, g^2_w\, \Big( \Phi_1\Phi^\dagger_1\, -\,
\Phi_2 \Phi^\dagger_2\Big)\nonumber\\
&& +\, \bigg(\, \frac{1}{4}\, g^2_w\, -\, \frac{1}{12}\, g'^2\,\bigg)\,
 \Big( \Phi^\dagger_1 \Phi_1\, -\, \Phi^\dagger_2\Phi_2 \Big)\, {\bf
1}_2\, ,\nonumber\\ 
(\widetilde{\cal M}^2)_{\widetilde{U}
\widetilde{Q}} &=& (\widetilde{\cal
M}^2)^\dagger_{\widetilde{Q}^\dagger \widetilde{U}^*}\ =\ - h_t A_t\,
\Phi^T_2 i\tau_2\ +\ h_t\mu^*\, \Phi^T_1 i\tau_2\, ,\nonumber\\
(\widetilde{\cal M}^2)_{\widetilde{D}\widetilde{Q}} &=&
(\widetilde{\cal M}^2)^\dagger_{\widetilde{Q}^\dagger
\widetilde{D}^*}\ =\ h_b A_b\, \Phi^\dagger_1\ -\ h_b\mu^*\,
\Phi^\dagger_2\, ,\nonumber\\ (\widetilde{\cal M}^2)_{\widetilde{U}
\widetilde{U}^*} &=& \widetilde{M}^2_t\ +\
|h_t|^2\,\Phi^\dagger_2\Phi_2\ +\ \frac{1}{3}\, g'^2\, \Big(
\Phi^\dagger_1 \Phi_1\, -\, \Phi^\dagger_2 \Phi_2 \Big)\,,\nonumber\\
(\widetilde{\cal M}^2)_{\widetilde{D} \widetilde{D}^*} &=&
\widetilde{M}^2_b\ +\ |h_b|^2\,\Phi^\dagger_1\Phi_1\ -\ \frac{1}{6}\,
g'^2\, \Big( \Phi^\dagger_1 \Phi_1\, -\, \Phi^\dagger_2 \Phi_2
\Big)\,,\nonumber\\ (\widetilde{\cal M}^2)_{\widetilde{U}
\widetilde{D}^*} &=& (\widetilde{\cal M}^2)^*_{\widetilde{D}
\widetilde{U}^*}\ =\ h_t h_b^*\, \Phi^T_1 i\tau_2\Phi_2\, .
\end{eqnarray} 
Here  and  in the   following,    we denote by    $\widetilde{M}^2_Q$,
$\widetilde{M}^2_t$ and $\widetilde{M}^2_b$  the  soft
supersymmetry-breaking
masses of the third generation of squarks.
It  is    rather    difficult  to   express   the   four   eigenvalues
$\widetilde{m}^2_{q_k}$    (${q_k}  =       t_1,b_1,t_2,b_2$)       of
$\widetilde{\cal M}^2$ in a simple form. However, as we detail in Appendix
A.3,
it     is     not  necessary   to     know     the analytic   form  of
$\widetilde{m}^2_{q_k}$  in  order to evaluate the  Higgs-boson masses
and     mixing angles~\cite{BERZ}.   Of course,   for
$\phi^\pm_{1,2} =0$,
the
field-dependent squark eigenvalues simplify to
\begin{eqnarray}
  \label{msqud}
\widetilde{m}^2_{t_1 (t_2)} & =& \frac{1}{2}\, \bigg[\,
\widetilde{M}^2_Q\, +\, \widetilde{M}^2_t\, +\, 2|h_t|^2\, |\phi^0_2|^2\, +\,
\frac{g^2_w +g'^2}{4}\, \big( |\phi^0_1|^2 - |\phi^0_2|^2 \big)\nonumber\\
&&+(-)\, \sqrt{ \big[\,\widetilde{M}^2_Q\, -\, \widetilde{M}^2_t\, +\, 
x_t \big( |\phi^0_1|^2 - |\phi^0_2|^2 \big)\big]^2\ +\ 
4|h_t|^2\,\big| A_t\phi^0_2\, -\, \mu^* \phi^0_1 \big|^2 }\ \bigg]\,,\nonumber\\
\widetilde{m}^2_{b_1 (b_2)} & =& \frac{1}{2}\, \bigg[\,
\widetilde{M}^2_Q\, +\, \widetilde{M}^2_b\, +\, 2|h_b|^2\, |\phi^0_1|^2\, -\,
\frac{g^2_w +g'^2}{4}\, \big( |\phi^0_1|^2 - |\phi^0_2|^2 \big)\nonumber\\
&&+(-)\, \sqrt{ \big[\,\widetilde{M}^2_Q\, -\, \widetilde{M}^2_b\, -\, 
x_b \big( |\phi^0_1|^2 - |\phi^0_2|^2 \big)\big]^2\ +\ 
4|h_b|^2\,\big| A^*_b\phi^0_1\, -\, \mu \phi^0_2 \big|^2 }\ \bigg]\,,
\end{eqnarray}
where  $x_t =  \frac{1}{4}  (g^2_w  - \frac{5}{3}  g'^2)$  and $x_b  =
\frac{1}{4}  (g^2_w - \frac{1}{3}  g'^2)$.

After  having set  the stage,  we now derive  the  minimization conditions
governing the  ground state of  the MSSM one-loop  effective potential
and  then  determine  the  Higgs-boson  mass matrices.  As  usual,  we
consider the following linear expansion of the Higgs doublets $\Phi_1$
and $\Phi_2$ around the ground state:
\begin{equation}
  \label{Phi12}
\Phi_1\ =\ \left( \begin{array}{c}
\phi^+_1 \\ \frac{1}{\sqrt{2}}\, ( v_1\, +\, \phi_1\, +\, ia_1)
\end{array} \right)\, ,\qquad
\Phi_2\ =\ e^{i\xi}\, \left( \begin{array}{c}
\phi^+_2 \\  \frac{1}{\sqrt{2}}\, ( v_2 \, +\, \phi_2\, +\, ia_2 )
 \end{array} \right)\, ,
\end{equation}
where $v_1$ and $v_2$ are  the moduli of the vacuum expectation values
(VEVs)  of the  Higgs  doublets and  $\xi$  is their  relative phase.  
Following  \cite{APLB,PW},  we  require  the vanishing  of  the  total
tadpole contributions
\begin{eqnarray}
  \label{Tphi12}
T_{\phi_1 (\phi_2)} &\equiv& \bigg<\frac{\partial {\cal L}_V}{\partial
  \phi_{1(2)} }\, \bigg> \ =\ 
v_{1 (2)} \bigg[\, \mu^2_{1(2)}\ +\ \frac{v_1 v_2}{v^2_{1 (2)}}\, 
{\rm Re} (m^2_{12}e^{i\xi})\,
  \, +\, \lambda_{1(2)} v^2_{1 (2)}\, +\,
  \frac{1}{2}\, (\lambda_3 + \lambda_4) v^2_{2 (1)}\, \bigg] \nonumber\\
&&-\, \frac{3}{16\pi^2}\, \sum\limits_{q=t,b} 
\bigg[\, \sum\limits_{i =1,2} 
\bigg<\frac{\partial \widetilde{m}^2_{q_i}}{\partial \phi_{1(2)} }\bigg>\,  
m^2_{\tilde{q}_i}\, \bigg( \ln\frac{m^2_{\tilde{q}_i}}{Q^2}\ 
-\ 1\,\bigg) \: -\: 2\,
\bigg<\frac{\partial \bar{m}^2_q}{\partial \phi_{1(2)} } \bigg>\,
m^2_q\nonumber\\
&&\times\, \bigg( \ln\frac{m^2_q}{Q^2}\ -\ 1\,\bigg)\, \bigg]\, ,\\
  \label{Ta12}
T_{a_1 (a_2)} &\equiv& \bigg< \frac{\partial {\cal L}_V}{\partial 
 a_{1(2)} }\bigg>
\ =\ +(-)\, v_{2 (1)}\, {\rm Im} (m^2_{12}e^{i\xi})\, 
-\, \frac{3}{16\pi^2}  \sum\limits_{q=t,b}\ \sum\limits_{i =1,2}
\bigg<\frac{\partial \widetilde{m}^2_{q_i}}{\partial a_{1(2)} }\bigg>\,  
m^2_{\tilde{q}_i}\nonumber\\
&&\times\, \bigg( \ln\frac{m^2_{\tilde{q}_i}}{Q^2}\ -\ 1\,\bigg)\,,
\end{eqnarray}
where  $\big< \widetilde{m}^2_{q_i}\big> = m^2_{\tilde{q}_i}$, and the
tadpole  derivatives $\big< \partial \bar{m}^2_q/\partial  \phi_{1(2)}
\big>$,  $\big< \partial  \widetilde{m}^2_{q_k}/\partial   \phi_{1(2)}
\big>$ and     $\big< \partial \widetilde{m}^2_{q_k}/\partial a_{1(2)}
\big>$ are given  in (\ref{tadf}) and  (\ref{tads}).  Moreover,
from (\ref{tads}), we  readily  see that $T_{a_1} =  -\tan\beta\,
T_{a_2}$,  with $\tan\beta =  v_2/v_1$.  This last fact \cite{APLB,PW}
allows us to  perform an orthogonal  rotation in the  space spanned by
the `CP-odd' scalars $a_1$ and $a_2$,
\begin{equation}
  \label{G0A}
\left( \begin{array}{c} a_1 \\ a_2\end{array}\right)\ =\
\left( \begin{array}{cc} \cos\beta & -\sin\beta \\ 
\sin\beta & \cos\beta\end{array}\right)\,
\left( \begin{array}{c} G^0 \\ a\end{array}\right)\, .
\end{equation}
The  Higgs potential then  has a  flat direction  with respect  to the
$G^0$  field, {\em  i.e.}, $\langle  \partial {\cal  L}_V/\partial G^0
\rangle = 0$, and the  $G^0$ field becomes the true would-be Goldstone
boson eaten by the longitudinal component of the $Z$ boson.

We observe from  (\ref{Ta12}) that a relative phase  $\xi$ between the
two  Higgs vacuum  expectation values  is induced  radiatively  in the
$\overline{\rm MS}$ scheme  \cite{APLB,PW}.  However, we should stress
that the phase $\xi$  is renormalization-scheme dependent.  For
example, one  may adopt  a renormalization scheme,  slightly different
from  the $\overline{\rm  MS}$  one, in  which  $\xi$ is  set to  zero
order-by-order  in  perturbation  theory  \cite{APLB}.   This  can  be
achieved by requiring the  bilinear Higgs-mixing mass $m^2_{12}$ to be
real at the tree level, but to receive an imaginary counter-term (CT),
${\rm Im}\,  m^2_{12}$, at higher  orders, which is determined  by the
vanishing of the CP-odd tadpole parameters $T_{a_1}$ and $T_{a_2}$ for
$\xi = 0$.  As we detail below, the scheme of renormalization of ${\rm
  Im}   (m^2_{12}   e^{i\xi})$    does   not   directly   affect   the
renormalization scheme  of other physical kinematic  parameters of the
theory to  one loop, such  as Higgs-boson masses and  $\tan\beta$.  In
fact, it has been explicitly demonstrated in \cite{APLB} that physical
CP-violating   transition  amplitudes,  such   as  scalar-pseudoscalar
transitions, are independent  of the renormalization subtraction point
$Q^2$ and the  choice of phase $\xi$.  In the  following, we adopt the
$\xi  = 0$  scheme of  renormalization, as  irrelevant $\xi$-dependent
phases  in the  effective chargino  and neutralino  mass  matrices can
thereby be completely avoided.

In  the remaining  part  of  this section,  we  evaluate the  one-loop
effective Higgs-boson mass matrices.  Employing the tadpole conditions
$T_{\phi_1} = T_{\phi_2}  = 0$ and $T_{a_1} = T_{a_2}  = 0$ allows one
to substitute  the mass parameters $\mu^2_1$ and  $\mu^2_2$, and ${\rm
  Im}\, m^2_{12}$  into the effective  potential (\ref{LVeff}).  After
performing   the    above   substitutions,   we    can   express   the
charged-Higgs-boson mass matrix as follows:
\begin{eqnarray}
  \label{Mcharged}
({ \cal M }^2_\pm )_{ij} &=& (-1)^{i+j}\, \frac{v_1 v_2}{v_i
  v_j}\, \bigg( {\rm Re}\, m^2_{12}\, +\, 
\frac{1}{4}\, g^2_w v_1v_2\, \bigg)\, +\, \frac{3}{16\pi^2} 
 \sum\limits_{q =t,b} \bigg[\, \sum\limits_{k=1,2}
\bigg(\, \bigg<\frac{\partial^2 \widetilde{m}^2_{q_k}}{\partial \phi^+_i\,
\partial \phi^-_j} \bigg> \nonumber\\
&&-\, \frac{\delta_{ij}}{v_i}\, 
\bigg<\frac{\partial \widetilde{m}^2_{q_k}}{\partial \phi_j}\bigg>\,
-\, \frac{i (1-\delta_{ij} )}{v_i}\, \bigg<\frac{\partial 
\widetilde{m}^2_{q_k}}{\partial a_j}\bigg>\, \bigg)\, 
m^2_{\tilde{q}_k}\,
\bigg( \ln\frac{m^2_{\tilde{q}_k}}{Q^2}\ -\ 1\,\bigg)\, \nonumber\\
&&-2\,\bigg(\, \bigg<\frac{\partial^2 
\bar{m}^2_q}{\partial \phi^+_i\,\partial \phi^-_j  } \bigg>\,
-\, \frac{\delta_{ij}}{v_i}\, \bigg<\frac{\partial 
\bar{m}^2_q}{\partial \phi_j}\bigg>\, \bigg)\,
m^2_q\, \bigg( \ln\frac{m^2_q}{Q^2}\ -\ 1\,\bigg)\, \bigg]\, .
\end{eqnarray}
Since  ${\rm det}\,  { \cal  M  }^2_\pm =0$  in (\ref{Mcharged}),  the
square of  the charged Higgs-boson mass $M^2_{H^+}$  may be determined
by the matrix element $({ \cal M }^2_\pm )_{12}$:
\begin{eqnarray}
  \label{MHplus}
M^2_{H^+} \!&=&\! \frac{v^2}{v_1 v_2}\, \bigg\{ 
\bigg( {\rm Re}\, m^2_{12}\, +\, 
\frac{1}{4}\, g^2_w v_1v_2\, \bigg)\ -\ \frac{3}{16\pi^2} 
 \sum\limits_{q =t,b} \bigg[\, \sum\limits_{k=1,2} \bigg(\,
\bigg<\frac{\partial^2 \widetilde{m}^2_{q_k}}{\partial \phi^+_1\,
\partial \phi^-_2} \bigg>\, -\, \frac{i}{v_1}\, \bigg<\frac{\partial 
\widetilde{m}^2_{q_k}}{\partial a_2}\bigg>\, \bigg) \nonumber\\
&&\times\, 
m^2_{\tilde{q}_k}\,
\bigg( \ln\frac{m^2_{\tilde{q}_k}}{Q^2}\ -\ 1\,\bigg)\  
-2\, \bigg<\frac{\partial^2 
\bar{m}^2_q}{\partial \phi^+_1\,\partial \phi^-_2  } \bigg>\,
\, m^2_q\, \bigg( \ln\frac{m^2_q}{Q^2}\ -\ 1\,\bigg)\, \bigg]\,\bigg\}\,.
\end{eqnarray}
The   self-energy  derivatives   appearing  in   (\ref{Mcharged})  and
(\ref{MHplus}):   $\big<\partial^2   \bar{m}^2_q/  \partial   \phi^+_1
\partial  \phi^-_2 \big>$ and  $\big<\partial^2 \widetilde{m}^2_{q_k}/
\partial  \phi^+_1 \partial \phi^-_2  \big>$, as  well as  the tadpole
terms $\big< \partial \bar{m}^2_{q_k} / \partial \phi_j \big>$, $\big<
\partial  \widetilde{m}^2_{q_k} /  \partial \phi_j  \big>$  and $\big<
\partial \widetilde{m}^2_{q_k} / \partial a_j \big>$, are exhibited in
Appendix A.

By analogy,  in the  weak basis $\{  \phi_1,\phi_2, a_1, a_2  \}$, the
neutral-Higgs-boson mass  matrix takes on  the form\footnote[3]{Notice
  that our  convention differs  from that given  in \cite{PW},  as the
  neutral-Higgs-boson  mass  matrix ${\cal  M}^2_0$  in  that work  is
  expressed in the weak basis $\{ a_1, a_2, \phi_1,\phi_2 \}$.}
\begin{equation}
  \label{NHiggs}
{\cal M}^2_0 \ =\ 
\left(\begin{array}{cc} {\cal M}^2_S  & {\cal M}^2_{SP} \\
         ({\cal M}^2_{SP})^T &  {\cal M}^2_P \end{array} \right)\, ,
\end{equation}
where ${\cal M}^2_S$, ${\cal  M}^2_P$ and ${\cal M}^2_{SP}$ denote the
two-by-two    matrices    of     the    scalar,    pseudoscalar    and
scalar-pseudoscalar squared  mass terms of  the neutral Higgs  bosons. 
Observe that  the presence of CP-violating self-energy  terms leads to
mass eigenstates  with no  well-defined CP quantum  numbers. Therefore
the CP-odd Higgs-boson mass $M_A$ cannot be identified with any of the
neutral Higgs-boson  masses. The individual matrix  elements of ${\cal
  M}^2_0$ are given by
\begin{eqnarray}
  \label{calMS}
({\cal M}^2_S)_{ij} &=& (-1)^{i+j}\, \frac{v_1 v_2}{v_i
  v_j}\, {\rm Re}\, m^2_{12}\, +\, 
\frac{1}{4}\, (g^2_w + g'^2) v_iv_j\ +\ \frac{3}{16\pi^2} 
\sum\limits_{q=t,b} \bigg\{ \sum\limits_{k =1,2} \bigg[\,\bigg(\,
\bigg<\frac{\partial^2 \widetilde{m}^2_{q_k}}{\partial \phi_i\,
\partial \phi_j} \bigg> \nonumber\\
&&-\, \frac{\delta_{ij}}{v_i}\, 
\bigg<\frac{\partial \widetilde{m}^2_{q_k}}{\partial \phi_j}\bigg>\,\bigg)\, 
m^2_{\tilde{q}_k}\,
\bigg( \ln\frac{m^2_{\tilde{q}_k}}{Q^2}\ -\ 1\,\bigg)\ +\
\bigg<\frac{\partial \widetilde{m}^2_{q_k}}{\partial \phi_i}\bigg>\,
\bigg<\frac{\partial \widetilde{m}^2_{q_k}}{\partial \phi_j}\bigg>\,
\ln\frac{m^2_{\tilde{q}_k}}{Q^2}\, \bigg] \nonumber\\
&&-\, 2\, \bigg[\, \bigg(\,
\bigg<\frac{\partial^2 \bar{m}^2_q}{\partial \phi_i\,
\partial \phi_j} \bigg>\ -\, \frac{\delta_{ij}}{v_i}\, 
\bigg<\frac{\partial \bar{m}^2_q}{\partial \phi_j}\bigg>\,\bigg)\, 
m^2_q\,
\bigg( \ln\frac{m^2_q}{Q^2}\ -\ 1\,\bigg)\nonumber\\
&&+\,\bigg<\frac{\partial \bar{m}^2_q}{\partial \phi_i}\bigg>\,
\bigg<\frac{\partial \bar{m}^2_q}{\partial \phi_j}\bigg>\,
\ln\frac{m^2_q}{Q^2}\,\bigg]\, \bigg\}\, ,\\
  \label{calMSP}
({\cal M}^2_{SP})_{ij} &=& \frac{3}{16\pi^2} 
\sum\limits_{q=t,b}\ \sum\limits_{k =1,2} \bigg[\, \bigg(\,
\bigg<\frac{\partial^2 \widetilde{m}^2_{q_k}}{\partial \phi_i\,
\partial a_j} \bigg>\, -\, \frac{(1-\delta_{ij})}{v_i}\, 
\bigg<\frac{\partial \widetilde{m}^2_{q_k}}{\partial a_j}\bigg>\,
 \bigg)\, m^2_{\tilde{q}_k}\, 
\bigg( \ln\frac{m^2_{\tilde{q}_k}}{Q^2}\ -\ 1\,\bigg)\nonumber\\ 
&&+\,
\bigg<\frac{\partial \widetilde{m}^2_{q_k}}{\partial \phi_i}\bigg>\,
\bigg<\frac{\partial \widetilde{m}^2_{q_k}}{\partial a_j}\bigg>\,
\ln\frac{m^2_{\tilde{q}_k}}{Q^2}\, \bigg]\, ,\\
  \label{calMP}
({ \cal M }^2_P )_{ij} &=& (-1)^{i+j}\, \frac{v_1 v_2}{v_i
  v_j}\, {\rm Re}\, m^2_{12}\ +\ \frac{3}{16\pi^2} 
\sum\limits_{q=t,b} \bigg\{
\sum\limits_{k =1,2} \bigg[\, \bigg(\,
\bigg<\frac{\partial^2 \widetilde{m}^2_{q_k}}{\partial a_i\,
\partial a_j} \bigg>\, -\, \frac{\delta_{ij}}{v_i}\, 
\bigg<\frac{\partial \widetilde{m}^2_{q_k}}{\partial \phi_j}\bigg>\,
 \bigg)\nonumber\\
&&\times\, m^2_{\tilde{q}_k}\,
\bigg( \ln\frac{m^2_{\tilde{q}_k}}{Q^2}\ -\ 1\,\bigg)\ +\
\bigg<\frac{\partial \widetilde{m}^2_{q_k}}{\partial a_i}\bigg>\,
\bigg<\frac{\partial \widetilde{m}^2_{q_k}}{\partial a_j}\bigg>\, 
\ln\frac{m^2_{\tilde{q}_k}}{Q^2}\, \bigg]\nonumber\\
&&- 2\, \bigg(\, \bigg<\frac{\partial^2 
\bar{m}^2_q}{\partial a_i\,\partial a_j  } \bigg>\,
-\, \frac{\delta_{ij}}{v_i}\, \bigg<\frac{\partial 
\bar{m}^2_q}{\partial \phi_j}\bigg>\, \bigg)\,
m^2_q\, \bigg( \ln\frac{m^2_q}{Q^2}\ -\ 1\,\bigg)\, \bigg\}\, .
\end{eqnarray}
Again,  the  analytic  expressions  for the  self-energy  and  tadpole
derivatives with respect  to the background Higgs fields  are given in
Appendix A.

Since $G^0$ does not mix  with the other neutral fields, the $(4\times
4)$ matrix  ${\cal  M}^2_0$   reduces  to   a  $(3\times
3)$ matrix, which we denote by ${\cal M}^2_N$. In the weak
basis $\{ \phi_1,\phi_2, a\}$,  the reduced neutral mass-squared matrix
${\cal  M}^2_N$ may be expressed by
\begin{equation}
  \label{calMN}
{\cal M}^2_N\ =\ \left( \begin{array}{ccc}
({\cal M}^2_{S})_{11} & ({\cal M}^2_{S})_{12} & 
\frac{1}{\cos\beta}\, ({\cal M}^2_{SP})_{12} \\
({\cal M}^2_{S})_{21} & ({\cal M}^2_{S})_{22} & 
-\, \frac{1}{\sin\beta}\, ({\cal M}^2_{SP})_{21} \\
\frac{1}{\cos\beta}\,({\cal M}^2_{SP})_{12} &
-\, \frac{1}{\sin\beta}\, ({\cal M}^2_{SP})_{21} 
&-\, \frac{1}{\sin\beta\,\cos\beta}\, ({\cal M}^2_P)_{12} 
\end{array} \right) .
\end{equation}
In  writing  ${\cal  M}^2_N$   in  (\ref{calMN}),  we  have  used  the
properties  of  the  matrix  elements of  ${\cal  M}^2_{SP}$:  $({\cal
  M}^2_{SP})_{11} = -  \tan\beta\, ({\cal M}^2_{SP})_{12}$ and $({\cal
  M}^2_{SP})_{22}  =   -  \cot\beta\,  ({\cal   M}^2_{SP})_{21}$,  and
likewise for ${\cal M}^2_P$.

Using the  expressions (\ref{MHplus}) and  (\ref{calMN}), we determine
the analytic forms of  the RG-improved charged and neutral Higgs-boson
masses in the next Section.

\setcounter{equation}{0}
\section{RG-Improved Higgs-Boson Mass Matrices}

In this Section,  we perform a one-loop RG  improvement of the squared
charged  Higgs-boson  mass  $M^2_{H^+}$  and of  the  squared  neutral
Higgs-boson   mass  matrix   ${\cal  M}^2_N$.    The   RG  improvement
incorporates  all  leading  two-loop  logarithmic corrections  to  the
Higgs-boson  mass-matrix elements,  which  were already  found in  the
CP-conserving case  to give rise  to significant contributions  to the
Higgs-boson masses  and couplings.  In particular, the  upper bound on
the lightest CP-even  Higgs mass was found to  be strongly affected by
the  two-loop  logarithmic  corrections~\cite{KYS,CEQR,CEQW,CQW}.   In
carrying  out the  RG improvement,  we follow  the  procedure outlined
in~\cite{CQW}, in which the improvement of the Higgs-boson mass-matrix
elements was performed by carefully applying the process of decoupling
of the third-generation squarks.

Within the framework of the RG approach, the dominant contributions to
the Higgs-boson  mass matrix ${\cal M}^2$ may  
be written conceptually as a sum of two terms:
\begin{equation}
  \label{RGappr} 
{\cal M}^2 (m_t) \ =\ \overline{\cal M}^2 (m_t)\: +\: 
{\cal M}^{2,{\rm th}}(m_t)\, .
\end{equation}
The  first term,  $\overline{\cal  M}^2 (m_t)$,  contains the  genuine
logarithmic contributions  which determine the  whole scale dependence
of  the one-loop  effective potential.   These contributions  would be
present, even if the left-right  mixing of the stop and sbottom states
were absent. The second  term, ${\cal M}^{2,{\rm th}}(m_t)$, describes
the threshold effect of the decoupling of the heavier stop and sbottom
squarks and their  respective mixing with the lighter  states.  At the
one-loop level,  the second term is manifestly  scale independent.  In
(\ref{RGappr}),  we   are  interested  in   evaluating  the  effective
potential at $m_t$,  since it has been shown~\cite{CEQR}  that this is
the scale at which two-loop  corrections are minimized.  As we explain
below, the renormalization of the above two contributions must proceed
in different ways.

Let us denote  by $Q_{tb}$ the scale of  the heaviest third-generation
squark, which we  assume to be higher than  the electroweak scale.  In
the language  of the RG approach,  we have first to  consider that the
aforementioned  threshold  contribution   is  `frozen'  at  the  scale
$Q_{tb}$,   ${\cal  M}^{2,{\rm   th}}(Q_{tb})   \equiv  \Delta   {\cal
  M}^2(Q_{tb})$, with all the involved kinematic parameters defined at
this  particular  scale.   Then,  we  have to  rescale  the  threshold
contribution  with the  anomalous  dimension factors  of the  relevant
Higgs fields:
\begin{equation}
  \label{Mth}
{\cal M}_{ij}^{2,{\rm th}}(m_t)\ =\ 
\Delta {\cal M}_{ij}^2(Q_{tb})\, 
\xi_i^{-1}(m_t)\, \xi_j^{-1}(m_t)\, ,
\end{equation}
where $\xi_i  (m_t)$ is  the anomalous dimension  factor of  the $H_i$
state to  be determined below.   The one-loop matrix  elements $\Delta
{\cal  M}^2_{ij}(Q_{tb})$ depend on  the running  quark masses  at the
scale  $Q_{tb}$,  which  have   to  be  conveniently  re-expressed  as
functions  of the corresponding  running masses  at $m_t$.   Thus, the
anomalous-dimension factors in  combination with the one-loop relation
between the  quark masses at  scales $Q_{tb}$ and $m_t$  yield sizeable
two-loop corrections to the  mass-matrix elements originating from the
one-loop threshold effects.

As was already mentioned, the contribution $\overline{\cal M}^2 (m_t)$
of the  third-generation squarks to the  effective potential describes
the genuine one- and two-loop leading-logarithmic running of the Higgs
quartic couplings.  In this context, there are two important technical
details that should be mentioned.  First, we notice that, in the MSSM,
the  tree-level  Higgs  quartic   couplings  $\lambda_i$,  with  $i  =
1,2,3,4$, are all proportional  to the squared gauge couplings $g^2_w$
and  $g'^2$  (cf.   (\ref{LVpar})).   However,  the  one-loop  $\beta$
functions  of $\lambda_i$  can generally  have appreciable  values, as
they are proportional to the fourth power of the top- and bottom-quark
Yukawa couplings.   As a result, the low-energy  values of $\lambda_i$
differ  significantly  from  their  tree-level ones.  The  RG-improved
approach  followed  here  is  crucial for  implementing  properly  the
potentially  large  logarithmic   corrections  to  the  Higgs  quartic
couplings.

The second technical remark pertains  to the RG evolution of the Higgs
quartic couplings $\lambda_i$,  with $i = 5, 6,  7$ (for the notation,
see \cite{HH,PW}), which  are absent in the Born  approximation to the
MSSM  Higgs  potential.   On  field-theoretic grounds,  these  quartic
couplings must  have vanishing one-loop $\beta$  functions, and cannot
be  generated by  RG running.   However, these  quartic  couplings are
radiatively induced by threshold  effects, and have already been taken
into account in $\Delta {\cal M}^2 (Q_{tb})$, given by (\ref{Mth}).

Following the above discussion, we now proceed with the RG improvement
of the Higgs-boson mass-matrix elements. To this end, we first need to
compute the one-loop values of the quartic couplings $\lambda_i$, with
$i=1,2,3,4$,   where  the   decoupling   of  the   stop  and   sbottom
contributions at  their appropriate thresholds is  properly taken into
account.  The best way to  calculate the latter effects is to consider
the logarithmic  part of the effective potential  (\ref{LVeff}) in the
limit  where the  squark mixing  parameters vanish,  {\em i.e.},
$\mu  =  A_t =  A_b  = 0$.   The  pertinent  one-loop running  quartic
couplings, denoted by $\lambda^{(1)}_i$, may then be obtained by
\begin{eqnarray}
  \label{beta1}
\lambda^{(1)}_1 \!\!&=&\!\! -\, \frac{3}{32\pi^2}\, \bigg[\, \bigg(
\frac{g^2_w}{4}\, -\, \frac{g'^2}{12} \bigg)^2\, 
\ln\bigg(\frac{\widetilde{M}^2_Q + m^2_t}{Q^2}\bigg)\: +\:
\bigg( |h_b|^2\,-\,\frac{g^2_w}{4}\, -\, \frac{g'^2}{12}\bigg)^2\,
\ln\bigg(\frac{\widetilde{M}^2_Q + m^2_b}{Q^2}\bigg) \nonumber\\
&&+\, \frac{g'^4}{9}\,
\ln\bigg(\frac{\widetilde{M}^2_t + m^2_t}{Q^2}\bigg)\: +\:
\bigg( |h_b|^2\,-\,\frac{g'^2}{6}\bigg)^2\,
\ln\bigg(\frac{\widetilde{M}^2_b + m^2_b}{Q^2}\bigg)\, \bigg]\,,\\ 
  \label{beta2}
\lambda^{(1)}_2 \!\!&=&\!\! -\, \frac{3}{32\pi^2}\, \bigg[\, 
\bigg( |h_t|^2\,-\,\frac{g^2_w}{4}\, +\, \frac{g'^2}{12}\bigg)^2\,
\ln\bigg(\frac{\widetilde{M}^2_Q + m^2_t}{Q^2}\bigg)\: +\:
\bigg(
\frac{g^2_w}{4}\, +\, \frac{g'^2}{12} \bigg)^2\, 
\ln\bigg(\frac{\widetilde{M}^2_Q + m^2_b}{Q^2}\bigg)\nonumber\\
&&+\, \bigg( |h_t|^2\,-\,\frac{g'^2}{3}\bigg)^2\,
\ln\bigg(\frac{\widetilde{M}^2_t + m^2_t}{Q^2}\bigg)\: +\:
\frac{g'^4}{36}\, 
\ln\bigg(\frac{\widetilde{M}^2_b + m^2_b}{Q^2}\bigg)\,\bigg]\,,\\
  \label{beta3}
\lambda^{(1)}_3 \!\!&=&\!\! -\, \frac{3}{16\pi^2}\, \bigg\{\,
|h_t|^2|h_b|^2\,\bigg[\, \ln\bigg(\frac{\widetilde{M}^2_Q +
  m^2_t}{Q^2}\bigg)\, +\, \ln \bigg(\frac{ {\rm max}\, 
(\widetilde{M}^2_t + m^2_t\, ,\ \widetilde{M}^2_b + m^2_b )}{Q^2}
          \bigg)\,\bigg]\nonumber\\
&& -\, \bigg[\,\bigg(\frac{g^2_w}{4}\, +\, \frac{g'^2}{12}\bigg)
\bigg(|h_t|^2\, -\, \frac{g^2_w}{4}\bigg)\, -\, 
\frac{g'^2}{12}\, \bigg(\frac{g^2_w}{4}\, -\, \frac{g'^2}{12}\bigg)\, 
\bigg]\, \ln\bigg(\frac{\widetilde{M}^2_Q +
  m^2_t}{Q^2}\bigg)\nonumber\\
&&-\, \bigg[\,\bigg(\frac{g^2_w}{4}\, -\, \frac{g'^2}{12}\bigg)
\bigg(|h_b|^2\, -\, \frac{g^2_w}{4}\bigg)\, +\, 
\frac{g'^2}{12}\, \bigg(\frac{g^2_w}{4}\, +\, \frac{g'^2}{12}\bigg)
\, \bigg]\, \ln\bigg(\frac{\widetilde{M}^2_Q +
  m^2_b}{Q^2}\bigg)\nonumber\\
&&+\, \frac{g'^2}{3}\, \bigg( |h_t|^2\, -\,\frac{g'^2}{3}\bigg)\,
\ln\bigg(\frac{\widetilde{M}^2_t + m^2_t}{Q^2}\bigg)\: +\:
\frac{g'^2}{6}\, \bigg( |h_b|^2\, -\,\frac{g'^2}{6}\bigg)\,
\ln\bigg(\frac{\widetilde{M}^2_b + m^2_b}{Q^2}\bigg)\, \bigg\}\,,\\
  \label{beta4}
\lambda^{(1)}_4 &=& \frac{3}{16\pi^2}\, \bigg\{\, 
|h_t|^2|h_b|^2\,\bigg[\, \ln\bigg(\frac{\widetilde{M}^2_Q +
  m^2_t}{Q^2}\bigg)\, +\, \ln \bigg(\frac{ 
{\rm max}\, (\widetilde{M}^2_t + m^2_t\, ,\
\widetilde{M}^2_b + m^2_b )}{Q^2}\bigg)\,\bigg]\nonumber\\
&&-\, \frac{g^2_w}{2}\,\bigg( |h_t|^2\, -\, \frac{g^2_w}{4}\bigg)\,
\ln\bigg(\frac{\widetilde{M}^2_Q + m^2_t}{Q^2}\bigg)\: -\:
\frac{g^2_w}{2}\,\bigg( |h_b|^2\, -\, \frac{g^2_w}{4}\bigg)\,
\ln\bigg(\frac{\widetilde{M}^2_Q + m^2_b}{Q^2}\bigg)\, \bigg\}\, .
\end{eqnarray}
Moreover,  we  need   to  know  the  one-loop  running   of  the  soft
supersymmetry-breaking parameter ${\rm Re}\, m^2_{12}$.  Gathering the
relevant
logarithmic terms present in the effective potential (\ref{LVeff}), we
find
\begin{eqnarray}
  \label{rem12}
{\rm  Re}\, m^{2(1)}_{12} &=& \frac{3}{16\pi^2}\, \bigg[\, |h_t|^2\, 
{\rm Re}\, (\mu A_t)\, \ln \bigg(\frac{ {\rm max}\, 
(\widetilde{M}^2_Q + m^2_t\, ,\ \widetilde{M}^2_t + 
   m^2_t)}{Q^2}\bigg)\nonumber\\
&& +\, |h_b|^2\, {\rm Re}\, (\mu A_b)\, 
\ln \bigg(\frac{ {\rm max}\, (\widetilde{M}^2_Q + m^2_b\, ,\ 
\widetilde{M}^2_b + m^2_b )}{Q^2}\bigg)\, \bigg]\ .
\end{eqnarray}
The analytic form  of $\Delta {\cal M}^2 (Q_{tb})$  in (\ref{Mth}) can
now be obtained by  subtracting the one-loop Born-improved mass matrix
${\cal  M}^{2\,  (0)}$ from  its  total  one-loop contribution  ${\cal
  M}^{2\,  (1)}$.  Here, we  have  in  mind  the charged  and  neutral
Higgs-boson mass matrices ${\cal  M}^{2\, (1)}_\pm$ and ${\cal M}^{2\,
  (1)}_N$ calculated in Section 2. More explicitly, $\Delta {\cal M}^2
(Q_{tb})$ is given by
\begin{equation}
  \label{DeltaM}
\Delta  {\cal  M}^2 (Q_{tb})\ =\ {\cal M}^{2\,  (1)}(Q_{tb})\: -\:
{\cal   M}^{2\,  (0)}\,\Big[\,{\rm Re}\,  m^{2(1)}_{12} (Q_{tb}), 
\lambda^{(1)}_i (Q_{tb})\,\Big]\, , 
\end{equation}
where ${\cal  M}^{2\, (0)}$ represents the  tree-level functional form
of  ${\cal M}^2$, expressed  in terms  of $\lambda^{(1)}_i$  and ${\rm
  Re}\, m^{2(1)}_{12}$.  Furthermore, it  is essential to stress again
that  the kinematic  parameters  involved in  (\ref{DeltaM}), such  as
masses and couplings, are evaluated at the scale $Q_{tb}$.

Another important ingredient in  the RG improvement of the Higgs-boson
mass matrices  is the analytic  two-loop result for the  Higgs quartic
couplings  $\lambda_1,   ...,  \lambda_4$.    As  has  been   done  in
(\ref{beta1})--(\ref{beta4}),  we  have  to include  two-loop  leading
logarithms,  by   appropriately  considering  the   stop  and  sbottom
thresholds.  These  two-loop leading logarithmic  contributions to the
Higgs quartic couplings, which  we denote by $\lambda^{(2)}_i$, can be
determined by solving iteratively the RG equations~\cite{CQW}. In this
way, we obtain
\begin{eqnarray}
  \label{2loopl1}
\lambda^{(2)}_1 \!\!&=&\!\! -\, \frac{6|h_b|^4}{(32\pi^2)^2}\, \bigg(
\frac{3}{2}\, |h_b|^2\: +\: \frac{1}{2}\,|h_t|^2\: -\: 8g^2_s\, \bigg)\,
\bigg[\,\ln^2\bigg(\frac{\widetilde{M}^2_Q + m^2_b}{Q^2}\bigg)\: +\:
\ln^2\bigg(\frac{\widetilde{M}^2_b + m^2_b}{Q^2}\bigg)\, \bigg]\,,\qquad\\
  \label{2loopl2}
\lambda^{(2)}_2 \!\!&=&\!\! -\, \frac{6|h_t|^4}{(32\pi^2)^2}\, \bigg(
\frac{3}{2}\, |h_t|^2\: +\: \frac{1}{2}\,|h_b|^2\: -\: 8g^2_s\, \bigg)\,
\bigg[\,\ln^2\bigg(\frac{\widetilde{M}^2_Q + m^2_t}{Q^2}\bigg)\: +\:
\ln^2\bigg(\frac{\widetilde{M}^2_t + m^2_t}{Q^2}\bigg)\, \bigg]\,,\quad\\
  \label{2loopl3}
\lambda^{(2)}_3 \!\!&=&\!\! -\, \frac{3|h_t|^2|h_b|^2}{(16\pi^2)^2}\, \bigg(
|h_t|^2\: +\: |h_b|^2\: -\: 8g^2_s\, \bigg)\nonumber\\
&&\times\, \bigg[\,\ln^2\bigg(\frac{\widetilde{M}^2_Q +
  m^2_t}{Q^2}\bigg)\: +\: 
\ln^2 \bigg(\frac{ {\rm max}\, (\widetilde{M}^2_t + m^2_t\, ,\ 
\widetilde{M}^2_b + m^2_b)}{Q^2}\bigg)\, \bigg]\,,\\
  \label{2loopl4}
\lambda^{(2)}_4 \!\!&=&\!\! -\,\lambda^{(2)}_3\, .
\end{eqnarray}
For later  convenience, we  define collectively the  sum of  the tree,
one-loop and two-loop quartic couplings as follows:
\begin{equation}
  \label{barlam}
\bar{\lambda}_i\  =\ \lambda_i\: +\: \lambda^{(1)}_i\: +\: \lambda^{(2)}_i\,,
\end{equation}
with  $i=1,2,3,4$.   Similarly, the  sum  of  the  tree, one-loop  and
two-loop  contributions  to  the  soft-bilinear Higgs  mixing  may  be
defined as
\begin{equation}
  \label{barm12}
{\rm  Re}\, \bar{m}^2_{12}\ =\ {\rm  Re}\, m^{2}_{12}\: +\:
{\rm  Re}\, m^{2(1)}_{12}\: +\: {\rm  Re}\, m^{2(2)}_{12}\ .
\end{equation}
As we  see below, knowledge  of the two-loop contribution  ${\rm Re}\,
m^{2(2)}_{12}$  is not  required in  the one-loop  RG
improvement of the MSSM Higgs potential.

Given  the above  definitions of  the quartic  couplings and  the soft
Higgs-mixing  parameter in (\ref{barlam})  and (\ref{barm12}),  we can
express  the  one-  and  two-loop  leading  logarithmic  contributions
$\overline{\cal  M}^2 (m_t)$  to  ${\cal M}^2(m_t)$  by  means of  the
two-loop Born-improved mass matrix:
\begin{equation}
  \label{barM}
\overline{\cal M}^2 (m_t)\ =\ 
{\cal   M}^{2\,  (0)}\,\Big[\,{\rm Re}\,  \bar{m}^2_{12} (m_t), 
\bar{\lambda}_i (m_t)\,\Big]\, .
\end{equation}
Note  that $\overline{\cal  M}^2 (m_t)$  also includes  the tree-level
terms.   As   has  also   been  stated  explicitly   in  (\ref{barM}),
$\overline{\cal M}^2 (m_t)$ is expressed in terms of mass and coupling
parameters evaluated at the top-quark-mass scale.

The last ingredient for completing  the programme of RG improvement of
the Higgs-boson mass matrices is knowledge of the analytic expressions
for the anomalous dimension  factors that occur in (\ref{Mth}).  These
analytic  expressions are  given  below for  the  charged and  neutral
Higgs-boson cases separately.

Adopting the framework outlined  above~\cite{CQW}, it is not difficult
to compute the RG-improved  charged Higgs-boson mass $M^2_{H^+} (m_t)$
at the top-mass scale through the relation:
\begin{equation}
   \label{MHmt}
M^2_{H^+} (m_t)\ =\ \overline{M}^2_{H^+} (m_t)\ +\ 
\Big[ \xi^+_1 (m_t)\,\xi^-_2 (m_t)\Big]^{-1}\, 
(\Delta M^2_{H^+})^{\tilde{t}\tilde{b}} (Q_{tb} )\ +\ 
(M^{2\,(1)}_{H^+})^{tb} (m_t)\, ,
\end{equation}
where  $Q^2_{tb}  =  {\rm  max}\, \Big(  \widetilde{M}^2_Q  +  m^2_t,\ 
\widetilde{M}^2_t + m^2_t,\ \widetilde{M}^2_b + m^2_b \Big)$, 
and $\xi^+_1
(m_t)$ and $\xi^-_2 (m_t)$ are  the anomalous dimension factors of the
charged Higgs fields $\phi^+_1$ and $\phi^-_2$, respectively:
\begin{equation}
  \label{xi12}
\xi^+_1 (m_t)\ =\ 1\: +\: \frac{3|h_b|^2}{32\pi^2}\,
\ln\frac{Q^2_{tb}}{m^2_t}\ ,\qquad
\xi^-_2 (m_t)\ =\ 1\: +\: \frac{3|h_t|^2}{32\pi^2}\,
\ln\frac{Q^2_{tb}}{m^2_t}\ .
\end{equation}
Further,   $\overline{M}^2_{H^+}  (m_t)$   is  the   squared  two-loop
Born-improved charged Higgs-boson mass given by
\begin{equation}
  \label{Mbarplus}
\overline{M}^2_{H^+}   (m_t)\ =\ \frac{{\rm Re}\, \bar{m}^2_{12}
  (m_t)}{\sin\beta (m_t)\, \cos\beta (m_t)}\ +\ 
\frac{1}{2}\, \bar{\lambda}_4 (m_t)\,  v^2 (m_t)\, 
\end{equation}
and   $(\Delta   M^2_{H^+})^{\tilde{t}\tilde{b}}$   is  the   one-loop
scale-invariant part that contains the stop and sbottom contributions:
\begin{eqnarray}
  \label{deltamhp}
(\Delta M^2_{H^+})^{\tilde{t}\tilde{b}} (Q_{tb} )\ & = & 
-\, \frac{1}{\sin\beta  (m_t) \cos\beta (m_t)}\, \Big[\,
({\cal M}^{2\,(1)}_{\pm})^{\tilde{t}\tilde{b}}_{12} (Q_{tb})\: +\: 
{\rm Re}\, m_{12}^{2(1)}(Q_{tb}) \nonumber\\
&&+\, \frac{1}{2}\, \lambda^{(1)}_4 (Q_{tb})\,  
                            v_1 (Q_{tb})\, v_2 (Q_{tb})\, \Big]\, ,
\end{eqnarray}
where the scale at which  the kinematic parameters are to be evaluated
has been indicated explicitly.   In (\ref{deltamhp}), the term between
brackets  is the  threshold  contribution to  the off-diagonal  matrix
element  of   the  charged-Higgs-boson  mass   matrix,  where  $({\cal
  M}^{2\,(1)}_{\pm})^{\tilde{t}\tilde{b}}_{12}$  denotes  the one-loop
contribution    of   the    third-generation   squarks    to   $({\cal
  M}^2_{\pm})_{12}$.  Finally, $(M^{2\,(1)}_{H^+})^{tb}$ describes the
one-loop quark contribution to $M^2_{H^+}$ (see (\ref{MHmt})).

We remark that the charged-Higgs-boson mass matrix receives the common
anomalous dimension factor $\xi_1^+  (m_t) \xi_2^- (m_t)$, even though
different matrix  elements of ${\cal M}^2_\pm$ are  involved.  This is
because ${\cal  M}^2_\pm$ must possess a vanishing  determinant at any
RG scale $Q^2$, as one of  its mass eigenstates must correspond to the
massless would-be  Goldstone boson  $G^+$ that forms  the longitudinal
component  of  the  $W^+$  boson.   As a  consequence,  the  following
relations  among the  matrix elements  of the  RG-frozen  part $\Delta
{\cal M}^2_\pm$ of the charged-Higgs-boson mass matrix are obtained:
\begin{eqnarray}
  \label{MHdet}
(\Delta {\cal M}^2_\pm )_{11} (Q_{tb}) \!&=&\! -\tan\beta (Q_{tb})\
(\Delta {\cal M}^2_\pm )_{12} (Q_{tb})\, ,\nonumber\\
(\Delta {\cal M}^2_\pm )_{22} (Q_{tb})\!&=&\! -\cot\beta (Q_{tb})\  
(\Delta {\cal M}^2_\pm )_{21} (Q_{tb})\, .\quad
\end{eqnarray}
After including  the    RG running due to  the   Higgs-boson anomalous
dimensions, we find
\begin{eqnarray}
  \label{MHHdet}
\frac{(\Delta {\cal M}^2_\pm )_{11} (Q_{tb})}{[\xi^+_1 (m_t)]^2} \!&=&\!
-\tan\beta (m_t)\  
\frac{(\Delta {\cal M}^2_\pm )_{12} (Q_{tb})}{\xi^+_1 (m_t)\, \xi^-_2 (m_t)}
\, ,\nonumber\\
\frac{(\Delta {\cal M}^2_\pm )_{22} (Q_{tb})}{[\xi^-_2 (m_t)]^2} \!&=&\!
-\cot\beta (m_t)\ 
\frac{(\Delta {\cal M}^2_\pm )_{12} (Q_{tb})}{\xi^+_1 (m_t)\, \xi^-_2
  (m_t)}\, ,
\end{eqnarray}
where we have used the RG relation: 
\begin{equation}
  \label{tbmat}
\tan\beta (Q_{tb})\ =\ \frac{\xi^+_1 (m_t)}{\xi^-_2 (m_t)}\ \tan\beta (m_t)\, .
\end{equation}
As a  consequence of  this last  relation, it is  evident that  the RG
running of  the different matrix  elements of $\Delta  {\cal M}^2_\pm$
may be  expressed in terms of  the running of  $(\Delta {\cal M}^2_\pm
)_{12}$ and the value of $\tan\beta$ at the scale $m_t$.

Correspondingly,  the  RG-improved  neutral-Higgs-boson  mass  matrix
${\cal M}^2_N$ may be computed by
\begin{eqnarray}
  \label{MN}
({\cal M}^2_N)_{ij} (m_t) \!&=&\! (\overline{{\cal M}}^2_N)_{ij} (m_t)\ +\
\Big[\, \xi^{\tilde{t}}_{ij} (m_t)\, \Big]^{-1}\, 
(\Delta {\cal M}^{2}_N)^{\tilde{t}}_{ij} (Q_t)\ +\ 
\Big[\, \xi^{\tilde{b}}_{ij} (m_t)\,\Big]^{-1}\, 
(\Delta {\cal M}^{2}_N)^{\tilde{b}}_{ij} (Q_b)\nonumber\\
&& +\, ({\cal M}^{2\,(1)}_N)^{tb}_{ij} (m_t)\, ,
\end{eqnarray}
with $i,j  = 1,2,3$,  $Q^2_t = {\rm  max}\, \Big(  \widetilde{M}^2_Q +
m^2_t,\  \widetilde{M}^2_t +  m^2_t \Big)$  and $Q^2_b  =  {\rm max}\,
\Big( \widetilde{M}^2_Q  + m^2_b,\  \widetilde{M}^2_b + m^2_b  \Big)$. 
Notice that, unlike the charged-Higgs-boson case, one has to introduce
here  two decoupling  scales  $Q_t$  and $Q_b$,  as  the stop/top  and
sbottom/bottom  loop   effects  occur  separately   in  the  threshold
contributions.   The parameters $\xi^{\tilde{q}}_{ij}$  (with $q  = t,
b$)  are  the  anomalous-dimension  factors  related  to  the  neutral
Higgs-boson fields
\begin{eqnarray}
  \label{xij}
\xi^{\tilde{q}}_{ij} (m_t)\ =\ \xi^{\tilde{q}}_{ji} (m_t)\ =\ \left\{
\begin{array}{l} \xi^{\tilde{q}}_i (m_t)\, \xi^{\tilde{q}}_j (m_t)\, ,
\quad {\rm for}\ \ i,j=1,2\\
\xi^{\tilde{q}}_1 (m_t)\, \xi^{\tilde{q}}_2 (m_t)\,,
\quad {\rm for}\ \ i = 1,2,3\ \ {\rm and}\ \ j=3 \end{array} \right.\, ,
\end{eqnarray}
with 
\begin{eqnarray}
  \label{xitb12}
\xi^{\tilde{t}}_1 (m_t) \!&=&\! 1\: +\: \frac{3 |h_b|^2}{32\pi^2}\,
\ln\frac{Q^2_t}{m^2_t}\ ,\qquad   
\xi^{\tilde{t}}_2 (m_t) \ =\ 1\: +\: \frac{3 |h_t|^2}{32\pi^2}\,
\ln\frac{Q^2_t}{m^2_t}\ ,\nonumber\\
\xi^{\tilde{b}}_1 (m_t) \!&=&\! 1\: +\: \frac{3 |h_b|^2}{32\pi^2}\,
\ln\frac{Q^2_b}{m^2_t}\ ,\qquad   
\xi^{\tilde{b}}_2 (m_t) \ =\ 1\: +\: \frac{3 |h_t|^2}{32\pi^2}\,
\ln\frac{Q^2_b}{m^2_t}\ . 
\end{eqnarray}
We  should  observe  that,  for  the  very  same  reasons  as  in  the
charged-Higgs-boson case, the vanishing  of the determinants of ${\cal
  M}^2_P$ and ${\cal M}^2_{SP}$ at any $Q^2$ scale leads to the common
anomalous dimension  factor $\xi_{i3} (m_t)  = \xi_{3i} (m_t)  = \xi_1
(m_t)\, \xi_2  (m_t)$ in  the calculation of  ${\cal M}^2_N  (m_t)$ in
(\ref{MN}). Since this last  fact involves the matrix elements $({\cal
  M}^2_P)_{12}$,     $({\cal      M}^2_{SP})_{12}$     and     $({\cal
  M}^2_{SP})_{21}$,  the  corresponding  matrix elements  of  $(\Delta
{\cal M}^2_N)_{i3}^{\tilde{q}}$ in (\ref{MN}) are given by
\begin{eqnarray}
  \label{DMNqi3}
(\Delta  {\cal M}^2_N)_{13}^{\tilde{q}}(Q_q) \!&=&\!  
\frac{({\cal M}^2_{SP})_{12}^{\tilde{q}}(Q_q)}
{\cos\beta (m_t)}\ , \quad
(\Delta  {\cal M}^2_N)_{23}^{\tilde{q}}(Q_q)\ =\ -\, 
\frac{({\cal M}^2_{SP})_{21}^{\tilde{q}}(Q_q)}
{\sin\beta (m_t)}\ , \nonumber\\
(\Delta  {\cal M}^2_N)_{33}^{\tilde{q}} (Q_q) \!&=&\! -\, 
\frac{(\Delta  {\cal M}^2_P)_{12}^{\tilde{q}}(Q_q)}{\sin\beta (m_t)
  \cos\beta (m_t)}\ , 
\end{eqnarray}
where  the RG-scale  dependence of  the involved  quantities  has been
displayed explicitly.

In  (\ref{MN}), the $(3\times  3)$ matrices  $\overline{{\cal M}}^2_N$
and  $(\Delta  {\cal  M}^{2}_N)^{\tilde{q}}$ (with  $q=t,b$)  describe
respectively  the  two-loop  Born-improved  effects and  the  one-loop
threshold contributions  associated with  the decoupling of  the heavy
squark states:
\begin{eqnarray}
  \label{barMN}
\overline{{\cal M}}^2_N (m_t) &=& {\cal M}^{2\, (0)}_N\,
\Big[\, {\rm Re}\,\bar{m}_{12}^2 (m_t), 
\bar{\lambda}_1 (m_t), \bar{\lambda}_2 (m_t), 
\bar{\lambda}_{34}(m_t)\,\Big]\, ,\\
  \label{DMNq}
(\Delta {\cal M}^{2}_N)^{\tilde{q}}(Q_q) &=& \nonumber\\
&&\hspace{-1.5cm}
({\cal M}^{2\, (1)}_N)^{\tilde{q}}(Q_q)\: -\:
{\cal M}^{2\, (0)}_N
\Big[\, {\rm Re}\,m_{12}^{2(1),\tilde{q}}(Q_q), 
\lambda_1^{(1),\tilde{q}} (Q_q), \lambda_2^{(1),\tilde{q}} (Q_q), 
\lambda_{34}^{(1),\tilde{q}} (Q_q)\,\Big]\ ,\qquad
\end{eqnarray}
where   $\lambda^{(1)}_{34}  =   \lambda^{(1)}_3   +  \lambda^{(1)}_4$
(likewise  $\bar{\lambda}_{34} =  \bar{\lambda}_3  + \bar{\lambda}_4$)
and ${\cal M}^{2\, (0)}_N$ is the tree-level functional form of ${\cal
  M}^2_N$:
\begin{eqnarray}
  \label{MN0}
{\cal M}^{2\, (0)}_N &=&\\
&&\hspace{-2.23cm} \frac{{\rm Re}\, m^2_{12} }
{\sin\beta\, \cos\beta}\, \left( \begin{array}{ccc}
\sin^2\beta & -\sin\beta\cos\beta & 0\\
-\sin\beta\cos\beta & \cos^2\beta & 0 \\
0 & 0 & 1 \end{array} \right)\: -\: v^2 \left( \begin{array}{ccc}
2\lambda_1 \cos^2\beta & \lambda_{34}\sin\beta\cos\beta & 0\\
\lambda_{34} \sin\beta\cos\beta & 2\lambda_2\sin^2\beta & 0 \\
0 & 0 & 0 \end{array} \right),\nonumber
\end{eqnarray}
with     $\lambda_{34}=\lambda_3    +\lambda_4$.      As     in    the
charged-Higgs-boson     case,     we     write     $({\cal     M}^{2\,
  (1)}_N)^{\tilde{q}}$ to  denote the one-loop part  of ${\cal M}^2_N$
containing  the  contributions of  the  third-generation squarks,  and
$({\cal   M}^{2\,(1)}_N)^{tb}$  to   denote  its   fermionic  one-loop
counterpart.

The resulting RG-improved Higgs-boson mass matrix ${\cal M}^2_N (m_t)$
is  a  symmetric,  positive-definite  $(3\times 3)$  matrix,  and  can
therefore be diagonalized by an orthogonal transformation as follows:
\begin{equation}
  \label{Odiag}
O^T\, {\cal M}^2_N (m_t)\, O\ =\ {\rm diag}\, \Big[\, M^2_{H_1} (m_t),\ 
M^2_{H_2} (m_t),\ M^2_{H_3} (m_t)\, \Big]\ ,   
\end{equation}
where we  have defined  the Higgs fields  such that  their RG-improved
masses satisfy the inequality:
\begin{equation}
  \label{masdef}
M_{H_1}(m_t)\ \leq\ M_{H_2}(m_t)\ \leq\ M_{H_3}(m_t)\, .
\end{equation}
Notice that  our convention in (\ref{Odiag}) differs  from that chosen
in \cite{PW},  as we assign  the Higgs fields  in the reversed  order. 
Analytic  expressions for  $M_{H_i} (m_t)$  and $O$  are  presented in
Appendix B.

Before  closing this  Section, two  important  remarks are  in order.  
First, we observe that the free kinematic parameters of the MSSM Higgs
sector are
\begin{eqnarray}
  \label{freepar}
&&M_{H^+} (m_t)\,,\quad \tan\beta (m_t)\,,\quad \mu (Q_{tb})\,,
\quad A_t(Q_{tb})\,,\quad A_b(Q_{tb})\,,\nonumber\\  
&&\widetilde{M}^2_Q (Q_{tb})\, ,\quad \widetilde{M}^2_t (Q_{tb})\, ,\quad
\widetilde{M}^2_b (Q_{tb})\, .
\end{eqnarray}
In fact,  the soft  Higgs-mixing parameter ${\rm  Re}\, \bar{m}^2_{12}
(m_t)$ may  be substituted by the squared  RG-improved mass $M^2_{H^+}
(m_t)$   of   the   charged   Higgs  boson   (cf.   (\ref{MHmt})   and
(\ref{Mbarplus})) in the neutral Higgs-boson mass matrix ${\cal M}^2_N
(m_t)$ in (\ref{MN}).

Secondly,  we reiterate  the fact  that ${\rm  Im}\, m^2_{12}$  can be
renormalized independently,  without affecting the  renormalization of
the physical parameters of the theory~\cite{APLB}.  As was stressed in
Section 3,  the $\xi =  0$ scheme of  renormalization gives rise  to a
considerable   simplification,   since  we   can   get   rid  of   the
radiatively-induced   phase  $\xi$  between   the  two   Higgs  vacuum
expectation  values in  the  analytic expressions  of the  Higgs-boson
masses  and mixing  angles.   For example,  within  the above  $\xi=0$
scheme, the  mass renormalization of $H^+$ may  be entirely reabsorbed
by  a  corresponding  renormalization  of ${\rm  Re}\,  m^2_{12}$  and
$\lambda_4$.   In other  words,  it  can be  shown  that $M_{H^+}$  is
$Q^2$-independent,  after  including the  RG  running  of ${\rm  Re}\,
m^2_{12}$ and  $\lambda_4$, denoted as  $\gamma_{{\rm Re}\, m^2_{12}}$
and  $\beta_{\lambda_4}$.  For  simplicity,  we assume  that only  the
third  generation  of   squarks  contributes  to  $\gamma_{{\rm  Re}\,
m^2_{12}}$,  since fermions  do not  contribute to  the RG  running of
${\rm  Im}\, m^2_{12}$.   The  analytic forms  of $\gamma_{{\rm  Re}\,
m^2_{12}}$ and $\beta_{\lambda_4}$ are given by
\begin{eqnarray}
  \label{gamma12}
\gamma_{{\rm Re}\, m^2_{12}} \!&=&\! -\, \frac{3}{16\pi^2}\,
\bigg[\, |h_t|^2\, {\rm Re}\, (\mu A_t) \: +\: |h_b|^2\, {\rm Re}\, (\mu
A_b)\, \bigg]\, ,\\
  \label{betl4}
\beta_{\lambda_4} \!&\equiv&\! \frac{d\lambda^{(1)}_4}{d\ln Q^2}\ =\ 
-\, \frac{3}{16\pi^2}\, \bigg[\, 2|h_t|^2|h_b|^2\: -\:
\frac{g^2_w}{2}\, \Big( |h_t|^2\, +\,|h_b|^2 \Big) \: +\:
\frac{g^4_w}{4}\, 
\bigg]\, .
\end{eqnarray}
Obviously, the RG running of  ${\rm Re}\, m^2_{12}$ due to $\tilde{t}$
and $\tilde{b}$ is  only relevant for non-zero values  of $\mu A_t$ and
$\mu A_b$.  Employing (\ref{MHplus}) and examining only the $\ln
Q^2$-dependent part, one can verify that
\begin{equation}
  \label{MHrun}
\frac{dM^2_{H^+}}{d\ln Q^2}\ 
\propto\ -\, \gamma_{{\rm Re}\, m^2_{12}}\: -\: \frac{1}{2}\,
\beta_{\lambda_4} v_1v_2\: +\: \frac{3}{32\pi^2}\, \bigg(\,
\bigg< \frac{\partial^2\, {\rm Tr}\, \widetilde{\cal M}^4 }{\partial 
\phi^+_1\, \partial \phi^-_2}\bigg>\: -\: \frac{i}{v_1}\, 
\bigg< \frac{\partial\, {\rm Tr}\, \widetilde{\cal M}^4 }{\partial
  a_2}\bigg>\, \bigg)\ =\ 0\, , 
\end{equation}
as it should be. As can  also be seen from (\ref{MHrun}), an important
role  in this proof  is played  by the  necessary CP-odd  tadpole term
$\big< \partial\, {\rm  Tr}\, \widetilde{\cal M}^4/\partial a_2 \big>$
\cite{APLB}.

\setcounter{equation}{0}
\section{Effective Top and Bottom Yukawa Couplings}

In addition  to the  RG improvement of  the Higgs-boson  mass matrices
discussed  in  the  previous  Section,  we  consider  here  a  further
improvement  related to the  non-logarithmic threshold  corrections to
the top- and bottom-quark  Yukawa couplings.  Specifically, apart from
the  usual RG  running,  the effective  top-  and bottom-quark  Yukawa
couplings  obtain additional non-logarithmic  threshold contributions,
which are induced by the decoupling of the heavy SUSY states at a high
scale, {\em  e.g.}, $Q_{tb}$.  For  the bottom-quark Yukawa  case, the
one-loop  RG relation  between the  bottom mass  and  the bottom-quark
Yukawa  coupling at  the scale  $Q_{tb}$ receives  quantum corrections
that       also        include       terms       proportional       to
$\tan\beta$~\cite{EMa,dmb,sola}.   Since  these   last  terms  can  be
significant  for  large  values  of  $\tan\beta$,\footnote[4]{For  the
$\tau$-lepton Yukawa coupling,  the corresponding enhanced $\tan\beta$
terms  are much  smaller, because  they are  proportional to  the weak
gauge  couplings \cite{dmb,CMW}.}  we  must resum  them within  the RG
approach, so that the actual  size of the radiative corrections to the
Higgs-boson     masses    and     couplings     can    properly     be
extracted~\cite{CMW,CGNW}.  For the  top-quark case, instead, although
one-loop suppressed, the respective corrections can still give rise to
an  enhancement of  up to  4 GeV  in the  prediction for  the lightest
Higgs-boson mass~\cite{CHHHWW,CHWW}, and  therefore should be included
in the computation.

There may  also be important CP-violating one-loop  corrections to the
bottom-   and  top-quark   Yukawa  couplings,   in  addition   to  the
CP-violating  effects induced  by the  radiative mixing  of  the Higgs
states, which were  considered in Sections 2 and 3  in detail.  In the
leptonic sector,  these CP-violating vertex  corrections are generally
small \cite{BKW}.  However,  the CP-violating radiative corrections to
the  couplings of  the  Higgs  bosons to  $b$  quarks are  significant
\cite{PW},  because  of  the  large  Yukawa  and  colour-enhanced  QCD
interactions \cite{EMa}.  In particular,  the radiative effects of the
Higgs-boson couplings to the bottom quarks can be further enhanced, if
the  respective  Higgs-mass eigenstate  couples  predominantly to  the
Higgs  doublet $\Phi_2$ \cite{dmb,sola},  as the  tree-level $b$-quark
Yukawa coupling is  suppressed in this case. For  a general discussion
of the form  and the origin of these finite  Yukawa corrections to the
third-generation quark masses, the  reader is referred to the original
literature  \cite{dmb,sola}.   In  the  following,  we  give  a  brief
discussion of  the non-logarithmic corrections  to the top  and bottom
Yukawa couplings, and pay special attention to the CP-violating vertex
effects.

We start our discussion by considering the effective Lagrangian of
the $b$-quark Yukawa coupling~\cite{CMW,PW}:
\begin{equation}
  \label{phibb}
-\,{\cal L}_{\phi^0 \bar{b}b}\ =\
 (h_b + \delta h_b)\, \phi_1^{0*}\, \bar{b}_R b_L\: +\:
\Delta h_b\, \phi^{0*}_2\, \bar{b}_R b_L\ +\  {\rm  h.c.}\, ,
\end{equation}
with 
\begin{eqnarray}
  \label{ddhb}
\frac{\delta h_b}{h_b} &=& 
-\frac{2 \alpha_s}{3\pi} m^*_{\tilde{g}} A_b I(m_{\tilde{b}_1}^2,
m_{\tilde{b}_2}^2,|m_{\tilde{g}}|^2)\ 
-\ \frac{|h_t|^2}{16\pi^2} |\mu|^2
I(m_{\tilde{t}_1}^2,m_{\tilde{t}_2}^2,|\mu|^2)\, ,\\
  \label{Dhb}
\frac{\Delta h_b}{h_b} &=& 
\frac{2 \alpha_s}{3\pi} m^*_{\tilde{g}} \mu^* I(m_{\tilde{b}_1}^2,
m_{\tilde{b}_2}^2,|m_{\tilde{g}}|^2)\ +\ \frac{|h_t|^2}{16\pi^2} A^*_t \mu^*
I(m_{\tilde{t}_1}^2,m_{\tilde{t}_2}^2,|\mu|^2)\, ,
\end{eqnarray}
where  $\alpha_s  =  g^2_s/(4\pi)$  is the  SU(3)$_c$ coupling
strength, and $I(a,b,c)$ is the one-loop function
\begin{equation}
I(a,b,c)\ =\ \frac{ a b \ln (a/b) + b c \ln (b/c) + a c \ln (c/a)}
{(a-b)(b-c)(a-c)}\ .
\end{equation}
The $b$-quark Yukawa coupling  $h_b(Q_{tb})$  is then related  to  the
running $b$-quark mass $m_b (Q_{tb})$ by
\begin{equation}
  \label{yukb}
h_b\ =\ 
\frac{g_w m_b}{\sqrt{2} M_W \cos\beta\, 
[\,1 + \delta h_b/h_b + (\Delta h_b/h_b) \tan\beta\, ]}\ ,
\end{equation}
where $\delta h_b/h_b$ and  $\Delta h_b/h_b$ are given in (\ref{ddhb})
and  (\ref{Dhb}),  respectively.   The  running  $b$-quark  mass  $m_b
(Q_{tb})$ is obtained by means of the RG running of the $b$-quark mass
from  the  scale  $m_b$.   In  (\ref{yukb}),  we  have  redefined  the
right-handed $b$-quark superfield, so that the physical $b$-quark mass
is  positive.   Under  such  a  field redefinition,  only  the  Yukawa
coupling $h_b$  becomes complex, while the phases  of $\delta h_b/h_b$
and $\Delta h_b/h_b$ as well  as those of the supersymmetry-breaking
parameters
do not change. Moreover, since only the moduli of the Yukawa couplings
$h_b$ and $h_t$  enter the field-dependent quark and  squark masses in
(\ref{mudbar})  and   (\ref{msqud}),  the  neutral   Higgs-boson  mass
matrices remain  unaffected by the above field  redefinition. Also, we
have  checked  that  the   very  same  property  of  invariance  under
rephasings  of $h_t$ and  $h_b$ persists  for the  charged Higgs-boson
mass matrix as well.  At this point, it is interesting to observe that
the sum $\delta  h_b + \Delta h_b \tan\beta$  in (\ref{yukb}) receives
two sorts of quantum corrections, one originating from QCD effects and
another  from  a  chargino-mediated  graph.   The  QCD  correction  is
proportional  to   the  hermitean  conjugate   of  the  sbottom-mixing
parameter $X_b  = A_b - \mu^* \tan\beta$,  whilst the chargino-induced
diagram~\cite{dmb} depends linearly  on the stop-mixing parameter $X_t
= A_t - \mu^* \cot\beta$.

The effective Lagrangian  describing the $t$-quark Yukawa  coupling is
given by
\begin{equation}
  \label{phitt}
-\,{\cal L}_{\phi^0 \bar{t}t}\ =\
\Delta h_t\, \phi^0_1\, \bar{t}_R t_L\: +\: 
 (h_t + \delta h_t)\, \phi^0_2\, \bar{t}_R t_L\ +\  {\rm  h.c.}\, .
\end{equation}
The corresponding relation for $h_t$ as  a function of $m_t$ may easily
be   determined analogously by  the  effective    Lagrangian
(\ref{phitt}), and reads
\begin{equation}
  \label{yukt}
h_t\ =\ \frac{g_w m_t}{\sqrt{2} M_W \sin\beta\, 
[\,1 + \delta h_t/h_t + (\Delta h_t/h_t) \cot\beta\, ]}\ ,
\end{equation}
with
\begin{eqnarray}
  \label{Dht}
\frac{\Delta h_t}{h_t} &=& 
\frac{2 \alpha_s}{3\pi} m^*_{\tilde{g}} \mu^* I(m_{\tilde{t}_1}^2,
m_{\tilde{t}_2}^2,|m_{\tilde{g}}|^2)\ +\ \frac{|h_b|^2}{16\pi^2} 
A_b^* \mu^* I(m_{\tilde{b}_1}^2,m_{\tilde{b}_2}^2,|\mu|^2)\, ,\\
  \label{ddht}
\frac{\delta h_t}{h_t} &=& 
-\frac{2 \alpha_s}{3\pi} m^*_{\tilde{g}} A_t I(m_{\tilde{t}_1}^2,
m_{\tilde{t}_2}^2,|m_{\tilde{g}}|^2)\ 
-\ \frac{|h_b|^2}{16\pi^2} |\mu|^2
I(m_{\tilde{b}_1}^2,m_{\tilde{b}_2}^2,|\mu|^2)\, .
\end{eqnarray}
As in  the case of  the $b$-quark Yukawa  coupling, we have  to make a
judicious phase  rotation of  the right-handed  $t$-quark  superfield,
such that  the physical top-quark mass   becomes positive.  Again, one
can show that  such a field redefinition does  not change the analytic
results of the RG analysis.

At  this   stage,  it  is   important  to  remark  that,   within  the
RG-resummation  approach described in  Section 3,  the non-logarithmic
corrections must be treated as threshold effects and hence they should
only  contribute  to  the  RG-frozen  part  of  the  Higgs-boson  mass
matrices,  generically  denoted  as  $\Delta  {\cal  M}^2  (Q_{tb})$.  
Therefore,  the  decoupling  procedure  for the  heavy  squark  states
requires that the effective  $b$- and $t$-quark Yukawa couplings given
by (\ref{yukb}) and (\ref{yukt}) are  evaluated at the scale $Q_{tb}$. 
As we  discuss in Section 5,  these additional Yukawa
corrections can lead to observable effects in Higgs-boson searches.

It is now straightforward to obtain the interaction Lagrangians of the
Higgs-boson mass eigenstates $H_i$ to  the up- and down-type quarks,
collectively  denoted  as $u$    and  $d$. Taking into  account   both
CP-violating self-energy and vertex effects, we find
\begin{equation}
  \label{Hff}
{\cal L}_{H\bar{f}f}\ =\ - \sum_{i=1}^3\, H_i\,
\bigg[\,\frac{g_w m_d}{2 M_W}\, \bar{d}\,\Big( g^S_{H_idd}\, +\,
ig^P_{H_idd}\gamma_5 \Big)\, d\: +\: \frac{g_w m_u}{2 M_W}
\, \bar{u}\,\Big( g^S_{H_iuu}\, +\,
ig^P_{H_iuu}\gamma_5 \Big)\, u\, \bigg]\, ,
\end{equation}
with
\begin{eqnarray}
  \label{gSHdd}
g^S_{H_idd} & =& \frac{1}{h_d\, +\, \delta h_d\, +\, \Delta h_d
\tan\beta }\ \bigg\{\,
 {\rm Re}(h_d + \delta h_d)\, \frac{O_{1i}}{\cos\beta}\:
+\: {\rm Re}(\Delta h_d)\, \frac{O_{2i}}{\cos\beta}  
\nonumber\\
&&-\,  \Big[{\rm Im}(h_d + \delta h_d) \tan\beta  
-  {\rm Im} (\Delta h_d)\Big]\, O_{3i}\, \bigg\}\, ,\\
  \label{gPHdd}
g^P_{H_idd} & =&  \frac{1}{h_d\, +\, \delta h_d\, +\, \Delta h_d
\tan\beta }\ \bigg\{\, \Big[\, {\rm Re}\left(\Delta h_d\right)\, -\,
{\rm Re}(h_d + \delta h_d) \tan\beta\, \Big]\, O_{3i} \nonumber\\
&&-\, {\rm Im}(h_d + \delta h_d)\, \frac{O_{1i}}{\cos\beta}\:  
-\:  {\rm Im}(\Delta h_d)\, \frac{O_{2i}}{\cos\beta}\  \bigg\}\ ,\\  
  \label{gSHuu}
g^S_{H_iuu} & = & \frac{1}{h_u\, +\, \delta h_u\, +\, \Delta h_u
\cot\beta }\ \bigg\{\,
 {\rm Re}(h_u + \delta h_u)\, \frac{O_{2i}}{\sin\beta}
+\: {\rm Re}(\Delta h_u )\, \frac{O_{1i}}{\sin\beta} \nonumber\\
&&-\, \Big[\, {\rm Im}(h_u + \delta h_u) \cot\beta\, -\, 
{\rm Im}(\Delta h_u)\, \Big]\, O_{3i}\, \bigg\}\, ,\\
  \label{gPHuu} 
g^P_{H_iuu} & = & \frac{1}{h_u\, +\, \delta h_u\, +\, \Delta h_u
\cot\beta }\ \bigg\{\, \Big[\, {\rm Re}(\Delta h_u)\, -\,
{\rm Re}(h_u + \delta h_u) \cot\beta\, \Big]\, O_{3i} \nonumber\\
&&-\, {\rm Im}(h_u + \delta h_u)\, \frac{O_{2i}}{\sin\beta}\: -\: 
{\rm Im}(\Delta h_u)\, \frac{O_{1i}}{\sin\beta}\, \bigg\}\, ,
\end{eqnarray}
where the Higgs scalar and pseudoscalar couplings are normalized with
respect to their SM values.

Finally,  it  is     interesting to  investigate   the   behaviour  of
self-energy- and vertex-type CP  violation in the decoupling limit of
a  heavy charged Higgs  boson in  the  MSSM.  Thus,  for values of the
charged Higgs  mass $M_{H^+}  \gg M_Z$, one  has $O_{31}\to  0$, while
$O_{11}\to \cos\beta$ and $O_{21}\to  \sin\beta$.  In this limit,  the
scalar components of the $H_1 dd$ and  $H_1 uu$ couplings acquire the
known SM form,  given    by $g_w m_d/(2M_W)$  and    $g_w m_u/(2M_W)$,
respectively, where
\begin{eqnarray}
  \label{mtb}
m_d &=& \frac{1}{\sqrt{2}}\, \Big(\, h_d\: +\: \delta h_d\: +\: \Delta h_d
\tan\beta\, \Big)\, v_1\, ,\nonumber\\
m_u &=& \frac{1}{\sqrt{2}}\, \Big(\, h_u\: +\: \delta h_u\: +\: \Delta h_u
\cot\beta\, \Big)\, v_2\, 
\end{eqnarray}
have  already  been  defined   to  be  positive  in  (\ref{yukb})  and
(\ref{yukt}).  For similar reasons, the pseudoscalar parts of the $H_1
dd$  and  $H_1 uu$  couplings  vanish,  as  they are  proportional  to
$O_{31}$ and ${\rm Im}\, m_{d,u} = 0$.  On the other hand, in the same
large $M_{H^+}$  limit, the scalar and  pseudoscalar couplings of both
the  two heaviest  Higgs bosons  $H_2$ and  $H_3$ to  the up  and down
fermions do  not vanish. We  can therefore conclude  that CP-violating
self-energy and  vertex effects do  not decouple in the  heavy neutral
Higgs  sector. In  the  next section,  we  demonstrate explicitly  the
aforementioned (non-)decoupling features  of CP violation by analyzing
specific phenomenological examples.

\setcounter{equation}{0}
\section{Phenomenological Discussion}

In  this  Section  we  discuss the  phenomenological  implications  of
radiative  Higgs-sector  CP  violation  in the  MSSM  for  Higgs-boson
searches  at high-energy  colliders.  We  focus our  attention  on the
physics potential for discovering  Higgs bosons with mixed CP parities
at LEP~2 and  the upgraded Tevatron collider, and  also comment on the
enhanced search capabilities offered by the LHC.

At  the  LEP~2  and  Tevatron  colliders,  neutral  Higgs  bosons  are
predominantly produced via the  Higgs-strahlung processes in $e^+ e^-$
and $q\bar{q}$ collisions, such as $e^+e^-\to Z^*\to Z H_i$~\cite{ZH},
$q\bar{q}\to  Z^*\to  Z   H_i$  and  $q\bar{q}\to  W^{\pm^*}\to  W^\pm
H_i$~\cite{Yildiz}, with  $i =  1,2,3$. If the  next-to-lightest Higgs
boson is not too heavy, Higgs bosons can also be produced copiously in
pairs  through   the  reactions:   $e^+e^-\to  Z^*\to  H_i   H_j$  and
$q\bar{q}\to Z^*\to  H_i H_j$. In addition to  the Higgs-boson masses,
the Higgs-boson couplings  to the gauge fields play  an essential role
in  our forthcoming discussion.   The effective  Lagrangians governing
the interactions of  the Higgs bosons with the  $W^\pm$ and $Z$ bosons
are given by \cite{PW}
\begin{eqnarray}
  \label{HVV}
{\cal L}_{HVV} &=& g_wM_W\, \sum\limits_{i=1}^3\,
 g_{H_iVV}\, \Big(\, H_i W^+_\mu W^{-,\mu}\ +\
\frac{1}{2\cos^2\theta_w}\, H_i Z_\mu Z^\mu\, \Big)\, ,\\
  \label{HpHW}
{\cal L}_{H H^\mp W^\pm} &=& \frac{g_w}{2}\, \sum\limits_{i=1}^3\, 
 g_{H_iH^-W^+}\,  ( H_i\, i\!\!
\stackrel{\leftrightarrow}{\vspace{2pt}\partial}_{\!\mu} H^- )\, 
W^{+,\mu}\,\ +\ {\rm  h.c.},\\
  \label{HHZ}
{\cal L}_{HHZ} &=& \frac{g_w}{2\cos\theta_w}\, \sum\limits_{j>i=1}^3\,
g_{H_iH_jZ}\, ( H_i\, \!\! 
\stackrel{\leftrightarrow}{\vspace{2pt}\partial}_{\!\mu} H_j )\,Z^\mu\,  ,
\end{eqnarray}
where              $\cos\theta_w              =              M_W/M_Z$,
$\stackrel{\leftrightarrow}{\vspace{2pt}  \partial}_{\!  \mu}\ \equiv\ 
\stackrel{\rightarrow}{\vspace{2pt}      \partial}_{\!      \mu}     -
\stackrel{\leftarrow}{\vspace{2pt} \partial}_{\! \mu}$, and
\begin{eqnarray}
  \label{gHZZ}
g_{H_iVV} &=& \cos\beta\, O_{1i}\: +\: \sin\beta\, O_{2i}\,,\\  
 \label{gHHZ}
g_{H_i H_j Z} &=& O_{3i}\, \Big( 
\cos\beta\, O_{2j}\, -\, \sin\beta\, O_{1j}  \Big)\ -\
 O_{3j}\, \Big(  \cos\beta\, O_{2i}\,  -\,  \sin\beta\, O_{1i} \Big)\, ,\\
  \label{gHplus}
g_{H_iH^-W^+} &=& \cos\beta\, O_{2i}\: -\: \sin\beta\, O_{1i}\: +\: iO_{3i}\, .
\end{eqnarray}
For  completeness,  we  have   included  in (\ref{HpHW})  the
interactions of the charged    Higgs bosons $H^\pm$ with the   neutral
Higgs  and $W^\mp$ bosons. Note  that the couplings  $H_i ZZ$ and $H_i
W^+W^-$ are related to the $H_i H_j Z$ couplings through
\begin{equation}
  \label{Orel}
g_{H_k V V}\ =\ \varepsilon_{ijk}\, g_{H_i H_j Z}\, . 
\end{equation}
Moreover, unitarity provides the constraint
\begin{equation}
\sum\limits_{i=1}^3\, g_{H_i VV}^2\ =\ 1\, .
\end{equation}
Evidently,  if  two Higgs-boson couplings to   gauge bosons are known,
this  is  sufficient to determine the   complete set of  the couplings
$g_{H_i VV}$ and $g_{H_i H_j Z}$ \cite{Alex}. 

In the above calculation of the effective Higgs-gauge-boson couplings,
we   have  assumed   that  the   dominant  contributions   arise  from
Higgs-mixing  effects.   As  opposed  to  the  $b$-quark  Yukawa  case
discussed in Section  4, proper vertex corrections to  the $H_iZZ$ and
$H_iH_jZ$    couplings   do    not    contain   strong-coupling-    or
$\tan\beta$-enhanced  diagrams. Therefore, naive  dimensional analysis
suggests  that  these corrections  are  suppressed  relative to  their
tree-level  values by  loop factors  of the  kind: $(3\alpha_w/4\pi)\,
(|\mu A_t|  / m^2_{\tilde{t}_1})\, ,\  (3\alpha_w/4\pi)\, (|\mu A_t|^2
v^2   /  m^6_{\tilde{t}_1})   \stackrel{<}{{}_\sim}   10^{-2}$,  where
$\tilde{t}_1$ is the heaviest stop squark (see (\ref{msqud})).  In the
following,  we neglect proper  vertex corrections  to the  $H_iZZ$ and
$H_iH_jZ$ couplings.

For our  phenomenological discussion of CP violation,  we consider the
following two representative values  for $\tan\beta$: (i) $\tan\beta =
4$  and  (ii)  $\tan\beta  =  20$.  For  definiteness,  unless  stated
otherwise, the soft supersymmetry-breaking and  $\mu$ parameters are set
to the
values
\begin{eqnarray}
  \label{scenario}
M_{\rm SUSY} \!&=&\! \widetilde{M}_Q\ =\ \widetilde{M}_t\ =\ 
\widetilde{M}_b\ =\ 0.5\ {\rm TeV}\,,\quad
\mu \ =\ 2\ {\rm TeV}\,,\quad 
|A_t| \ =\  |A_b|\ =\ 1\ {\rm TeV}\,, \nonumber\\   
|m_{\tilde{B}}| \!&=&\! |m_{\tilde{W}}|\ =\ 0.3\ {\rm TeV}\,,\qquad
 |m_{\tilde{g}}|\ =\ 1\ {\rm TeV}\, ,   
\end{eqnarray}
As can  be seen in  (\ref{scenario}), we have chosen  relatively large
values for  the stop  and sbottom mixing  parameters $A_t$,  $A_b$ and
$\mu$, as well as a  common left- and right-handed squark mass $M_{\rm
  SUSY}$, which  leads to enhanced CP-violating effects  of the CP-odd
phases {\rm arg}\,$(\mu A_{t,b})$ on the Higgs sector.

As was mentioned in the  Introduction, large CP-odd phases may lead to
rather  large EDM  contributions,  thereby violating  the known  upper
bounds on the electron and neutron EDMs $d_e$ and $d_n$: $d_e/e < 0.5\ 
10^{-26}$ cm \cite{commins} and  $d_n/e < 0.6\ 10^{-25}$ cm \cite{PGH}
at   the   2-$\sigma$   level.  One   phenomenologically   interesting
possibility for avoiding  the possible CP crisis is  to make the first
two generations of squarks rather heavy with masses much above the TeV
scale \cite{PN,GD}, keeping the third generation relatively light with
masses of  order 0.5 TeV.  In  such a scenario, CP  violation may only
reside in the third generation.   For our illustrations, we shall take
the $\mu$ parameter to be real  and assume that the only CP-odd phases
in  the theory  are ${\rm  arg}\,(A_t) =  {\rm arg}\,(A_b)$  and ${\rm
  arg}\, (m_{\tilde{g}})$.   Again, as one  way to avoid  the one-loop
EDM constraints  \cite{IN}, we have  taken a gluino  mass of 1  TeV in
(\ref{scenario}).  However, in  such a scheme, one has  to worry about
the  fact   that  Higgs-boson  two-loop  contributions   to  the  EDMs
\cite{CKP}  might  still  become  sizeable.  For  the  low-$\tan\beta$
scenario in  (\ref{scenario}), the  two-loop EDM contributions  are of
the order of the experimental EDM upper bounds mentioned above.  Since
these two-loop EDM effects  depend almost linearly on $\tan\beta$, one
might then need to arrange  for cancellations among the different one-
and two-loop EDM terms \cite{IN} at the level of 10\% for the scenario
with $\tan\beta =  20$.  We believe that this  can be achieved without
excessive fine-tuning of the CP-violating parameters of the theory.

As was already noticed  in \cite{HH,CQW,PW}, the radiative corrections
to the lightest Higgs boson  $H_1$ depend crucially on the stop mixing
parameter  $|X_t|  =  |A_t  - \mu^*  \cot\beta|$.   Specifically,  the
radiatively-corrected $H_1$-boson mass increases as $|X_t|$ increases,
reaching a  maximum when $|X_t|/M_{\rm SUSY} \approx  2.45$.  Then, as
$|X_t|$  further increases, the  radiative corrections  to $H_1$-boson
mass decrease and may even become negative, driving the latter to very
small, experimentally  excluded values.  A distinctive  feature of the
CP-violating SUSY  scenario compared to the CP-conserving  one is that
$|X_t|$ can be increased by  varying only the phase ${\rm arg}\,(A_t)$
from zero to higher values, but holding fixed $|A_t|$ and $|\mu|$. For
similar reasons,  high values  of $|X_t|$ induced  by large  values of
${\rm  arg}\,(A_t)$ can  make the  mass  of the  lightest stop  squark
$\tilde{t}_2$  very  low,  so   as  to  violate  present  experimental
constraints,  {\em i.e.}, $m_{\tilde{t}_2}  \stackrel{>}{{}_\sim} 100$
GeV.    Furthermore,   light   stop  quarks,   with   $m_{\tilde{t}_2}
\stackrel{<}{{}_\sim} 300$ GeV and large $|X_t|$ values, can give rise
to  observably  large   contributions  to  the  electroweak  precision
parameter   $\Delta\rho$  \cite{CCRW}.    For   the  scenarios   under
discussion, $m_{\tilde{t}_2}$ is always larger than about 300 GeV, for
all  the  parameter space  for  which  the  $H_1$-boson mass  acquires
acceptable values.   Therefore, apart from  the bounds derived  by EDM
constraints,  the requirement  that the  lightest Higgs-boson  mass is
positive can  be used naively  to set an  upper bound on the  phase of
$A_t$.

\begin{figure}
   \leavevmode
 \begin{center}
   \epsfxsize=16.2cm
    \epsffile[0 0 539 652]{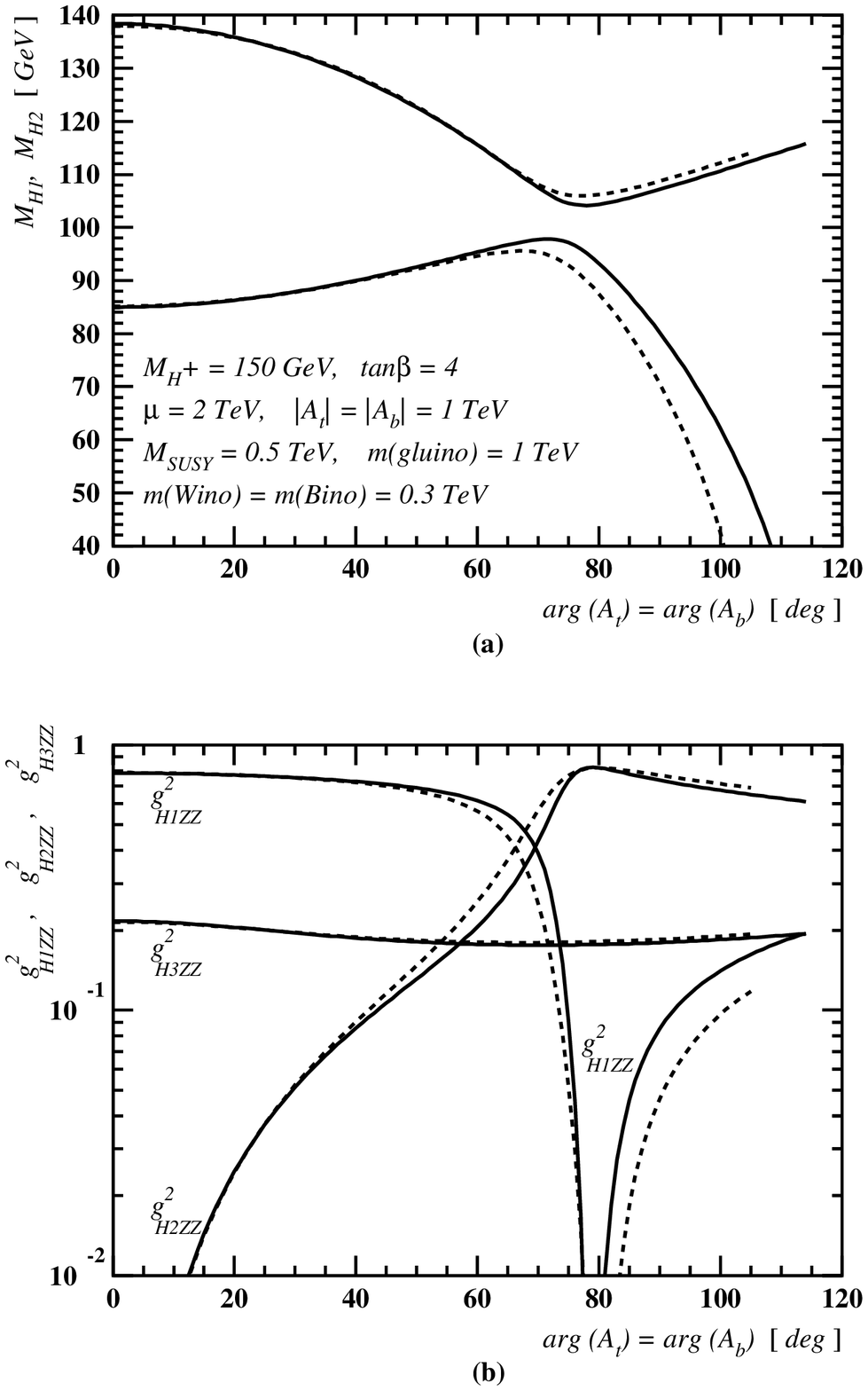}
 \end{center}
 \vspace{-0.5cm} 
\caption{\it Numerical estimates of (a) $M_{H_1}$ and $M_{H_2}$ and (b)
$g^2_{H_iZZ}$ as functions of ${\rm arg}\, (A_t)$, for the indicated
choices of MSSM parameters. Solid lines
correspond to ${\rm arg}\, (m_{\tilde{g}}) = 0$, dashed lines to
${\rm arg}\, (m_{\tilde{g}}) = 90^\circ$.}\label{fig:cph1}
\end{figure}

In  Fig.\ \ref{fig:cph1}  we give  numerical predictions  for  the two
lightest Higgs-boson masses $M_{H_1}$ and $M_{H_2}$, and for the three
relevant  $H_iZZ$ couplings  squared  as a  function  of ${\rm  arg}\,
(A_t)$, for two different values of ${\rm arg} (m_{\tilde{g}})$: ${\rm
arg} (m_{\tilde{g}}) = 0$ (solid lines) and ${\rm arg} (m_{\tilde{g}})
= \pi/2$ (dashed lines). We first discuss the scenario with $\tan\beta
=  4$,  for which  the  values  of  the remaining  soft
supersymmetry-breaking
parameters  and  $\mu$  are  given  in  (\ref{scenario}).   Since  our
interest is to analyze dominant CP-violating effects for a light Higgs
sector,    we   present   predictions    for   a    relatively   small
charged-Higgs-boson mass,  $M_{H^+} =  150$ GeV. In  the CP-conserving
limit of the  theory (${\rm arg}\, (A_t) = 0$),  the mass $M_{H_1}$ of
the lightest neutral Higgs boson is  close to 85 GeV, while the square
of  the $H_1ZZ$ coupling,  $g^2_{H_1 ZZ}$,  is approximately  equal to
0.8. These  values  of  masses  and  couplings  are  now  excluded  by
Higgs-boson  searches at  LEP~2~\cite{expMH}. However,  this situation
changes crucially  once CP-violating  phases become relevant.   As the
phase of $A_t$ increases, two important effects take place.  First, as
was mentioned above, the stop mixing parameter $|X_t|$ becomes larger,
giving  rise to  larger $H_1$-boson  masses.  Second,  the mass-matrix
terms describing the  scalar-pseudoscalar mixing are enhanced, thereby
effectively  leading to large  modifications in  the couplings  of the
Higgs bosons  to gauge bosons.   This second effect can  be attributed
entirely  to  CP  violation.  In  fact,  as  can  be seen  from  Fig.\
\ref{fig:cph1}(b), for ${\rm  arg}\,(A_t) \approx 80$ degrees, $g_{H_1
ZZ}^2$  gets very suppressed,  implying that  LEP~2 cannot  detect the
Higgs boson  $H_1$ via $e^+e^-\to Z^*  \to ZH_1$.  On  the other hand,
for the same range of values of ${\rm arg}\, (A_t)$, {\it i.e.}, ${\rm
arg}\, (A_t) = 80^\circ$--95$^\circ$,  the $H_2$ and $H_3$ bosons have
significant couplings to the $Z$  bosons.  Although the $H_3$ boson is
too heavy to be detected at LEP~2  in this case, the $H_2$ boson has a
mass of  $105$--$110$ GeV and  $g_{H_2 ZZ}^2 \approx  0.8$--0.6, which
may be probed  at LEP~2 in this year's run.  For  larger values of the
phase  of $A_t$,  $95^\circ  <  {\rm arg}\,  (A_t)  < 110^\circ$,  the
discovery of a Higgs boson at LEP~2 is more challenging.  The lightest
Higgs  boson $H_1$  acquires  a mass  below  90 GeV,  but the  $H_1ZZ$
coupling  is too  small to  allow experimental  detection  through the
reaction  $e^+e^-\to Z^*  \to  ZH_1$.  In  addition,  the $H_2$  boson
becomes too heavy to allow for discovery via $e^+e^-\to Z^* \to ZH_2$,
with $g_{H_2 ZZ}^2\stackrel{<}{{}_\sim} 0.7$.

As was discussed in \cite{PW}, the  $H_1$ and $H_2$ bosons may also be
searched  for in the  channel $e^+e^-\to  Z^* \to H_1 H_2$.   Since the
squared coupling
$g^2_{H_1H_2Z}  =  g^2_{H_3ZZ} \approx  0.2$  almost independently  of
${\rm arg}\, (A_t)$, a  careful experimental analysis of the parameter
region  of interest  will be  necessary to  determine whether  the two
lightest Higgs bosons can be  observed for such large mass differences
($M_{H_1}-M_{H_2}   \geq  40$  GeV)   and  such   small  $g_{H_1H_2Z}$
couplings.

In Fig.\ \ref{fig:cph1} we also present predictions for a gluino phase
of 90  degrees (dashed  lines). Since the vertex  corrections  are
generally  small for  low or moderate  values of
$\tan\beta$, they  are
expected to  induce only small  corrections to the  Higgs-boson masses
and  mixings.  This last  fact is  reflected in  Fig.\ \ref{fig:cph1},
even though  the coupling $g^2_{H_1ZZ}$  ($g^2_{H_2ZZ}$) gets slightly
smaller (larger) for larger values of the phase of $A_t$.

\begin{figure}
   \leavevmode
 \begin{center}
   \epsfxsize=16.2cm
    \epsffile[0 0 539 652]{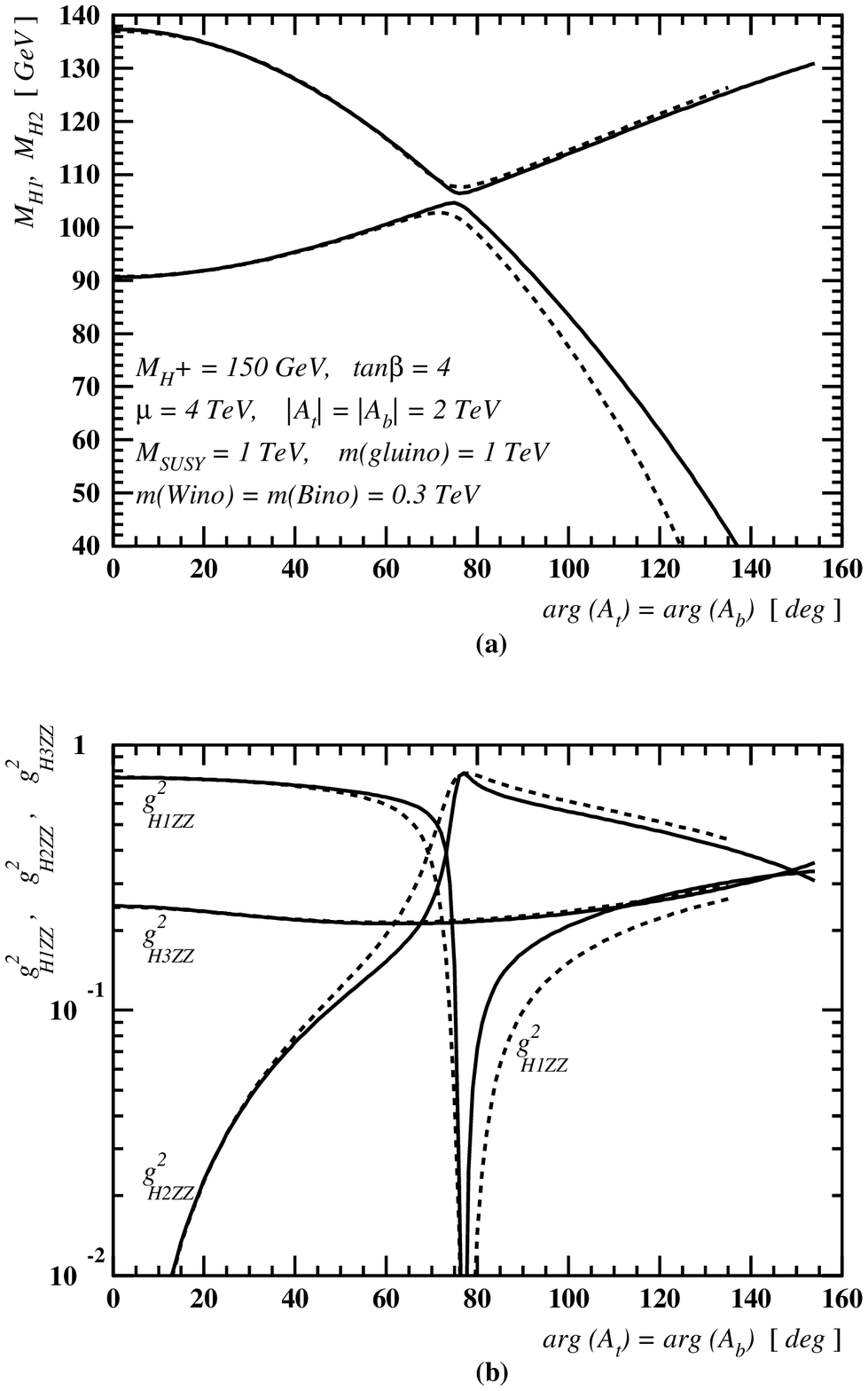}
 \end{center}
 \vspace{-0.5cm} 
\caption{\it As Fig.\ \ref{fig:cph1}, but with $\mu = 4$ TeV,
$|A_t| = |A_b| = 2$ TeV and $M_{\rm SUSY} = 1$~TeV.}\label{fig:cph2}
\end{figure}

Fig.\ \ref{fig:cph2} shows the changes in the predictions for the same
choice of parameters as  in Fig.\ \ref{fig:cph1}, but with $M_{\rm
  SUSY}$, $\mu$ and  $|A_t| = |A_b|$ rescaled by a  factor of 2. 
This rescaling  leads to  a  slight increase  (decrease) of  the
Higgs-boson mass $M_{H_1}$  ($M_{H_2}$), while $g_{H_1 ZZ}^2$ exhibits
a slightly  different quantitative dependence  on the phase  of $A_t$,
especially  for the region in which  $g_{H_1 ZZ}^2$  is very  small. 
Thus, although $M_{H_1}$  becomes small for ${\rm arg}\,  (A_t) > 115$
degrees,  the $H_1ZZ$ coupling  gets sizeable  again, well  within the
capabilities of LEP~2 to test.

\begin{figure}
   \leavevmode
 \begin{center}
   \epsfxsize=16.2cm
    \epsffile[0 0 539 652]{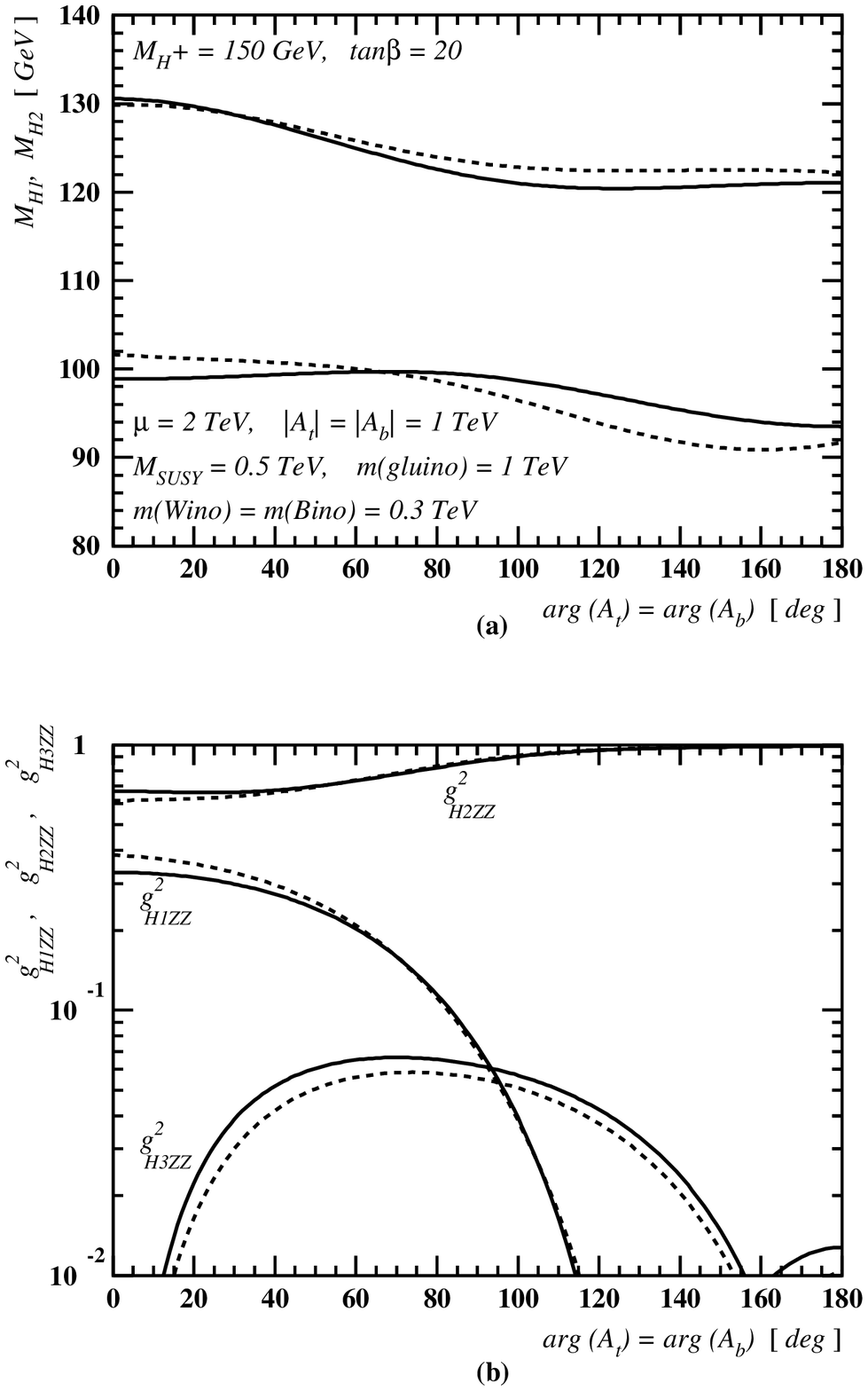}
 \end{center}
 \vspace{-0.5cm} 
\caption{\it As in Fig.\ \ref{fig:cph1}, but with $\tan\beta =
20$.}\label{fig:cph3}
\end{figure}

We  now   investigate  more   quantitatively  the  predictions   of  a
large-$\tan\beta$ scenario  for the Higgs-boson  masses and couplings.
We adopt the scenario given in (\ref{scenario}), with $\tan\beta = 20$
and $M_{H^+} =  150$ GeV. In this large-$\tan\beta$  scenario, one has
$|\mu| \cot\beta  \ll |A_t|$, and the effective  stop mixing parameter
$|X_t| \approx  |A_t|$ is almost  independent of ${\rm  arg}\, (A_t)$.
Therefore,  as is  seen  in Fig.\  \ref{fig:cph3}(a), the  Higgs-boson
masses  $M_{H_1}$  and  $M_{H_2}$   do  not  exhibit  any  significant
variation as  a function of ${\rm  arg}\, (A_t)$.  In  contrast to the
Higgs-boson  masses, Fig.\  \ref{fig:cph3}(b)  shows that  there is  a
non-trivial dependence of the squared couplings $g^2_{H_iZZ}$ on ${\rm
arg}\, (A_t)$.  Furthermore, the next-to-lightest Higgs boson $H_2$ is
heavy enough to render its search through the $e^+e^-\to Z^* \to ZH_2$
reaction kinematically inaccessible at LEP~2.  For similar reasons, we
find that, for all values of  ${\rm arg}\, (A_t)$, the $H_1$ and $H_2$
bosons are rather too heavy to  be produced via the $e^+e^-\to Z^* \to
H_1 H_2$ channel at LEP~2.  As a result, Higgs-boson searches at LEP~2
tend to  be more efficient for  small values of the  $A_t$ phases, for
which the  lightest neutral Higgs-boson mass  is close to  100 GeV and
$g_{H_1ZZ}^2$  is  non-negligible  ($g_{H_1ZZ}^2  \approx$  0.3).   In
addition,    for   large    $A_t$   phases,    ${\rm    arg}\,   (A_t)
\stackrel{>}{{}_\sim} 80^\circ$, the $H_1VV$ coupling (with $V = Z,W$)
is  rather suppressed,  so that  the  $H_1$ Higgs  boson, although  it
becomes lighter with  a mass in the range 90--95  GeV, will be elusive
at LEP~2, and may also  escape detection via the corresponding channel
at the  upgraded Tevatron.  However, the  next-to-lightest Higgs boson
$H_2$ has couplings of order unity to the $Z$ and $W$ bosons.  Present
simulations  show that  a neutral  Higgs  boson, such  as $H_2$,  with
$M_{H_2} \approx  180$ GeV and  a SM-like coupling strength  to vector
gauge  bosons can  be tested  at the  Tevatron collider  with  a total
integrated luminosity  of 10 fb$^{-1}$.  However, discovery of  such a
Higgs boson  at the 5-$\sigma$  level would demand a  total integrated
luminosity  of 30 fb$^{-1}$,  and would  have a  reach up  to $M_{H_2}
\approx 120$  GeV \cite{Tevrep}.   Finally, even though  the $H_1H_3Z$
coupling is  close to unity ($g^2_{H_1H_3Z} =  g^2_{H_2ZZ}$) for ${\rm
arg}\,  (A_t) >  100^\circ$, the  Higgs-pair production  of  $H_1$ and
$H_3$ is  not kinematically allowed  at LEP~2, since  $M_{H_3} \approx
M_{H^+}$.   Further  studies  will  be necessary  to  investigate  the
potential of this production mechanism at the Tevatron.

\begin{figure}
   \leavevmode
 \begin{center}
   \epsfxsize=16.2cm
    \epsffile[0 0 539 652]{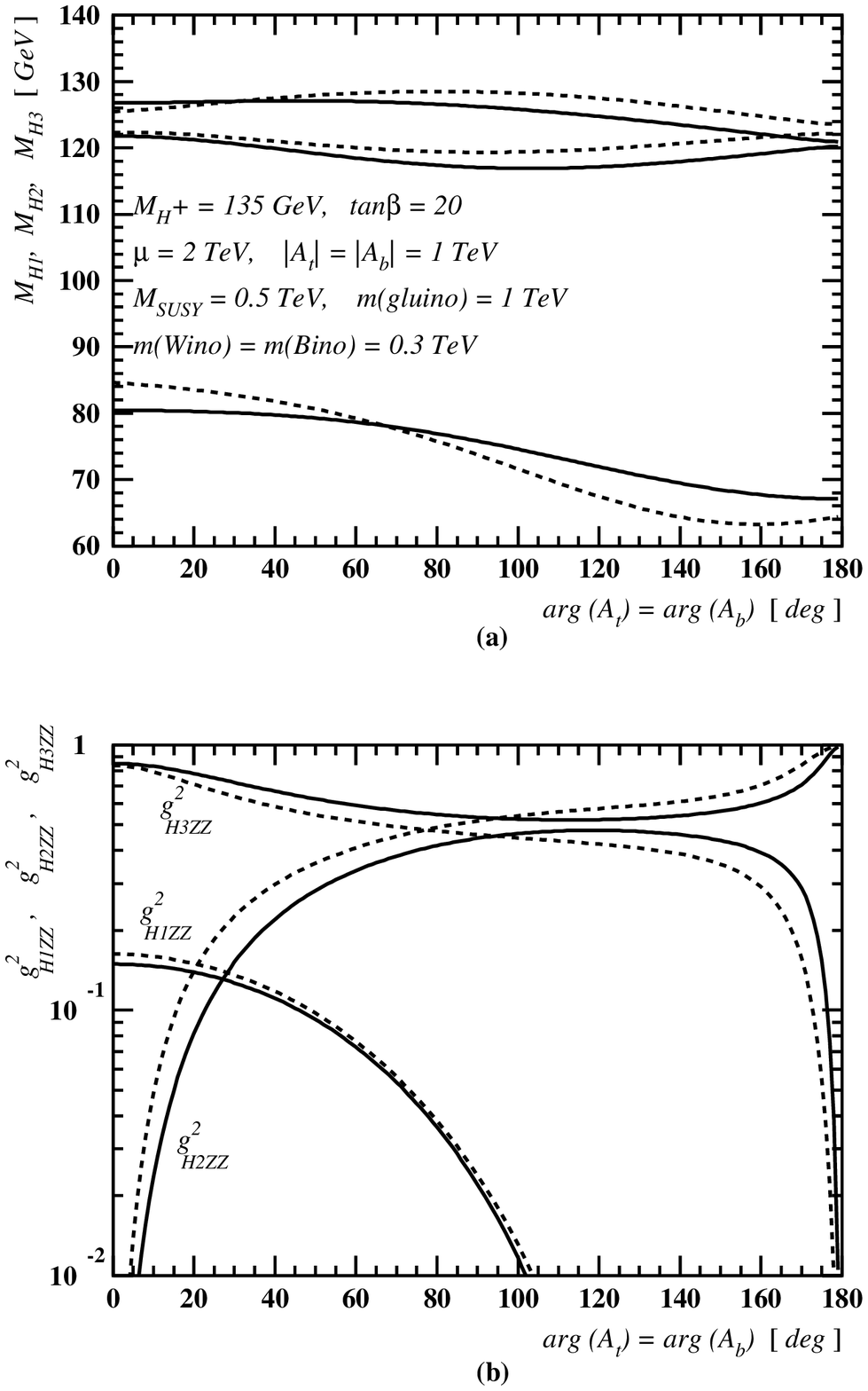}
 \end{center}
 \vspace{-0.5cm} 
\caption{\it As in Fig.\ \ref{fig:cph3}, but with $M_{H^+} = 135$
GeV.}\label{fig:cph4}
\end{figure}

It  is interesting  to  present  predictions  for  the  neutral
Higgs-boson  masses and  their  couplings to  gauge  bosons for  lower
values  of  the  charged  Higgs-boson  mass $M_{H^+}$ in the  above
large-$\tan\beta$   scenario.   In   Fig.\  \ref{fig:cph4},   we  plot
numerical  estimates for  the same  kinematic parameters  as  in Fig.\ 
\ref{fig:cph3},  but with  $M_{H^+} =  135$  GeV.  In  this case,  the
$H_1$-boson mass varies  approximately between 80 and 65  GeV, and the
$H_1ZZ$ coupling rapidly  decreases as the  phase of $A_t$  increases. 
The two heaviest  neutral Higgs bosons $H_2$ and  $H_3$ have masses in
the  range between 120  and 130  GeV.  Hence,  these two  Higgs bosons
cannot  be produced via $e^+e^-\to  Z^* \to ZH_2$ or
$e^+e^-\to  Z^* \to ZH_3$ at LEP~2.  
However,  the $H_2$  and $H_3$  bosons may  still be  accessed 
via $e^+e^-\to  Z^* \to H_1 H_2$ or $H_1 H_3$. Interestingly,  the
squared
couplings   $g^2_{H_1H_2Z}  =   g^2_{H_3ZZ}$   and  $g^2_{H_1H_3Z}   =
g^2_{H_2ZZ}$  exhibit a  cross-over as  a  function of  ${\rm
  arg}\, (A_t)$.  The  crossing point of the two  squared couplings is
when ${\rm arg}\,  (A_t) \approx 90^\circ$.  For ${\rm  arg}\, (A_t) =
180^\circ$, one of the squared couplings goes to 0 and the other to 1,
depending on  the phase  of the  gluino mass.  In  this case,  the two
heaviest neutral  Higgs bosons become almost degenerate  in mass.  For
the whole range of values of  ${\rm arg}\, (A_t)$, either the $H_2$ or
$H_3$  Higgs boson  can be  tested at  the upgraded  Tevatron collider
provided a total integrated luminosity of 10 fb$^{-1}$ per detector is
available \cite{Tevrep}

\begin{figure}
   \leavevmode
 \begin{center}
   \epsfxsize=16.2cm
    \epsffile[0 0 539 652]{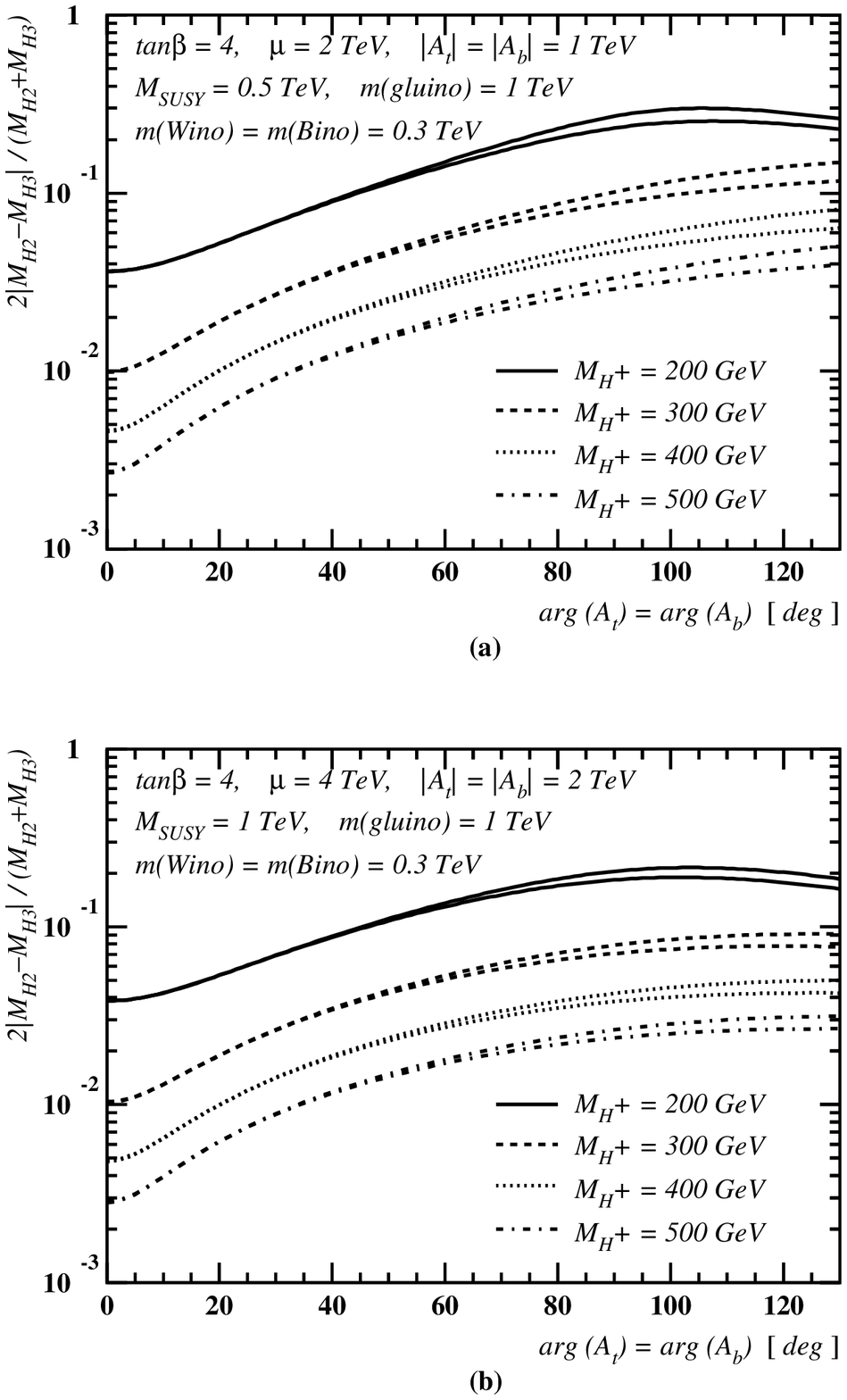}
 \end{center}
 \vspace{-0.8cm} 
\caption{\it Numerical estimates of (a) $M_{H_1}$ and (b)
$2|M_{H_2}-M_{H_3}|/(M_{H_2} + M_{H_3})$ as functions of the
CP-violating phase arg($A_t$). Lower values of the same line type
correspond to ${\rm arg}\, (m_{\tilde{g}}) = 0$, the higher ones
to ${\rm arg}\, (m_{\tilde{g}}) = 90^\circ$.}\label{fig:cph5}
\end{figure}

Figs.\  \ref{fig:cph5}(a) and (b)  show the  degree of  mass splitting
between the two heaviest neutral Higgs bosons $H_2$ and $H_3$, for the
same   choice  of   parameters   as  in   Figs.\  \ref{fig:cph1}   and
\ref{fig:cph2}, but for $M_{H^{\pm}} =$ 200, 300, 400, 500 GeV.  As was
already observed  in \cite{APLB,PW}, even  though the $H_2$  and $H_3$
bosons are almost degenerate in the CP-conserving limit of the theory,
they  can  have  a degree  of  splitting  up  to  30\% for  a  maximal
CP-violating  phase   ${\rm  arg}\,  (A_t)   \approx  90^\circ$.   The
comparison of  the Fig.\ \ref{fig:cph5}(a) with (b)  reveals that this
last  result is  almost  independent  of the  common  scale factor  of
$M_{\rm SUSY}$, $\mu$ and $|A_t|$.  Also, the degree of mass splitting
is not  much affected by  the value of  the gluino phase  ${\rm arg}\,
(m_{\tilde{g}})$:  the  results  for  ${\rm arg}\,  (m_{\tilde{g}})  =
90^\circ$   are   slightly  higher   than   those   of  ${\rm   arg}\,
(m_{\tilde{g}}) = 0$.  In this vein, it is interesting to mention that
large CP-violating scalar-pseudoscalar  mixings can lead to observable
phenomena   of  resonant   CP  violation   at   high-energy  colliders
\cite{APRL,APMN}.

\begin{figure}
   \leavevmode
 \begin{center}
   \epsfxsize=16.2cm
    \epsffile[0 0 539 652]{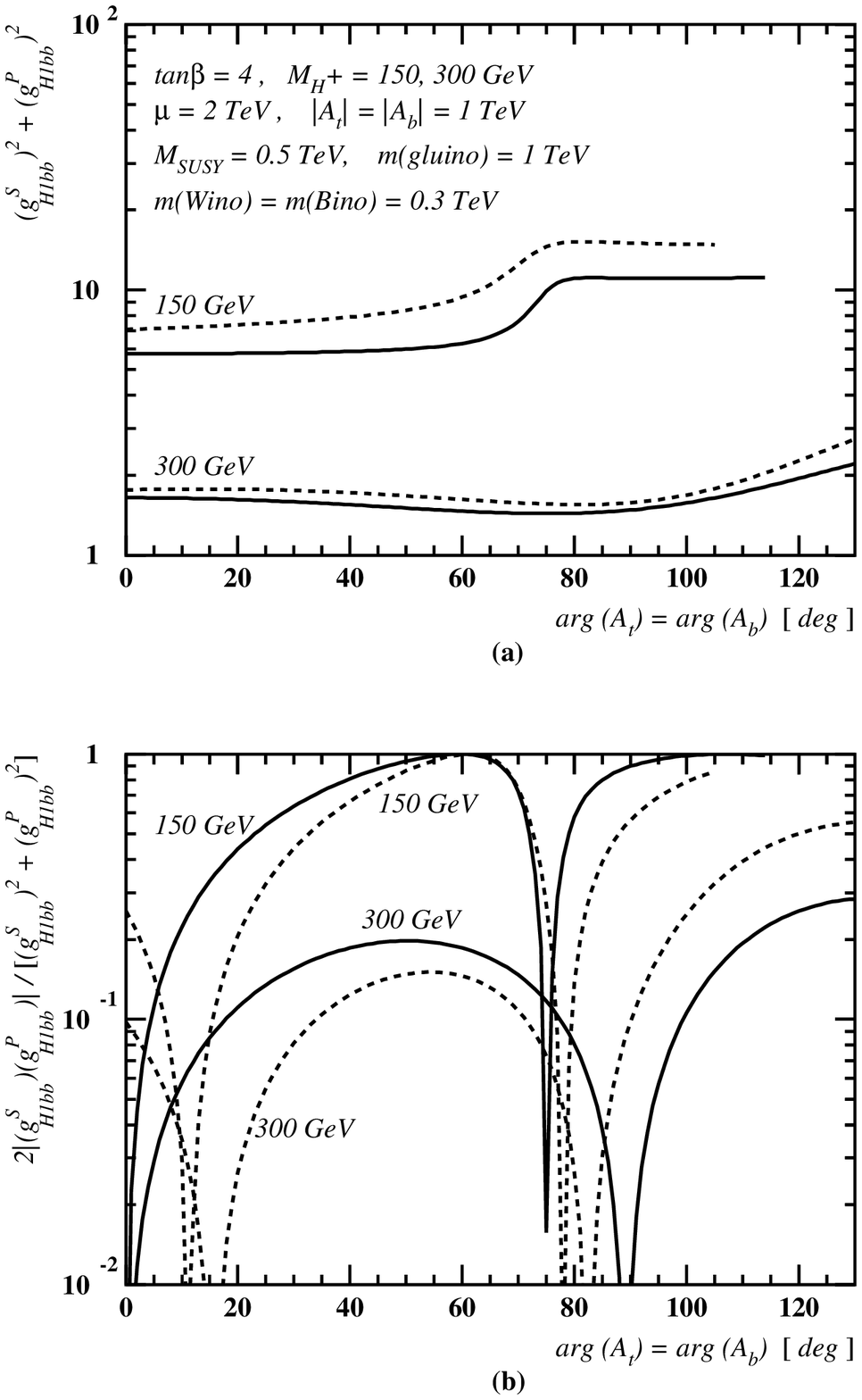}
 \end{center}
 \vspace{-0.8cm} 
\caption{\it Numerical estimates of (a) $(g^S_{H_1bb})^2 +
(g^P_{H_1bb})^2$ and (b) $2|(g^S_{H_1bb})\, (g^P_{H_1bb})| /
[(g^S_{H_1bb})^2 + (g^P_{H_1bb})^2]$ versus ${\rm arg} (A_t)$.  Solid
lines correspond to ${\rm arg}\, (m_{\tilde{g}}) = 0$,
dashed ones to ${\rm arg}\, (m_{\tilde{g}}) =
90^\circ$.}\label{fig:cph6}
\end{figure}

\begin{figure}
   \leavevmode
 \begin{center}
   \epsfxsize=16.2cm
    \epsffile[0 0 539 652]{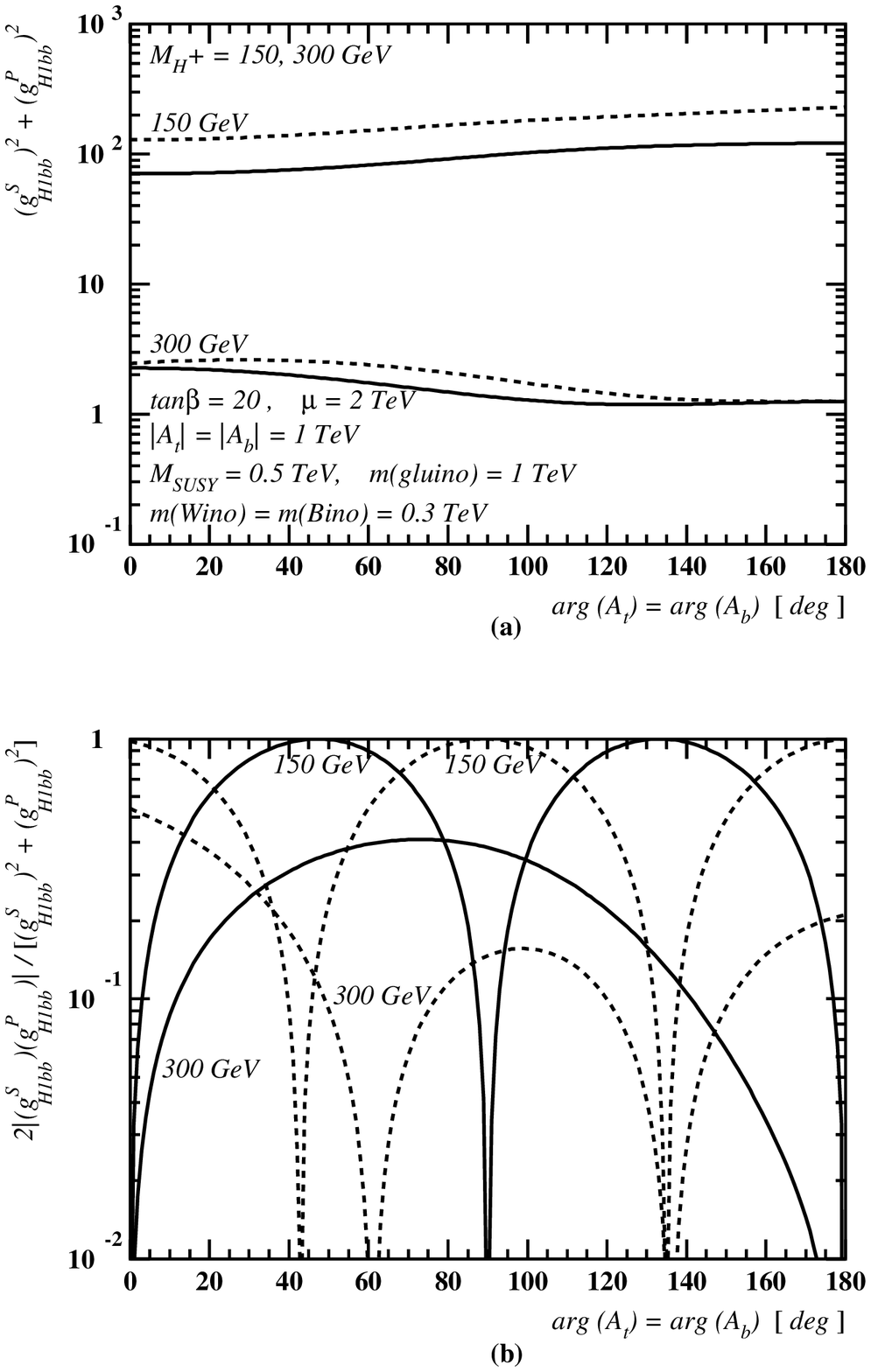}
 \end{center}
 \vspace{-0.5cm} 
\caption{\it As in Fig. \ref{fig:cph6}, but with $\tan\beta =
20$.}\label{fig:cph7}
\end{figure}

In Figs.\ \ref{fig:cph6} and \ref{fig:cph7}, we examine the behaviours
of  the scalar  and  pseudoscalar  parts of  the  $H_1bb$ coupling  as
functions of the  CP-odd phase ${\rm arg}\, (A_t)$,  for two different
charged Higgs-boson masses,  $M_{H^+} = 150$ and 300  GeV. As was done
in \cite{PW}, we find that the  best way of analyzing such a behaviour
is in terms of the  CP-even and CP-odd quantities: $[(g^S_{H_1bb})^2 +
(g^P_{H_1bb})^2]$ and $2 g^S_{H_1bb}\, g^P_{H_1bb}/ [(g^S_{H_1bb})^2 +
(g^P_{H_1bb})^2]$.   For  example,  Higgs-boson branching  ratios  are
proportional to  the first  quantity, while the  second one  will only
occur in  CP-violating observables.  In other  words, $2 g^S_{H_1bb}\,
g^P_{H_1bb}/ [(g^S_{H_1bb})^2  + (g^P_{H_1bb})^2]$ gives  a measure of
the CP-violating component in the $H_1bb$ coupling.  If we compare the
predictions for $M_{H^+} = 150$ GeV with those for $M_{H^+} = 300$ GeV
in   Figs.~\ref{fig:cph6}  and  \ref{fig:cph7},   we  find   that  the
CP-violating component  of the $H_1bb$ coupling  reduces in magnitude,
for large  values of the  charged Higgs-boson mass. Such  a decoupling
behaviour of  the CP-violating $H_1bb$ component is  in agreement with
our  observation, which  we  already made  at  the end  of Section  4.
{}From Figs.~\ref{fig:cph6}(b) and  \ref{fig:cph7}(b), we see that the
impact  of the  gluino  phase  on the  CP-violating  component of  the
$H_1bb$ coupling is more important for the large-$\tan\beta$ scenario.
This may be  attributed to the fact that  the radiatively-induced term
$\Delta h_b  \tan\beta$, which crucially  depends on the  gluino phase
and $\tan\beta$, has a dominant contribution to the $H_1bb$ coupling.

\begin{figure}
   \leavevmode
 \begin{center}
   \epsfxsize=16.2cm
    \epsffile[0 0 539 652]{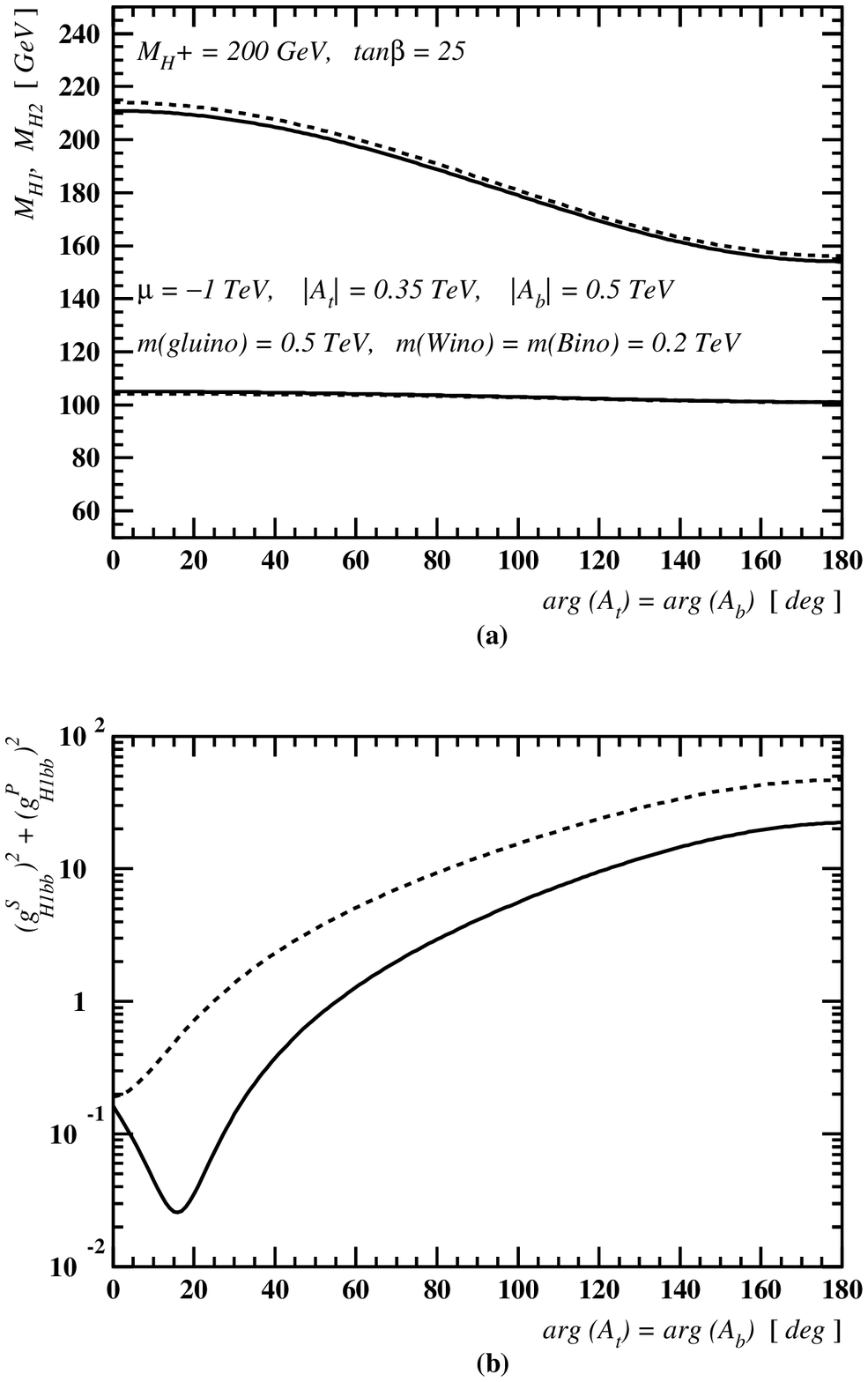}
 \end{center}
 \vspace{-0.5cm} 
\caption{\it Numerical estimates of (a) $M_{H_1}$ and $M_{H_2}$ and
(b) $[(g^S_{H_1bb})^2 + (g^P_{H_1bb})^2]$ versus ${\rm arg}\, (A_t)$,
with soft squark masses: $\widetilde{M}_Q =\widetilde{M}_b = 0.6$ TeV
and $\widetilde{M}_t$ = 0. Solid lines are for ${\rm arg}\, 
(m_{\tilde{g}}) = -90^\circ$ and dashed lines for ${\rm arg}\, 
(m_{\tilde{g}}) = 0^\circ$.}\label{fig:cph8}
\end{figure}

There can be a cancellation or a strong suppression of the coupling of
the lightest Higgs boson $H_1$ to  the bottom quarks, depending on the
magnitude  of the CP-violating phases  and of  the products $A_t \mu$,
$A_b \mu$ and $m_{\tilde{g}}  \mu$.   This cancellation usually  takes
place for moderate values  of the charged Higgs  mass and large values
of $\tan \beta$.  Such an effect  is also present in the CP-conserving
case,  for   specific  signs  and  magnitudes   of  the above products
involving the  trilinear terms $A_{t,b}$  and the gluino mass, and has
been discussed in  detail  in  \cite{H1bb}.  Figure  \ref{fig:cph8}(b)
illustrates such a cancellation for the  CP-violating SUSY model under
discussion.  For  example, we   observe   that for   the set  of  SUSY
parameters considered   in Fig.  \ref{fig:cph8},  the $H_1bb$ coupling
can be strongly suppressed for ${\rm arg}\,  (A_t) = {\rm arg}\, (A_b)
\approx 15^\circ$.  Moreover,  we have checked that  for this same set
of parameters  the $H_1 ZZ$ coupling  is almost SM-like.  In addition,
Fig.\  \ref{fig:cph8}(a) shows that the   mass  of the lightest  Higgs
boson is of order 105 GeV, practically independent of the CP-violating
phase.  This is therefore an  extremely interesting example, since the
lightest Higgs mass is in the mass range that may  be within the reach
of LEP~2, and its production cross section will  be SM-like.  However,
the main  decay   channel,  $H_1 \to  b   \bar{b}$,  can be   strongly
suppressed if the CP-violating phases ${\rm arg} (A_t)$ and ${\rm arg}
(m_{\tilde{g}})$ lie in  a   specific  range.  Therefore, in  such   a
scenario, the   detection  of the  $H_1$ boson  may   in principle  be
impossible, even  in the  final run  of LEP~2.   To make  a conclusive
statement  on this  possibility,   one   should study  in   detail the
capability of LEP~2 to detect such a  light $H_1$ boson via its decays
into $\tau$ pairs, or  into other hadronic modes.   Most intriguingly,
the  set  of parameters  considered in  this example also  allow for a
light  right-handed     stop squark and   moderate   mixing parameter,
$|X_t|/\widetilde{M}_Q$,   of  the  type     necessary to   allow  the
possibility of  electroweak baryogenesis \cite{EWbau}.  Hence, if such
a Higgs boson cannot be discovered at  LEP~2 via other decay channels,
the final phase of  LEP~2 will leave   an open window  for electroweak
baryogenesis.  A careful study of the CP-violating phases required for
electroweak  baryogenesis and the detection  capabilities of LEP~2 for
alternative decay modes becomes  essential  for testing this  exciting
scenario.

We shall briefly   comment   on the  enhanced  LHC   capabilities  for
Higgs-boson  searches~\cite{LHC}.   The LHC  has a considerably higher
reach than LEP~2 and the upgraded Tevatron collider  in the search for
heavier Higgs    bosons, and hence  has more   chances  to unravel the
complete Higgs-boson spectrum of the MSSM  with explicit CP violation. 
At the LHC, Higgs bosons may be copiously produced  via a wide variety
of processes  which depend in many  different ways on the couplings of
the neutral and charged Higgs bosons both to gauge bosons and fermions
\cite{LHC}.  In the case of the CP-violating version of the MSSM under
study, we  have  shown that  mixing  between states with different  CP
parities   can   dramatically modify   those    couplings  and, hence,
importantly affect the  associated  production and decay  mechanisms.  
Studies including  CP-violating effects on the gluon-fusion production
of  Higgs  bosons  at  the LHC  have already   been  considered in the
literature  \cite{DM}.  Our work  provide   the basic tool to  improve
further those studies,   and to  perform a  complete   analysis of the
CP-violating effects  on the many  other Higgs-boson search mechanisms
available at LHC.   The  LHC, together with the  information  gathered
from experiments at LEP~2 and  the upgraded Tevatron collider, will be
capable of providing a thorough test of the MSSM Higgs sector and shed
light   on the possibility  of    explicit  radiative breaking of   CP
invariance in supersymmetry.


Finally, it is interesting to make a comparative analysis between our
results  and  those  obtained previously  in~\cite{PW}.   In the
latter work, the  effective RG-improved potential  was expanded  up to
renormalizable operators of dimension 4.   The expansion was performed
in powers    of   the  stop-mass  splitting,   $m^2_{\tilde{t}_1}    -
m^2_{\tilde{t}_2}$, relative to the  arithmetic  average of the  squared
stop
masses,  $M^2_{\rm  SUSY}     =  \frac{1}{2}    (m^2_{\tilde{t}_1}   +
m^2_{\tilde{t}_2})$.  Moreover, the  two-loop effect originating  from
the one-loop radiative corrections to  the Yukawa couplings of the top
and bottom quarks was not taken into account in the computation of the
effective potential in  \cite{PW}.  Nevertheless,  for moderate values
of $\tan\beta$ and for all  soft supersymmetry-breaking masses equal to
$M_{\rm
  SUSY}$, the deviations of   the results presented in  \cite{PW} with
our  results  are  expected  to  be small,   for small  values  of the
stop-mixing    parameter $X_t =   |A_t   -  \mu^{*}  \cot\beta|$;  the
deviations will only grow for increasing  values of $X_t$.  For larger
values of $\tan\beta$, instead,  the impact of the bottom-mass quantum
corrections, which were  omitted in~\cite{PW}, is significant. In
fact, only in the limit of small values of  $|\mu|$, in which case the
bottom-mass corrections are  small, are both approaches guaranteed  to
give comparable numerical estimates.

\begin{figure}
   \leavevmode
 \begin{center}
   \epsfxsize=16.2cm
    \epsffile[0 0 539 652]{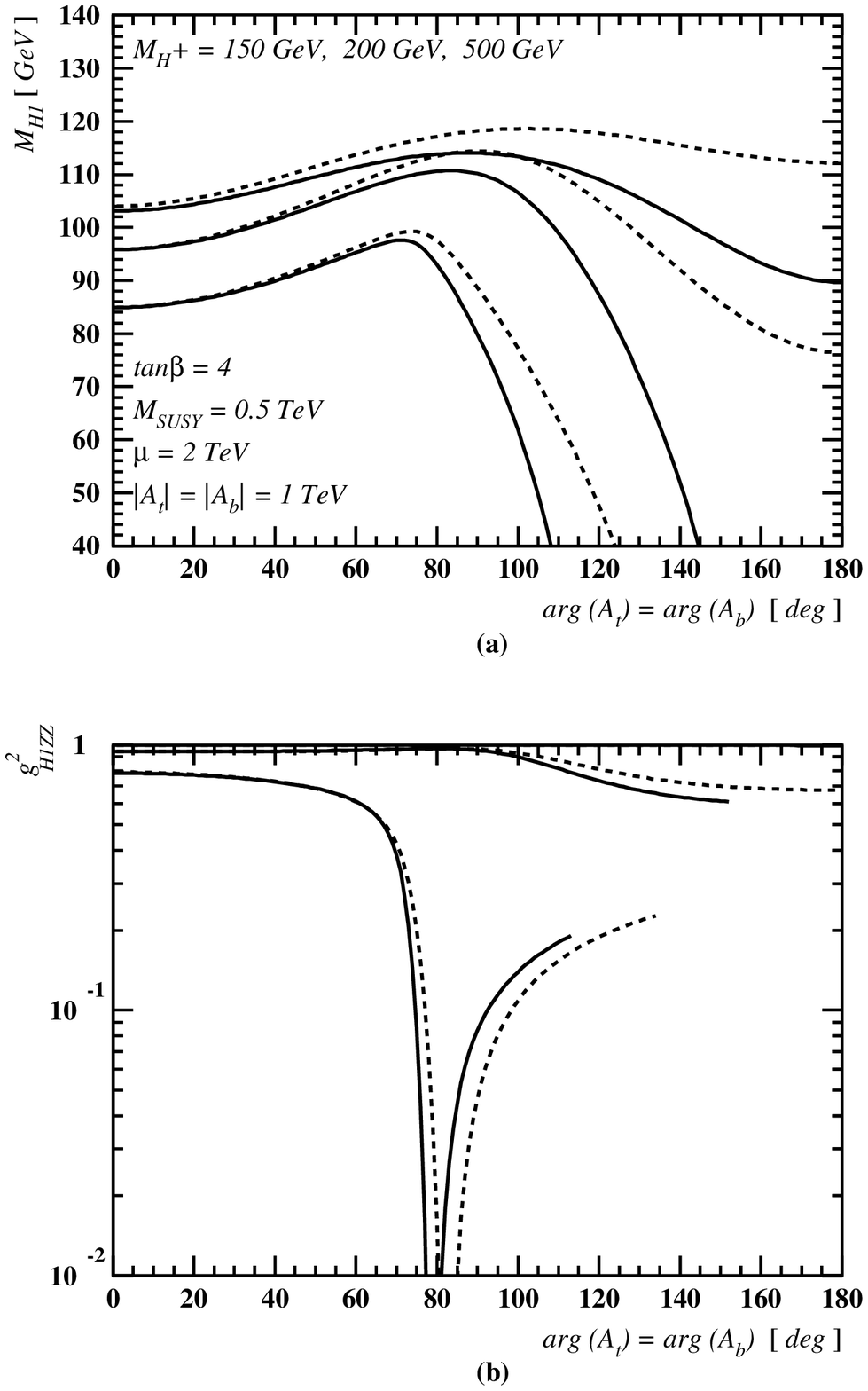}
 \end{center}
 \vspace{-0.5cm} 
\caption{\it Numerical predictions for $M_{H_1}$ and $g^2_{H_1ZZ}$
  as obtained by the present complete RG approach (solid lines) and
  the operator-expansion method of \cite{PW} (dashed lines).}\label{fig:cph9}
\end{figure}

In Fig.\ \ref{fig:cph9}, we  show the predictions for  the mass of the
lightest neutral Higgs boson $H_1$ and its coupling  to $Z$ bosons, as
obtained    obtained  by  our  RG    approach  (solid lines)  and  the
operator-expansion method of~\cite{PW} (dashed lines).  For  the
sake of comparison,  we consider the same  input parameters as those
chosen in Fig.\ \ref{fig:cph1},  for vanishing gluino phase  and three
different values of the charged Higgs-boson mass: $M_{H^+} = 150$, 200
and 500 GeV.  As was discussed above, we find  that the predictions of
the two  works are in excellent  agreement with one another  for small
values   of $X_t$.  For  large  values  of  $X_t$, instead, we observe
larger  quantitative  differences in the  results  obtained by the two
approaches,   even though   the   qualitative  behaviour  of  the  two
predictions exhibits a quite analogous functional dependence.

Our one-loop RG-improved approach overcomes the limitations present in
earlier analyses.  In particular, our RG approach  allows for a rather
precise    determination of  the   radiative   effects   on a  generic
Higgs-boson mass spectrum, even in cases  of large stop mixings and/or
large hierarchies  between  the left- and  right-handed  stop masses.  
Also,  within our RG approach, the  important effects  of the one-loop
corrections   to the quark  Yukawa couplings  are  incorporated in the
computation of  the Higgs-boson    masses  and in   their   respective
Higgs-boson couplings to gauge   and fermion fields.  A  Fortran  code
that computes the Higgs-boson masses  and couplings, including all the
CP-violating effects  as presented   in  this work,  may  be found  in
\cite{subroutine}.


\section{Conclusions}

We have performed a complete  one-loop RG improvement of the effective
Higgs  potential  in  the  MSSM,  in which  CP  violation  is  induced
radiatively by  soft CP-violating trilinear  interactions that involve
the Higgs  fields and the  stop and sbottom squarks.   Earlier studies
\cite{APLB,PW,Demir}  of the  neutral Higgs-boson  mass  spectrum were
based  on   a  number  of  particular   assumptions  and/or  kinematic
approximations. The  present work goes well beyond  those studies, and
extends the  most detailed  analysis~\cite{PW}, in which  the one-loop
RG-improved  effective  potential was  expanded  up to  renormalizable
operators of dimension 4, assuming a moderate mass splitting among the
stop squarks.  This assumption seemed to impose  a serious limitation,
given  that CP-violating  effects  exhibit an  enhanced behaviour  for
large values of the stop-mixing parameter $X_t$.  The results obtained
using  the  present  RG  approach confirm,  however,  the  qualitative
phenomenological  features  found  in~\cite{PW}, for  the  Higgs-boson
masses  and their couplings  to fermions  and to  the $W^\pm$  and $Z$
bosons.  It offers very accurate predictions, at the same level as the
most  accurate calculations  in  the CP-conserving  case (implying  an
uncertainty   of  order  3   GeV)  ~\cite{HHW,CHHHWW,CHWW},   for  the
Higgs-boson masses and for the whole range of the MSSM parameter space
in   the   presence   of   non-trivial  CP-violating   phases.    More
specifically,  the  present   study  also  includes  two-loop  leading
logarithms associated  with QCD effects and $t$-  and $b$-quark Yukawa
couplings.   It also  contains all  dominant  two-loop non-logarithmic
contributions to  the one-loop effective potential,  which are induced
by  one-loop  threshold  effects  on  the $t$-  and  $b$-quark  Yukawa
couplings due  to the decoupling  of the third-generation  squarks (as
considered in~\cite{CHHHWW,CHWW}  in the CP-conserving  limit).  These
one-loop  threshold  terms,  $\delta  h_{u,d}$ and  $\Delta  h_{u,d}$,
strongly depend on  the phase of the gluino mass  and so introduce new
CP violation into the MSSM effective potential at the two-loop level.

Large radiative  effects of  CP violation in  the Higgs sector  of the
MSSM can  have important phenomenological  consequences on Higgs-boson
searches  at LEP~2,  the Tevatron  and  the LHC.   We have  explicitly
demonstrated  that the  radiatively-induced CP  violation in  the MSSM
Higgs potential  can lead  to important effects  of mass  and coupling
level  crossing  among  the  three  neutral  Higgs  particles.   These
CP-violating  effects of  level crossing  in the  Higgs  sector modify
drastically the Higgs-boson couplings to the up- and down-type quarks,
and to the $W^\pm$ and $Z$ bosons.  In particular, CP violation in the
lightest Higgs sector becomes  relevant for a relatively light charged
Higgs  boson,  with  $M_{H^+}  \stackrel{<}{{}_\sim}  160$  GeV.   For
instance, for $M_{H^+} = 150$ GeV  and $\tan\beta = 4$, even a neutral
Higgs  boson  as  light as  60  GeV  may  escape detection  at  LEP~2.
However,  the upgraded  Tevatron  may have  the  physics potential  to
explore such  CP-violating scenarios at low  $\tan\beta$ values, which
may remain  at the edge of  accessibility even during  the final LEP~2
run.

We  have also  studied the  effects  induced by  a non-trivial  CP-odd
gluino  phase, which enters  the effective  potential at  the two-loop
level.   The  presence of  a  gluino phase  gives  rise  to small  but
non-negligible  changes in the  Higgs-boson mass  spectrum and  in the
couplings of the Higgs fields to  the $W^\pm$ and $Z$ bosons.  In this
context, we  have also found that  the product of the  scalar times the
pseudoscalar coupling of the lightest  Higgs boson $H_1$ to the bottom
quarks has a very strong  dependence on the gluino phase.
This product of couplings  gives a measure  of CP violation  in the
$H_1bb$ coupling,  which can  even be  of order  unity  for relatively
small charged Higgs-boson masses.

Finally, it  is worth stressing  that CP violation decouples  from the
lightest Higgs sector in the large-mass limit of a heavy charged Higgs
boson.   This decoupling  property, which  was known  to hold  for the
CP-violating self-energy effect \cite{APLB,PW},  has now been shown to
be valid  for the CP-violating vertex  effects as well.   As a result,
the predictions for the lightest Higgs-boson mass and its couplings to
gauge  bosons in the  above decoupling  regime of  the theory  will be
practically  identical   to  the  corresponding   predictions  in  the
CP-conserving  case. However, unlike  the lightest  Higgs sector,  CP
violation
does not decouple from the  heaviest Higgs sector in the MSSM, opening
up  new possibilities for  studying enhanced effects in CP-violating
Higgs  scalar-pseudoscalar  transitions  at  the LHC  or  future  muon
colliders \cite{APRL,APMN},  where the heavy MSSM Higgs  bosons can be
resonantly produced.   

In  conclusion: the  present analysis  has  shown that  the MSSM  with
explicit radiative  breaking of CP invariance constitutes  a very rich
theoretical framework,  introducing new  challenges in the  search for
fundamental Higgs scalars at LEP~2, the Tevatron and the LHC.

\subsection*{Acknowledgements}

We thank Manuel Drees for useful discussions.  A.P.  thanks the theory
group of Fermilab for the  kind hospitality extended to him while part
of this work  was done. Work supported in part  by the U.S. Department
of   Energy,   High    Energy   Physics   Division,   under   Contract
W-31-109-Eng-38.

\subsection*{Note Added} 

After completion of the  work described here, we saw~\cite{CDL}, which
also computes the one-loop RG-improved effective potential of the MSSM
with explicit CP  violation.  Here we further {\it  i)} perform a more
complete RG improvement of  the effective potential, {\it ii)} develop
a  self-consistent treatment of  the whole  MSSM Higgs  sector, taking
into  account  the  crucial  one-loop  relation  between  the  charged
Higgs-boson  and neutral  Higgs-boson  mass matrices,  and {\it  iii)}
include  the  dominant  two-loop  non-logarithmic corrections  to  the
effective potential,  which may also  have an important impact  on the
$H_1bb$  coupling.   In the  limits  where  a  comparison between  the
results was possible, we find reasonable agreement with~\cite{CDL}.

\newpage

\def\theequation{\Alph{section}.\arabic{equation}}
\begin{appendix}
\setcounter{equation}{0}
\section{Derivatives of Background-Field-Dependent \\
Masses}

In  this Appendix, we list  analytic expressions  pertaining to
derivatives  of quark and  squark masses  with respect  to their
background  Higgs  fields.   These  derivative  expressions  are  very
useful, as they constitute the building blocks of the general one-loop
effective potential  presented in Section  2. We have
divided the Appendix into three subsections.  In the first subsection,
we list the derivatives of  quark masses with respect to Higgs fields,
while the  next two subsections contain  the corresponding expressions
for  derivatives of squark  masses with  respect to  neutral and
charged Higgs fields, respectively.

\subsection{Quark Derivatives}

First,  we  give the  derivatives related  to  non-vanishing  tadpole
contributions:
\begin{equation}
  \label{tadf}
\frac{1}{v_2}\, \bigg< \frac{\partial \bar{m}^2_t}{\partial
  \phi_2}\bigg>\ =\ |h_t|^2\, ,\qquad   
\frac{1}{v_1}\, \bigg< \frac{\partial \bar{m}^2_b}{\partial
  \phi_1}\bigg>\ =\ |h_b|^2\, ,
\end{equation}
where  the  operation $\big<  \cdots  \big>$  denotes  that the  above
expressions  should be  evaluated in the  ground state  of the  Higgs
potential. 

Then, the  self-energy-type derivatives involving  neutral and charged
Higgs bosons may be listed as follows:
\begin{eqnarray}
  \label{fer0}
\bigg< \frac{\partial^2 \bar{m}^2_t}{\partial
  \phi^2_2}\bigg> \!&=&\! \bigg< \frac{\partial^2 \bar{m}^2_t}{\partial
  a^2_2}\bigg>\ =\ |h_t|^2\, ,\qquad 
\bigg< \frac{\partial^2 \bar{m}^2_b}{\partial
  \phi^2_1}\bigg>\ =\ \bigg< \frac{\partial^2 \bar{m}^2_b}{\partial
  a^2_1}\bigg>\ =\ |h_b|^2\, ,\nonumber\\
\bigg< \frac{\partial^2 \bar{m}^2_t}{\partial
  \phi^+_1\, \partial \phi^-_1}\bigg> \!&=&\! \frac{|h_b|^2 m^2_t}{m^2_t -
  m^2_b}\ ,\qquad
\bigg< \frac{\partial^2 \bar{m}^2_b}{\partial
  \phi^+_1\, \partial \phi^-_1}\bigg> \ =\ \frac{|h_b|^2 m^2_b}{m^2_b -
  m^2_t}\ ,\nonumber\\
\bigg< \frac{\partial^2 \bar{m}^2_t}{\partial
  \phi^+_2\, \partial \phi^-_2}\bigg> \!&=&\! \frac{|h_t|^2 m^2_t}{m^2_t -
  m^2_b}\ ,\qquad
\bigg< \frac{\partial^2 \bar{m}^2_b}{\partial
  \phi^+_2\, \partial \phi^-_2}\bigg>\ =\ \frac{|h_t|^2 m^2_b}{m^2_b -
  m^2_t}\ ,\\
\bigg< \frac{\partial^2 \bar{m}^2_t}{\partial
  \phi^+_1\, \partial \phi^-_2}\bigg> \!&=&\! 
\bigg< \frac{\partial^2 \bar{m}^2_t}{\partial
  \phi^+_2\, \partial \phi^-_1}\bigg>\ =\ 
-\,\bigg< \frac{\partial^2 \bar{m}^2_b}{\partial
  \phi^+_1\, \partial \phi^-_2}\bigg>\ =\  
-\, \bigg< \frac{\partial^2 \bar{m}^2_b}{\partial
  \phi^+_2\, \partial \phi^-_1}\bigg>\ =\ 
\frac{|h_t h_b| m_t m_b}{m^2_b -
  m^2_t}\ .\nonumber
\end{eqnarray}
Note that we have not listed derivatives that vanish.

\subsection{Derivatives of Squark Masses with Respect to\\ 
  Neutral Higgs Fields}

This section contains  the derivatives of the field-dependent squark
masses   $\widetilde{m}^2_{t_{1,2}}$  and  $\widetilde{m}^2_{b_{1,2}}$
with respect  to neutral Higgs  fields $\phi_{1,2}$ and  $a_{1,2}$. We
first give the tadpole terms:
\begin{eqnarray}
  \label{tads}
\frac{1}{v_1}\,\bigg< \frac{\partial \widetilde{m}^2_{t_1 (t_2)}}{\partial
  \phi_1}\bigg> \!&=&\! \frac{g^2_w + g'^2}{8}\, +(-)\, 
\frac{1}{2 \big(m^2_{\tilde{t}_1} - m^2_{\tilde{t}_2}\big)}\,
\Big[\, x_t \Big( \widetilde{M}^2_Q - \widetilde{M}^2_t 
+ {\textstyle \frac{1}{2}} x_t v^2 \cos 2\beta \Big)\nonumber\\
&&-\, 2|h_t|^2\Big( {\rm Re}\,(\mu A_t) \tan\beta - |\mu|^2
  \Big)\,\Big]\, ,\nonumber\\
\frac{1}{v_1}\,\bigg< \frac{\partial \widetilde{m}^2_{b_1 (b_2)}}{\partial
  \phi_1}\bigg> \!&=&\! |h_b|^2\, -\, \frac{g^2_w + g'^2}{8}\, -(+)\, 
\frac{1}{2 \big(m^2_{\tilde{b}_1} - m^2_{\tilde{b}_2}\big)}\,
\Big[\, x_b \Big( \widetilde{M}^2_Q - \widetilde{M}^2_b 
- {\textstyle \frac{1}{2}} x_b v^2 \cos 2\beta \Big)\nonumber\\
&&+\, 2|h_b|^2\Big( {\rm Re}\,(\mu A_b) \tan\beta - |A_b|^2
  \Big)\,\Big]\, ,\nonumber\\
\frac{1}{v_2}\,\bigg< \frac{\partial \widetilde{m}^2_{t_1 (t_2)}}{\partial
  \phi_2}\bigg> \!&=&\! |h_t|^2\, -\, \frac{g^2_w + g'^2}{8}\, -(+)\, 
\frac{1}{2 \big(m^2_{\tilde{t}_1} - m^2_{\tilde{t}_2}\big)}\,
\Big[\, x_t \Big( \widetilde{M}^2_Q - \widetilde{M}^2_t 
+ {\textstyle \frac{1}{2}} x_t v^2 \cos 2\beta \Big)\nonumber\\
&&+\, 2|h_t|^2\Big( {\rm Re}\,(\mu A_t) \cot\beta - |A_t|^2
  \Big)\,\Big]\, ,\nonumber\\
\frac{1}{v_2}\,\bigg< \frac{\partial \widetilde{m}^2_{b_1 (b_2)}}{\partial
  \phi_2}\bigg> \!&=&\! \frac{g^2_w + g'^2}{8}\, +(-)\, 
\frac{1}{2 \big(m^2_{\tilde{b}_1} - m^2_{\tilde{b}_2}\big)}\,
\Big[\, x_b \Big( \widetilde{M}^2_Q - \widetilde{M}^2_b 
- {\textstyle \frac{1}{2}} x_b v^2 \cos 2\beta \Big)\nonumber\\
&&-\, 2|h_b|^2\Big( {\rm Re}\,(\mu A_b) \cot\beta - |\mu|^2
  \Big)\,\Big]\, ,\nonumber\\
\frac{1}{v_2}\,\bigg< \frac{\partial \widetilde{m}^2_{t_1 (t_2)}}{\partial
  a_1}\bigg> \!&=&\! -\, \frac{1}{v_1}\,\bigg< \frac{\partial
  \widetilde{m}^2_{t_1 (t_2)}}{\partial  a_2}\bigg>\ =\ -(+)\, 
  \frac{|h_t|^2\, {\rm Im}\, (\mu A_t )}{m^2_{\tilde{t}_1} -
  m^2_{\tilde{t}_2}}\ ,\nonumber\\  
\frac{1}{v_2}\,\bigg< \frac{\partial \widetilde{m}^2_{b_1 (b_2)}}{\partial
  a_1}\bigg> \!&=&\! -\, \frac{1}{v_1}\,\bigg< \frac{\partial
  \widetilde{m}^2_{b_1 (b_2)}}{\partial  a_2}\bigg>\ =\ -(+)\, 
  \frac{|h_b|^2\, {\rm Im}\, (\mu A_b )}{m^2_{\tilde{b}_1} -
  m^2_{\tilde{b}_2}}\ ,
\end{eqnarray}
where the coupling  parameters $x_t$ and $x_b$ are  defined after
(\ref{msqud}).   Then, the  self-energy-type  terms $\big<  \partial^2
\widetilde{m}^2_{q_k}/\partial  a_i\partial a_j  \big>$,  with $q_k  =
t_1, b_1, t_2, b_2$ and $i,j = 1, 2$, are found to be
\begin{eqnarray}
  \label{selfaa}
\bigg< \frac{\partial^2 \widetilde{m}^2_{t_1 (t_2)}}{\partial
  a^2_1}\bigg> \!&=&\! \frac{g^2_w + g'^2}{8}\, +(-)\, 
\frac{1}{2 \big(m^2_{\tilde{t}_1} - m^2_{\tilde{t}_2}\big)}\,
\Big[\, x_t \Big( \widetilde{M}^2_Q - \widetilde{M}^2_t 
+ {\textstyle \frac{1}{2}} x_t v^2 \cos 2\beta \Big)\,
+\, 2|h_t|^2 |\mu|^2\,\Big]\, ,\nonumber\\
\bigg< \frac{\partial^2 \widetilde{m}^2_{b_1 (b_2)}}{\partial
  a^2_1}\bigg> \!&=&\! |h_b|^2\, -\, \frac{g^2_w + g'^2}{8}\, -(+)\, 
\frac{1}{2 \big(m^2_{\tilde{b}_1} - m^2_{\tilde{b}_2}\big)}\,
\Big[\, x_b \Big( \widetilde{M}^2_Q - \widetilde{M}^2_b 
- {\textstyle \frac{1}{2}} x_b v^2 \cos 2\beta \Big)\nonumber\\
&&-\, 2|h_b|^2 |A_b|^2\,\Big]\, ,\nonumber\\
\bigg< \frac{\partial^2 \widetilde{m}^2_{t_1 (t_2)}}{\partial
  a^2_2}\bigg> \!&=&\! |h_t|^2\, -\, \frac{g^2_w + g'^2}{8}\, -(+)\, 
\frac{1}{2 \big(m^2_{\tilde{t}_1} - m^2_{\tilde{t}_2}\big)}\,
\Big[\, x_t \Big( \widetilde{M}^2_Q - \widetilde{M}^2_t 
+ {\textstyle \frac{1}{2}} x_t v^2 \cos 2\beta \Big)\nonumber\\
&&-\, 2|h_t|^2 |A_t|^2\,\Big]\, ,\nonumber\\
\bigg< \frac{\partial^2 \widetilde{m}^2_{b_1 (b_2)}}{\partial
  a^2_2}\bigg> \!&=&\! \frac{g^2_w + g'^2}{8}\, +(-)\, 
\frac{1}{2 \big(m^2_{\tilde{b}_1} - m^2_{\tilde{b}_2}\big)}\,
\Big[\, x_b \Big( \widetilde{M}^2_Q - \widetilde{M}^2_b 
- {\textstyle \frac{1}{2}} x_b v^2 \cos 2\beta \Big)\, +\,
2|h_b|^2 |\mu|^2 \,\Big] ,\nonumber\\
\bigg< \frac{\partial^2 \widetilde{m}^2_{t_1 (t_2)}}{\partial
  a_1\,\partial a_2}\bigg> \!&=&\!  -(+)\, 
  \frac{|h_t|^2\, {\rm Re}\, (\mu A_t )}{m^2_{\tilde{t}_1} -
  m^2_{\tilde{t}_2}}\: +(-)\: \frac{2v_1v_2\,|h_t|^4  
{\rm Im}^2 (\mu A_t)}{(m^2_{\tilde{t}_1} -
  m^2_{\tilde{t}_2})^3}\ ,\nonumber\\   
\bigg< \frac{\partial^2 \widetilde{m}^2_{b_1 (b_2)}}{\partial
  a_1\, \partial a_2}\bigg> \!&=&\! -(+)\, 
  \frac{|h_b|^2\, {\rm Re}\, (\mu A_b )}{m^2_{\tilde{b}_1} -
  m^2_{\tilde{b}_2}}\: +(-)\: \frac{2v_1v_2\,|h_b|^4  
{\rm Im}^2 (\mu A_b)}{(m^2_{\tilde{b}_1} -
  m^2_{\tilde{b}_2})^3}\ .
\end{eqnarray}
In addition,  the non-vanishing CP-violating  self-energy terms $\big<
\partial^2  \widetilde{m}^2_{q_k}/\partial \phi_i\partial  a_j  \big>$ are
given by
\begin{eqnarray}
  \label{selfas}
\bigg< \frac{\partial^2 \widetilde{m}^2_{t_1 (t_2)}}{\partial
  \phi_1\,\partial a_2}\bigg> \!&=&\! 
+(-)\, \frac{|h_t|^2\, {\rm Im}\, (\mu A_t )}{m^2_{\tilde{t}_1} -
  m^2_{\tilde{t}_2}}\ \bigg\{\, 1\:
-\: \frac{v^2_1}{\big(m^2_{\tilde{t}_1} - m^2_{\tilde{t}_2}\big)^2}\,
\Big[\, x_t \Big( \widetilde{M}^2_Q - \widetilde{M}^2_t 
+ {\textstyle \frac{1}{2}} x_t v^2 \cos 2\beta \Big)\nonumber\\
&&-\, 2|h_t|^2\Big( {\rm Re}\,(\mu A_t) \tan\beta - |\mu|^2
  \Big)\,\Big]\, \bigg\}\, ,\nonumber\\
\bigg< \frac{\partial^2 \widetilde{m}^2_{b_1 (b_2)}}{\partial
  \phi_1\,\partial a_2}\bigg> \!&=&\! +(-)\, 
\frac{|h_b|^2\, {\rm Im}\, (\mu A_b )}{m^2_{\tilde{b}_1} -
  m^2_{\tilde{b}_2}}\ \bigg\{\, 1\:
+\: \frac{v^2_1}{\big(m^2_{\tilde{b}_1} - m^2_{\tilde{b}_2}\big)^2}\,
\Big[\, x_b \Big( \widetilde{M}^2_Q - \widetilde{M}^2_b 
- {\textstyle \frac{1}{2}} x_b v^2 \cos 2\beta \Big)\nonumber\\
&&+\, 2|h_b|^2\Big( {\rm Re}\,(\mu A_b) \tan\beta - |A_b|^2
  \Big)\,\Big]\, \bigg\}\, ,\nonumber\\
\bigg< \frac{\partial^2 \widetilde{m}^2_{t_1 (t_2)}}{\partial
  \phi_2\,\partial a_1}\bigg>  \!&=&\! -(+)\, 
\frac{|h_t|^2\, {\rm Im}\, (\mu A_t )}{m^2_{\tilde{t}_1} -
  m^2_{\tilde{t}_2}}\ \bigg\{\, 1\:
+\: \frac{v^2_2}{\big(m^2_{\tilde{t}_1} - m^2_{\tilde{t}_2}\big)^2}\,
\Big[\, x_t \Big( \widetilde{M}^2_Q - \widetilde{M}^2_t 
+ {\textstyle \frac{1}{2}} x_t v^2 \cos 2\beta \Big)\nonumber\\
&&+\, 2|h_t|^2\Big( {\rm Re}\,(\mu A_t) \cot\beta - |A_t|^2
  \Big)\,\Big]\, \bigg\}\, , \nonumber\\
\bigg< \frac{\partial^2 \widetilde{m}^2_{b_1 (b_2)}}{\partial
  \phi_2\,\partial a_1}\bigg> \!&=&\! -(+)\, 
\frac{|h_b|^2\, {\rm Im}\, (\mu A_b )}{m^2_{\tilde{b}_1} -
  m^2_{\tilde{b}_2}}\ \bigg\{\, 1\: -\:
\frac{v^2_2}{\big(m^2_{\tilde{b}_1} - m^2_{\tilde{b}_2}\big)^2}\,
\Big[\, x_b \Big( \widetilde{M}^2_Q - \widetilde{M}^2_b 
- {\textstyle \frac{1}{2}} x_b v^2 \cos 2\beta \Big)\nonumber\\
&&-\, 2|h_b|^2\Big( {\rm Re}\,(\mu A_b) \cot\beta - |\mu|^2
  \Big)\,\Big] \, \bigg\}\, .
\end{eqnarray}
Finally,   the  CP-conserving   self-energy-type   derivatives  $\big<
\partial^2  \widetilde{m}^2_{q_k}/\partial  \phi_i\partial  \phi_j  \big>$
have been calculated to be
\begin{eqnarray}
  \label{selfss}
\bigg< \frac{\partial^2 \widetilde{m}^2_{t_1 (t_2)}}{\partial
  \phi^2_1 }\bigg> \!&=&\! \frac{g^2_w + g'^2}{8}\,
+(-)\, 
\frac{1}{2 \big(m^2_{\tilde{t}_1} - m^2_{\tilde{t}_2}\big)}\,
\Big[\, x_t \Big( \widetilde{M}^2_Q - \widetilde{M}^2_t 
+ {\textstyle \frac{1}{2}} x_t v^2 (1+2\cos 2\beta) \Big)\nonumber\\
&&+\, 2|h_t|^2 |\mu|^2\,\Big]\, 
-(+)\, \frac{v^2_1}{2 \big(m^2_{\tilde{t}_1} - m^2_{\tilde{t}_2}\big)^3}\,
\Big[\, x_t \Big( \widetilde{M}^2_Q - \widetilde{M}^2_t 
+ {\textstyle \frac{1}{2}} x_t v^2 \cos 2\beta \Big)\nonumber\\
&&-\, 2|h_t|^2\Big( {\rm Re}\,(\mu A_t) \tan\beta - |\mu|^2
  \Big)\,\Big]^2\, ,\nonumber\\
\bigg< \frac{\partial^2 \widetilde{m}^2_{b_1 (b_2)}}{\partial
  \phi^2_1}\bigg> 
 \!&=&\! |h_b|^2\, -\, \frac{g^2_w + g'^2}{8}\, -(+)\, 
\frac{1}{2 \big(m^2_{\tilde{b}_1} - m^2_{\tilde{b}_2}\big)}\,
\Big[\, x_b \Big( \widetilde{M}^2_Q - \widetilde{M}^2_b 
- {\textstyle \frac{1}{2}} x_b v^2 (1+2\cos 2\beta) \Big)\nonumber\\
&&-\, 2|h_b|^2 |A_b|^2 \,\Big]\,
-(+)\, \frac{v^2_1}{2 \big(m^2_{\tilde{b}_1} - m^2_{\tilde{b}_2}\big)^3}\,
\Big[\, x_b \Big( \widetilde{M}^2_Q - \widetilde{M}^2_b 
- {\textstyle \frac{1}{2}} x_b v^2 \cos 2\beta \Big)\nonumber\\
&&+\, 2|h_b|^2\Big( {\rm Re}\,(\mu A_b) \tan\beta - |A_b|^2
  \Big)\,\Big]^2\, ,\nonumber\\
\bigg< \frac{\partial^2 \widetilde{m}^2_{t_1 (t_2)}}{\partial
  \phi^2_2}\bigg> \!&=&\! |h_t|^2\, -\, \frac{g^2_w + g'^2}{8}\,
-(+)\, 
\frac{1}{2 \big(m^2_{\tilde{t}_1} - m^2_{\tilde{t}_2}\big)}\,
\Big[\, x_t \Big( \widetilde{M}^2_Q - \widetilde{M}^2_t 
+ {\textstyle \frac{1}{2}} x_t v^2 (2\cos 2\beta -1) \Big)\nonumber\\
&&-\, 2|h_t|^2 |A_t|^2\,\Big]\,
-(+)\, \frac{v^2_2}{2 \big(m^2_{\tilde{t}_1} - m^2_{\tilde{t}_2}\big)^3}\,
\Big[\, x_t \Big( \widetilde{M}^2_Q - \widetilde{M}^2_t 
+ {\textstyle \frac{1}{2}} x_t v^2 \cos 2\beta \Big)\nonumber\\
&&+\, 2|h_t|^2\Big( {\rm Re}\,(\mu A_t) \cot\beta - |A_t|^2
  \Big)\,\Big]^2\, ,\nonumber\\
\bigg< \frac{\partial^2 \widetilde{m}^2_{b_1 (b_2)}}{\partial
  \phi^2_2}\bigg> \!&=&\! \frac{g^2_w + g'^2}{8}\, +(-)\, 
\frac{1}{2 \big(m^2_{\tilde{b}_1} - m^2_{\tilde{b}_2}\big)}\,
\Big[\, x_b \Big( \widetilde{M}^2_Q - \widetilde{M}^2_b 
- {\textstyle \frac{1}{2}} x_b v^2 (2\cos 2\beta - 1) \Big)\nonumber\\
&&+\, 2|h_b|^2 |\mu|^2 \,\Big]\, 
-(+)\, \frac{v^2_2}{2 \big(m^2_{\tilde{b}_1} - m^2_{\tilde{b}_2}\big)^3}\,
\Big[\, x_b \Big( \widetilde{M}^2_Q - \widetilde{M}^2_b 
- {\textstyle \frac{1}{2}} x_b v^2 \cos 2\beta \Big)\nonumber\\
&&-\, 2|h_b|^2\Big( {\rm Re}\,(\mu A_b) \cot\beta - |\mu|^2
  \Big)\,\Big]^2\, ,\nonumber\\
\bigg< \frac{\partial^2 \widetilde{m}^2_{t_1 (t_2)}}{\partial
  \phi_1\,\partial \phi_2 }\bigg> \!&=&\! 
-(+)\, \frac{1}{2 \big(m^2_{\tilde{t}_1} - m^2_{\tilde{t}_2}\big)}\,
\Big[\, {\textstyle \frac{1}{2}} x^2_t v^2 \sin 2\beta\, 
+\, 2|h_t|^2 {\rm Re}\, (\mu A_t)\,\Big]\, +(-)\, 
\frac{v_1 v_2}{2 \big(m^2_{\tilde{t}_1} -
  m^2_{\tilde{t}_2}\big)^3}\nonumber\\
&&\times\, \Big[\, x_t \Big( \widetilde{M}^2_Q - \widetilde{M}^2_t 
+ {\textstyle \frac{1}{2}} x_t v^2 \cos 2\beta \Big)\, -\, 
2|h_t|^2\Big( {\rm Re}\,(\mu A_t) \tan\beta - |\mu|^2
  \Big)\,\Big]\nonumber\\
&&\times\, \Big[\, x_t \Big( \widetilde{M}^2_Q - \widetilde{M}^2_t 
+ {\textstyle \frac{1}{2}} x_t v^2 \cos 2\beta \Big)\, 
+\, 2|h_t|^2\Big( {\rm Re}\,(\mu A_t) \cot\beta - |A_t|^2
  \Big)\,\Big]\, ,\nonumber\\
\bigg< \frac{\partial^2 \widetilde{m}^2_{b_1 (b_2)}}{\partial
  \phi_1\, \partial \phi_2}\bigg>  \!&=&\!
-(+)\, \frac{1}{2 \big(m^2_{\tilde{b}_1} - m^2_{\tilde{b}_2}\big)}\,
\Big[\, {\textstyle \frac{1}{2}} x^2_b v^2 \sin 2\beta\, 
+\, 2|h_b|^2 {\rm Re}\, (\mu A_b)\,\Big]\, 
+(-)\, \frac{v_1 v_2}{2 \big(m^2_{\tilde{b}_1} -
  m^2_{\tilde{b}_2}\big)^3}\nonumber\\
&&\times\, \Big[\, x_b \Big( \widetilde{M}^2_Q - \widetilde{M}^2_b 
- {\textstyle \frac{1}{2}} x_b v^2 \cos 2\beta \Big)\, +\, 
2|h_b|^2\Big( {\rm Re}\,(\mu A_b) \tan\beta - |A_b|^2
  \Big)\,\Big]\nonumber\\
&&\times\, \Big[\, x_b \Big( \widetilde{M}^2_Q - \widetilde{M}^2_b 
- {\textstyle \frac{1}{2}} x_b v^2 \cos 2\beta \Big)\,
-\, 2|h_b|^2\Big( {\rm Re}\,(\mu A_b) \cot\beta - |\mu|^2
  \Big)\,\Big]\, .
\end{eqnarray}

\subsection{Derivatives of Squark Masses with Respect to \\
Charged Higgs Fields}

Here we evaluate  the derivatives   of the  field-dependent
squark masses with respect to charged Higgs fields. To calculate
the    expressions  $\big<  \partial^2  \widetilde{m}^2_{q_k}/\partial
\phi^+_i\partial \phi^-_j \big>$ directly turns out to be a formidable
task.  The reason is that  $\widetilde{m}^2_{q_k}$ are the eigenvalues
of a  non-trivial $(4\times  4)$ squark  mass matrix
$\widetilde{\cal M}^2$ (cf. (\ref{Msquark})), and their analytic
form is very complicated.  Therefore,  we proceed  differently,
using a mathematical trick which was first applied in \cite{BERZ}.

First,  we  notice  that  $\big<  \partial  \widetilde{m}^2_{q_k}/\partial
\phi^\pm_i\big>  = 0$,  as a  consequence of  the fact  that  the true
ground state of the  effective potential should conserve charge. Then,
one may make use of the eigenvalue equation:
\begin{equation}
  \label{det}
{\rm det}\, (\widetilde{\cal M}^2 - \widetilde{m}^2_{q_k} {\bf 1}_4)\ =\ 
\widetilde{m}^8_{q_k}\, +\, A \widetilde{m}^6_{q_k}\, +\, B
\widetilde{m}^4_{q_k}\, +\, C \widetilde{m}^2_{q_k}\, +\, D\ =\ 0\,,
\end{equation}
with 
\begin{eqnarray}
  \label{ABCD}
A &=& - {\rm Tr}\, \widetilde{\cal M}^2\, ,\nonumber\\
B &=& \frac{1}{2}\, \Big(\, {\rm Tr}^2 \widetilde{\cal M}^2\, -\,
{\rm Tr}\, \widetilde{\cal M}^4\, \Big)\, ,\nonumber\\
C &=& \frac{1}{3}\, \Big(\, {\rm Tr}^3 \widetilde{\cal M}^2\, -\,
{\rm Tr}\, \widetilde{\cal M}^6\, \Big)\ -\ \frac{1}{2}\, 
{\rm Tr}\, \widetilde{\cal M}^2\, \Big(\, 
{\rm Tr}^2 \widetilde{\cal M}^2\, -\, {\rm Tr}\, \widetilde{\cal
  M}^4\, \Big)\, ,\nonumber\\
D &=& {\rm det}\, \widetilde{\cal M}^2\ =\ 
-\, \frac{1}{4}\, \Big(\, {\rm Tr}\, \widetilde{\cal M}^8\, +\, 
A\, {\rm Tr}\, \widetilde{\cal M}^6\, +\, B\, {\rm Tr}\,\widetilde{\cal M}^4\,
+\, C\, {\rm Tr}\, \widetilde{\cal M}^2\, \Big)\, ,
\end{eqnarray}
to obtain
\begin{equation}
  \label{phij}
\bigg< \frac{\partial^2 \widetilde{m}^2_{q_k}}{
\partial \phi^+_i \partial \phi^-_j }\bigg>\ =\ -\, \bigg<
\frac{A_{ij} \widetilde{m}^6_{q_k}\, +\, B_{ij} \widetilde{m}^4_{q_k}\,
+\, C_{ij} \widetilde{m}^2_{q_k}\, +\, D_{ij} }{
\prod\limits_{q_l \not= q_k}\, (\widetilde{m}^2_{q_k} - 
\widetilde{m}^2_{q_l} )}
\bigg>\, ,
\end{equation}
with $q_l,q_k = t_1,b_1,t_2,b_2$,  and $A_{ij} = \partial^2 A/\partial
\phi^+_i \partial \phi^-_j$,  $B_{ij} = \partial^2 B/\partial \phi^+_i
\partial  \phi^-_j$,  etc..  For  our  purposes,  it  is sufficient  to
calculate the  derivatives with respect to $\phi^+_1$  and $\phi^-_2$. 
In particular, it proves convenient  to use a representation in which
the  columns and  rows `2'  and `3'  of the  $(4\times 4)$
matrix $\widetilde{\cal  M}^2$ have  been interchanged.  With such a
reordering, we find
\begin{equation}
  \label{repr}
\big<\widetilde{\cal M}^2\big> \ \to \ 
\left( \begin{array}{cc} \widetilde{\cal M}^2_t & 0 \\
0 & \widetilde{\cal M}^2_b \end{array} \right)\, ,\quad
\bigg< \frac{\partial \widetilde{\cal M}^2}{\partial \phi^+_1}\bigg>\
\to \left( \begin{array}{cc} 0 & \widetilde{\cal M}_+ \\
0 & 0 \end{array} \right)\, ,\quad
\bigg< \frac{\partial \widetilde{\cal M}^2}{\partial \phi^-_2}\bigg>\
\to \left( \begin{array}{cc} 0 & 0 \\
\widetilde{\cal M}_- & 0 \end{array} \right)\, ,
\end{equation}
where  $\widetilde{\cal M}^2_t$ and  $\widetilde{\cal M}^2_b$  are the
usual  $\tilde{t}$-  and  $\tilde{b}$-  $(2\times  2)$-mass  matrices,
respectively, and
\begin{equation}
  \label{Mder}
\widetilde{\cal M}_+\ =\ \left( \begin{array}{cc}
\frac{1}{\sqrt{2}}\,  (|h_b|^2 - \frac{1}{2} g^2_w )\, v_1 & h_b^* A^*_b \\
h_t \mu^* & \frac{1}{\sqrt{2}}\, h_t h_b^* v_2 \end{array} \right)\, ,\quad
\widetilde{\cal M}_- \ =\ - \left( \begin{array}{cc}
\frac{1}{\sqrt{2}}\, (|h_t|^2 - \frac{1}{2} g^2_w )\, v_2  & h_t^* A^*_t \\
h_b \mu^* & \frac{1}{\sqrt{2}}\, h_t^* h_b v_1 \end{array} \right)\, .
\end{equation}
Then,  the  relevant  coefficients  $A_{12}$, $B_{12}$,  $C_{12}$  and
$D_{12}$ may be expressed in a compact form as follows:
\begin{eqnarray}
  \label{ABCD12}
\big< A_{12} \big>  &=& 0\, ,\nonumber\\
\big< B_{12} \big>  &=& -\, {\rm Tr}\, (\widetilde{\cal M}_+
\widetilde{\cal M}_- )\, ,\nonumber\\
\big< C_{12} \big>  &=& \Big(\, {\rm Tr}\, \widetilde{\cal M}^2_t\,
+\, {\rm Tr}\, \widetilde{\cal M}^2_b\,\Big)\, 
{\rm Tr}\, (\widetilde{\cal M}_+
\widetilde{\cal M}_- )\, -\, {\rm Tr}\, ( \widetilde{\cal M}^2_t
\widetilde{\cal M}_+ \widetilde{\cal M}_- )\, -\,
{\rm Tr}\, ( \widetilde{\cal M}^2_b 
\widetilde{\cal M}_- \widetilde{\cal M}_+ )\, ,\nonumber\\
\big< D_{12} \big>  &=& -\, {\rm Tr}\, ( \widetilde{\cal M}^2_t
\widetilde{\cal M}_+ \widetilde{\cal M}^2_b \widetilde{\cal M}_- )\,
-\,  {\rm Tr}\, ( \widetilde{\cal M}^4_t
\widetilde{\cal M}_+ \widetilde{\cal M}_- )\,
-\,  {\rm Tr}\, ( \widetilde{\cal M}^4_b
\widetilde{\cal M}_- \widetilde{\cal M}_+ )\nonumber\\
&&+\,  \Big(\, {\rm Tr}\,\widetilde{\cal M}^2_t\, +\,
{\rm Tr}\,\widetilde{\cal M}^2_b\, \Big)\, \Big[\, 
{\rm Tr}\, ( \widetilde{\cal M}^2_t
\widetilde{\cal M}_+ \widetilde{\cal M}_- )\, +\,
{\rm Tr}\, ( \widetilde{\cal M}^2_b
\widetilde{\cal M}_- \widetilde{\cal M}_+ )\, \Big]\nonumber\\
&&+\, \frac{1}{2}\, \Big[\, {\rm Tr}\, \widetilde{\cal M}^4_t\, +\,
{\rm Tr}\, \widetilde{\cal M}^4_b\, -\, 
\Big( {\rm Tr}\,\widetilde{\cal M}^2_t\, +\,
{\rm Tr}\,\widetilde{\cal M}^2_b\, \Big)^2\,\Big]\, {\rm Tr}\, 
(\widetilde{\cal M}_+ \widetilde{\cal M}_-)\, .
\end{eqnarray}
With the help of (\ref{ABCD12}), it is straightforward to obtain
the  derivatives   $\partial^2  \widetilde{m}^2_{q_k}  /\partial  \phi^+_1
\partial \phi^-_2$. More explicitly, we have
\begin{eqnarray}
  \label{derphi+}
\bigg< \frac{\partial^2 \widetilde{m}^2_{t_1}}{
\partial \phi^+_1 \partial \phi^-_2 }\bigg> &=& 
-\, \frac{ \big< B_{12} \big>\, m^4_{\tilde{t}_1}\: +\: 
\big< C_{12} \big>\, m^2_{\tilde{t}_1}\: +\: \big< D_{12} \big> }
{\big( m^2_{\tilde{t}_1} - m^2_{\tilde{b}_1}\big)\, 
\big( m^2_{\tilde{t}_1} - m^2_{\tilde{t}_2}\big)\, 
\big( m^2_{\tilde{t}_1} - m^2_{\tilde{b}_2}\big)}\ ,\nonumber\\
\bigg< \frac{\partial^2 \widetilde{m}^2_{t_2}}{
\partial \phi^+_1 \partial \phi^-_2 }\bigg> &=& 
-\, \frac{ \big< B_{12} \big>\, m^4_{\tilde{t}_2}\: +\: 
\big< C_{12} \big>\, m^2_{\tilde{t}_2}\: +\: \big< D_{12} \big> }
{\big( m^2_{\tilde{t}_2} - m^2_{\tilde{b}_1}\big)\, 
\big( m^2_{\tilde{t}_2} - m^2_{\tilde{t}_1}\big)\, 
\big( m^2_{\tilde{t}_2} - m^2_{\tilde{b}_2}\big)}\ ,\nonumber\\
\bigg< \frac{\partial^2 \widetilde{m}^2_{b_1}}{
\partial \phi^+_1 \partial \phi^-_2 }\bigg> &=& 
-\, \frac{ \big< B_{12} \big>\, m^4_{\tilde{b}_1}\: +\: 
\big< C_{12} \big>\, m^2_{\tilde{b}_1}\: +\: \big< D_{12} \big> }
{\big( m^2_{\tilde{b}_1} - m^2_{\tilde{t}_1}\big)\, 
\big( m^2_{\tilde{b}_1} - m^2_{\tilde{t}_2}\big)\, 
\big( m^2_{\tilde{b}_1} - m^2_{\tilde{b}_2}\big)}\ ,\nonumber\\
\bigg< \frac{\partial^2 \widetilde{m}^2_{b_2}}{
\partial \phi^+_1 \partial \phi^-_2 }\bigg> &=& 
-\, \frac{ \big< B_{12} \big>\, m^4_{\tilde{b}_2}\: +\: 
\big< C_{12} \big>\, m^2_{\tilde{b}_2}\: +\: \big< D_{12} \big> }
{\big( m^2_{\tilde{b}_2} - m^2_{\tilde{b}_1}\big)\, 
\big( m^2_{\tilde{b}_2} - m^2_{\tilde{t}_2}\big)\, 
\big( m^2_{\tilde{b}_2} - m^2_{\tilde{t}_1}\big)}\ .
\end{eqnarray}

\setcounter{equation}{0}
\section{Higgs-Boson Masses and Mixing Angles}

Here we present analytic expressions for the Higgs-boson masses
$M_{H_i}  (m_t)$  ($i=1,2,3$)  and  the  corresponding  $(3\times  3)$
orthogonal matrix $O$, after diagonalizing the RG-improved Higgs-boson
mass matrix $({\cal M}^2_N) (m_t)$.

For  notational   simplicity,  we  do  not display  explicitly  the
functional dependence of ${\cal M}^2_N$ on $m_t$. The mass eigenvalues
of  the  $(3\times  3)$  matrix ${\cal  M}^2_N$  are  then
obtained by solving the characteristic equation of cubic order:
\begin{equation}
  \label{lcubic}
x^3\ +\ r x^2 \ +\ s x\ +\ t\ =\ 0\, ,
\end{equation}
with 
\begin{eqnarray}
  \label{trace}
r & = & -\, {\rm Tr} ({\cal M}^2_N )\, ,\nonumber\\
s & = & \frac{1}{2}\, \Big[\, {\rm Tr}^2 ({\cal M}^2_N )\ -\
{\rm Tr} ({\cal M}^4_N )\, \Big]\, ,\nonumber\\
t &=& -\, {\rm det} ({\cal M}^2_N )\, . 
\end{eqnarray}
To  this   end, it proves  useful to    define the following auxiliary
parameters:
\begin{eqnarray}
  \label{p}
p &=& \frac{3s\, -\, r^2}{3}\ ,\nonumber\\
  \label{q}
q &=& \frac{2r^3}{27}\ -\ \frac{rs}{3}\, +\, t\, ,\nonumber\\
  \label{D}  
D &=& \frac{p^3}{27}\ +\ \frac{q^2}{4}\, .
\end{eqnarray}
To ensure that the three eigenvalues are positive, it is necessary and
sufficient to require that
\begin{equation}
  \label{condition}
D\ <\ 0\, ,\qquad r\ <\ 0\, ,\qquad s\ >\ 0\, ,\qquad t\ <\ 0\, .   
\end{equation}
Imposing these inequalities on the kinematic parameters of the theory,
we may express the three mass eigenvalues of ${\cal M}^2_N$ as
\begin{eqnarray}
  \label{e1}
M^2_{H_1} & =& -\frac{1}{3}\, r\ +\ 2\, \sqrt{-p/3}\, 
\cos\Big(\, \frac{\varphi}{3}\, +\, \frac{2\pi}{3}\, \Big)\, ,\nonumber\\
M^2_{H_2} & =& -\frac{1}{3}\, r\ +\ 2\, \sqrt{-p/3}\, 
\cos\Big(\, \frac{\varphi}{3}\, -\, \frac{2\pi}{3}\,\Big)\, ,\nonumber\\
M^2_{H_3} & =& -\frac{1}{3}\, r\ +\ 2\, \sqrt{-p/3}\, 
\cos\Big(\, \frac{\varphi}{3}\, \Big)\, ,
\end{eqnarray}
with
\begin{equation}
  \label{phi}
\varphi \ =\  {\rm arccos}\, \Big( - \frac{q}{2\sqrt{-p^3/27}}
\Big)\qquad {\rm and}\qquad 0 \le \varphi \le \pi\, .
\end{equation}
Since the Higgs-boson mass matrix  ${\cal M}^2_N$ is symmetric, we can
diagonalize it  by means  of an orthogonal  rotation $O$ as  stated in
(\ref{Odiag}).  Furthermore, one  can show \cite{costas} that the
Higgs-boson mass  eigenvalues in (\ref{e1}) satisfy  the desired
mass   hierarchy   in  accordance   with   the   inequality  of 
(\ref{masdef}).

If $M^2_{ij}$,  with   $i,j=1, 2,  3$, denote the   matrix elements of
${\cal   M}^2_N$,   the elements $O_{ij}$  can     then be obtained by
appropriately solving the underdetermined coupled system of equations,
$\sum_k M^2_{ik} O_{kj} = M^2_{H_j} O_{ij}$:
\begin{eqnarray}
  \label{system}
(M^2_{11} - M^2_{H_i} ) O_{1i}\, +\, M^2_{12} O_{2i}\, +\, 
M^2_{13} O_{3i} & =& 0\,, \nonumber\\ 
M^2_{21} O_{1i}\, +\, ( M^2_{22} - M^2_{H_i} ) O_{2i}\, +\, 
M^2_{23} O_{3i} & =& 0\,, \nonumber\\ 
M^2_{31} O_{1i}\, +\, M^2_{32}  O_{2i}\, +\, 
( M^2_{33} - M^2_{H_i} ) O_{3i} & =& 0\, . 
\end{eqnarray}
More explicitly, we have
\begin{equation}
  \label{Oij}
O \ =\   \left( \begin{array}{ccc}
|x_1|/\Delta_1 & x_2/\Delta_2 & x_3/\Delta_3 \\
y_1/\Delta_1 & |y_2|/\Delta_2 & y_3/\Delta_3 \\
z_1/\Delta_1 & z_2/\Delta_2 & |z_3|/\Delta_3 \end{array} \right)\, ,
\end{equation}
where
\begin{equation}
  \label{Deltas}
\Delta_i = \sqrt{ x^2_i\, +\, y^2_i\, +\, z^2_i} 
\end{equation}
and 
{\small \begin{eqnarray}
  \label{xyz}
|x_1| \!\!\!&=&\!\!\! \left|\!\left| \begin{array}{lr}
M^2_{22}-M^2_{H_1} & M^2_{23} \\
M^2_{32} & \hspace{-1.2cm} M^2_{33} - M^2_{H_1} \end{array}\right|\!\right|
,\
y_1 = {\rm s}_{x_1}\! \left| \begin{array}{cc}
M^2_{23} & M^2_{21} \\
M^2_{33} -M^2_{H_1} & M^2_{31} \end{array} \right|,\
z_1 = {\rm s}_{x_1}\! \left| \begin{array}{cc}
M^2_{21} & M^2_{22} - M^2_{H_1} \\
M^2_{31} & M^2_{32} \end{array} \right|,\nonumber\\
x_2 \!\!\!&=&\!\!\! {\rm s}_{y_2}\! \left| \begin{array}{cc}
M^2_{13} & M^2_{12} \\
M^2_{33} - M^2_{H_2} & M^2_{32} \end{array} \right|,\
|y_2| = \left|\!\left| \begin{array}{lr}
M^2_{11} - M^2_{H_2} & M^2_{13} \\
M^2_{31}  & \hspace{-1.2cm} M^2_{33} - M^2_{H_2} \end{array} 
\right|\!\right|,\ 
z_2 = {\rm s}_{y_2}\! \left| \begin{array}{cc}
M^2_{12} & M^2_{11} - M^2_{H_2} \\
M^2_{32} & M^2_{31} \end{array} \right|,\nonumber\\
x_3 \!\!\!&=&\!\!\! {\rm s}_{z_3}\! \left| \begin{array}{cc}
M^2_{12} & M^2_{13} \\
M^2_{22} - M^2_{H_3} & M^2_{23} \end{array} \right|,\
y_3 = {\rm s}_{z_3}\! \left| \begin{array}{cc}
M^2_{13} & M^2_{11} - M^2_{H_3} \\
M^2_{23} & M^2_{21} \end{array} \right|,\ 
|z_3| = \left|\!\left| \begin{array}{lr}
M^2_{11} - M^2_{H_3} & M^2_{12} \\
M^2_{21} & \hspace{-1.2cm} M^2_{22} - M^2_{H_3} \end{array}
\right|\!\right|.\nonumber\\ 
&&
\end{eqnarray}} 
In (\ref{xyz}), the abbreviation ${\rm s}_x  \equiv {\rm sign}\,
(x)$ is an  operation that simply gives the sign  of a real expression
$x$.

\end{appendix}

%
%
%
%
%
%
%
%
%
%
%
%
%
%
%
%

\end{document}